\documentclass[hyper]{JHEP3}

\usepackage{axodraw}
\usepackage[dvips]{graphicx}
\usepackage{epsfig}
\usepackage{multicol}
\usepackage{bbm}
\usepackage{amsmath}
\usepackage{cite}
\usepackage{lscape}
\usepackage{multirow}

\newcommand{\beq}{\begin{equation}}
\newcommand{\eeq}{\end{equation}}
\newcommand{\bea}{\begin{eqnarray}}
\newcommand{\eea}{\end{eqnarray}}

\newcommand{\lsim}{
\mathrel{\hbox{\rlap{\hbox{\lower4pt\hbox{$\sim$}}}\hbox{$<$}}}}
\newcommand{\gsim}{
\mathrel{\hbox{\rlap{\hbox{\lower4pt\hbox{$\sim$}}}\hbox{$>$}}}}

\title{Precise reconstruction of sparticle masses without ambiguities}

\author{
Konstantin T.~Matchev${}^\dagger$, Filip Moortgat${}^\ddagger$, Luc Pape${}^\ddagger$, Myeonghun Park${}^\dagger$ \\
${}^\dagger$ Physics Department, University of Florida, Gainesville, FL 32611, USA \\
${}^\ddagger$ETH Zurich, Zurich, Switzerland
}

\abstract{We critically reexamine the standard applications of
the method of kinematical endpoints for sparticle mass determination.
We consider the typical decay chain in supersymmetry (SUSY)
$\tilde q\to \tilde\chi^0_2\to \tilde\ell\to\tilde\chi^0_1$,
which yields a jet $j$, and two leptons $\ell_n^\pm$ and $\ell_f^\mp$.
The conventional approaches use the upper 
kinematical endpoints of the individual distributions $m_{j\ell\ell}$, 
$m_{j\ell(lo)}=\min\{m_{j\ell_n},m_{j\ell_f}\}$ and
$m_{j\ell(hi)}=\max\{m_{j\ell_n},m_{j\ell_f}\}$, 
all three of which suffer from parameter space region ambiguities
and may lead to multiple solutions for the SUSY mass spectrum.
In contrast, we do not use $m_{j\ell\ell}$, 
$m_{j\ell(lo)}$ and $m_{j\ell(hi)}$, and instead
propose a new set of (infinitely many) variables 
whose upper kinematic endpoints exhibit reduced sensitivity
to the parameter space region. We then outline an alternative, 
much simplified procedure for obtaining the SUSY mass spectrum. 
In particular, we show that the four endpoints observed in the three distributions
$m^2_{\ell\ell}$, $m^2_{j\ell_n}\cup m^2_{j\ell_f}$ and $m^2_{j\ell_n}+m^2_{j\ell_f}$
are sufficient to completely pin down the squark mass $m_{\tilde q}$ 
and the two neutralino masses $m_{\tilde \chi^0_2}$ and $m_{\tilde \chi^0_1}$,
leaving only a discrete 2-fold ambiguity for the slepton mass $m_{\tilde \ell}$.
This remaining ambiguity can be easily resolved in a number of different ways:
for example, by a single additional measurement of the kinematic endpoint of 
any one out of the many remaining 1-dimensional distributions 
at our disposal, or by exploring the correlations in the 2-dimensional
distribution of $m^2_{j\ell_n}\cup m^2_{j\ell_f}$ versus $m^2_{\ell\ell}$.  
We illustrate our 
method with two examples: the LM1 and LM6 CMS study points.
An additional advantage of our method is the expected improvement
in the accuracy of the SUSY mass determination, due to 
the multitude and variety of available measurements.
}


\preprint{UFIFT-HEP-09-04 \\
          August 17, 2009
          } 

\begin{document}

\section{Introduction}
\label{sec:introduction}

SUSY is a primary target of the LHC searches for new physics
beyond the Standard Model (BSM). In SUSY models with conserved $R$-parity 
the superpartners are produced in pairs and each one decays through a
cascade decay chain down to the lightest superpartner (LSP).
If the LSP is the lightest neutralino $\tilde\chi^0_1$, it
escapes detection, making it rather difficult to reconstruct
directly the preceding superpartners and thus measure their masses
and spins. In recognition of this fact, in recent years 
there has been an increased interest in developing new techniques for mass
\cite{Hinchliffe:1996iu,Lester:1999tx,
Bachacou:1999zb,Hinchliffe:1999zc,atlas,Nojiri:2000wq,Allanach:2000kt,Barr:2003rg,Nojiri:2003tu,lester,
Kawagoe:2004rz,Gjelsten:2004ki,Gjelsten:2005aw,Birkedal:2005cm,Lester:2005je,Miller:2005zp,
Meade:2006dw,Luc,Lester:2006yw,Lester:2006cf,Gjelsten:2006tg,Matsumoto:2006ws,Cheng:2007xv,Lester:2007fq,
Cho:2007qv,Gripaios:2007is,Barr:2007hy,Cho:2007dh,Ross:2007rm,Nojiri:2007pq,Huang:2008ae,
Nojiri:2008hy,Tovey:2008ui,Nojiri:2008ir,Cheng:2008mg,Cho:2008cu,Serna:2008zk,Bisset:2008hm,
Barr:2008ba,Kersting:2008qn,Nojiri:2008vq,Cheng:2008hk,Burns:2008va,Barr:2008hv,Konar:2008ei,
Costanzo:2009mq,Burns:2009zi,Alwall:2009zu,Cheng:2009fw}
and spin \cite{Barr:2004ze,Battaglia:2005zf,Smillie:2005ar,Battaglia:2005ma,Datta:2005zs,
Datta:2005vx,Barr:2005dz,Alves:2006df,Athanasiou:2006ef,Wang:2006hk,Athanasiou:2006hv,SA:2006jm,
Smillie:2006cd,Kong:2006pi,Kilic:2007zk,Alves:2007xt,Csaki:2007xm,Datta:2007xy,Buckley:2007th,
Buckley:2008pp,Kane:2008kw,Burns:2008cp,Cho:2008tj,Graesser:2008qi,Hallenbeck:2008hf,Gedalia:2009ym,Boudjema:2009fz}
measurements in such SUSY-like missing energy events.

Roughly speaking, there are three basic types of mass determination 
methods in SUSY\footnote{For a recent study representative of each method, 
see Refs.~\cite{Burns:2008va,Burns:2009zi,Cheng:2009fw}.}.
In this paper we concentrate on the classic method of
kinematical endpoints \cite{Hinchliffe:1996iu}. 
Following the previous SUSY studies,
for illustration of our results we shall use the generic decay chain 
$D\rightarrow{}jC\rightarrow{}j\ell_n^{\pm}B\rightarrow{}j\ell_n^{\pm}\ell_f^{\mp}A$
shown in Fig.~\ref{fig:chain}.
\FIGURE[t]{
{
\unitlength=1.5 pt
\SetScale{1.5}
\SetWidth{1.0}      
\normalsize    
{} \qquad\allowbreak
\begin{picture}(300,100)(0,50)
\SetColor{Gray}
\Line(100,80)(120,120)
\Line(150,80)(170,120)
\Line(200,80)(220,120)
\SetColor{Red}
\Text( 75, 86)[c]{\Red{$D$}}
\Text(125, 86)[c]{\Red{$C$}}
\Text(175, 86)[c]{\Red{$B$}}
\Text(225, 86)[c]{\Red{$A$}}
\SetWidth{1.2}      
\Line(50,80)(250,80)
\Text(100, 65)[c]{\Black{$R_{CD}=\frac{m_C^2}{m_D^2}$}}
\Text(150, 65)[c]{\Black{$R_{BC}=\frac{m_B^2}{m_C^2}$}}
\Text(200, 65)[c]{\Black{$R_{AB}=\frac{m_A^2}{m_B^2}$}}
\Text(120,129)[c]{\Black{$j$}}
\Text(170,129)[c]{\Black{$\ell_n$}}
\Text(220,129)[c]{\Black{$\ell_f$}}
\SetColor{Blue}
\Vertex(100,80){2}
\Vertex(150,80){2}
\Vertex(200,80){2}
\scriptsize
\end{picture}
}
\caption{The typical cascade decay chain under consideration in this paper.
Here $D$, $C$, $B$ and $A$ are new BSM particles, while the
corresponding SM decay products are: a QCD jet $j$, 
a ``near'' lepton $\ell_n^\pm$ and a ``far'' lepton $\ell_f^\mp$.
This chain is quite common in SUSY, with the identification
$D=\tilde q$, $C=\tilde\chi^0_2$, $B=\tilde \ell$ and $A=\tilde\chi^0_1$,
where $\tilde q$ is a squark, $\tilde \ell$ is a slepton, and
$\tilde\chi^0_1$ ($\tilde\chi^0_2$) is the first (second) 
lightest neutralino. In what follows we shall
quote our results in terms of the $D$ mass $m_D$ and the three
dimensionless squared mass ratios 
$R_{CD}$, $R_{BC}$ and $R_{AB}$ defined in eq.~(\ref{RABdef}).}
\label{fig:chain}
}
Here $D$, $C$, $B$ and $A$ are new BSM particles with masses
$m_D$, $m_C$, $m_B$ and $m_A$. Their
corresponding SM decay products are: a QCD jet $j$, 
a ``near'' lepton $\ell_n^\pm$ and a ``far'' lepton $\ell_f^\mp$.
This decay chain is quite common in SUSY, with the identification
$D=\tilde q$, $C=\tilde\chi^0_2$, $B=\tilde \ell$ and $A=\tilde\chi^0_1$,
where $\tilde q$ is a squark, $\tilde \ell$ is a slepton, and
$\tilde\chi^0_1$ ($\tilde\chi^0_2$) is the first (second) 
lightest neutralino. However, our analysis is not limited to SUSY only,
since the chain in Fig.~\ref{fig:chain} also appears in 
other BSM scenarios, e.g. Universal Extra Dimensions \cite{Cheng:2002ab}.
For concreteness, we shall assume that all three 
decays exhibited in Fig.~\ref{fig:chain} are two-body, i.e.
we shall consider the mass hierarchy
\beq 
m_D>m_C>m_B>m_A>0.
\label{masshierarchy}
\eeq
This presents the most challenging case, in which one has to determine 
all four masses $m_D$, $m_C$, $m_B$ and $m_A$.


The idea of the kinematic endpoint method is very simple.
Given the SM decay products $j$, $\ell_n$ and $\ell_f$ 
exhibited in Fig.~\ref{fig:chain},
form the invariant mass\footnote{We shall see below that the
formulas simplify considerably if we use invariant masses {\em squared} 
instead. This distinction is not central to our analysis.}
of every possible combination, 
$m_{\ell\ell}$, $m_{j\ell_n}$, $m_{j\ell_f}$, and $m_{j\ell\ell}$,
plot the resulting distributions and measure the corresponding 
upper kinematic endpoints \cite{Hinchliffe:1996iu,Allanach:2000kt,Gjelsten:2004ki}
\bea
\left(m_{\ell\ell}^{max}\right)^2  &=& m_D^2\, R_{CD}\, (1-R_{BC})\, (1-R_{AB}); 
\label{lldef}
\\ [4mm]
\left(m_{j\ell_n}^{max}\right)^2   &=& m_D^2\, (1-R_{CD})\, (1-R_{BC})\, ; \label{mjlnmax}
\\ [4mm]
\left(m_{j\ell_f}^{max}\right)^2   &=& m_D^2\, (1-R_{CD})\, (1-R_{AB})\, ; \label{mjlfmax}
\\ [4mm]
\left(m_{j\ell\ell}^{max}\right)^2 &=& 
\left\{ 
\begin{array}{lll}
m_D^2 (1-R_{CD})(1-R_{AC}),       & ~{\rm for}\ R_{CD}<R_{AC},       & {\rm case}\ (1,-),~~ \\[4mm] 
m_D^2 (1-R_{BC})(1-R_{AB}R_{CD}), & ~{\rm for}\ R_{BC}<R_{AB}R_{CD}, & {\rm case}\ (2,-),~~ \\[4mm] 
m_D^2 (1-R_{AB})(1-R_{BD}),       & ~{\rm for}\ R_{AB}<R_{BD},       & {\rm case}\ (3,-),~~ \\[4mm] 
m_D^2\left(1-\sqrt{R_{AD}}\,\right)^2,   & ~{\rm otherwise}, & {\rm case}\ (4,-).~~
\end{array}
\right.\ ,
\label{jlldef}
\eea
Here and below we follow the notation and conventions of Ref.~\cite{Burns:2009zi},
i.e. we write all results in terms of an overall mass scale
(given by the mass $m_D$ of the heaviest BSM particle $D$) 
and three dimensionless squared mass ratios
\begin{equation}
R_{ij} \equiv \frac{m_i^2}{m_j^2}\ ,
\qquad i,j\in\left\{A,B,C,D\right\}.
\label{RABdef}
\end{equation}
Note that there are only three independent ratios in (\ref{RABdef}).
We shall take those to be $R_{AB}$, $R_{BC}$, and $R_{CD}$ 
(see Fig.~\ref{fig:chain}), and their
definition domain will be the interval $(0,1)$.\footnote{As seen in
eq.~(\ref{jlldef}), at times we shall also utilize one or more of the 
other three ratios, $R_{AC}$, $R_{AD}$ and $R_{BD}$, whenever this 
will lead to a simplification of the formulas. Of course, the latter 
three ratios are related to our preferred set  
$\{R_{AB}, R_{BC}, R_{CD}\}$ due to the transitivity property
$R_{ij}R_{jk}=R_{ik}$.}

In spite of their transparent theoretical meaning, the set of four 
endpoints (\ref{lldef}-\ref{jlldef}) by themselves have (justifiably)
never been used as the sole basis for a SUSY mass determination analysis.
This is due to three generic problems, which are all very well known, and are 
separately reviewed in the next three subsections \ref{sec:near-far}, 
\ref{sec:lindep} and \ref{sec:regdep}. Our new approach to resolving 
these three problems, and the outline of the rest of the paper are 
presented in Sec.~\ref{sec:outline}.

\subsection{Near-far lepton ambiguity}
\label{sec:near-far}

The first problem is that one cannot differentiate between the ``near'' and 
``far'' leptons $\ell_n$ and $\ell_f$ on an event-by-event basis. 
Since all decays in Fig.~\ref{fig:chain} are prompt, both leptons 
point back to the primary interaction vertex and there is no way to tell
which came first and which came second. Consequently, one cannot
{\em separately} construct the individual $m_{j\ell_n}$ and $m_{j\ell_f}$ 
invariant mass distributions, whose upper endpoints would be 
given by (\ref{mjlnmax}) and (\ref{mjlfmax}). This problem has
motivated most of the previous invariant mass studies in the literature,
beginning with \cite{Allanach:2000kt}, to introduce an alternative 
definition of the two $j\ell$ distributions, simply by ordering the two
$m_{j\ell}$ entries in each event by invariant mass as follows
\begin{eqnarray}
m_{j\ell(lo)}
&\equiv& \min \left\{m_{j\ell_n}, m_{j\ell_f} \right\}  , \label{mjllodef} \\ [2mm]
m_{j\ell(hi)}
&\equiv& \max \left\{m_{j\ell_n}, m_{j\ell_f} \right\}  . \label{mjlhidef}
\end{eqnarray}
Both of the newly defined quantities $m_{j\ell(lo)}$ and $m_{j\ell(hi)}$ 
also exhibit upper kinematic endpoints ($m_{j\ell(lo)}^{max}$ and 
$m_{j\ell(hi)}^{max}$, correspondingly). Since the {\em individual} 
$m_{j\ell(lo)}$ and $m_{j\ell(hi)}$ distributions are observable,
their endpoints are experimentally measurable and can be related 
to the underlying SUSY mass spectrum as follows \cite{Allanach:2000kt,Gjelsten:2004ki}
\bea
\left(m_{j\ell(lo)}^{max}\right)^2 &=& \left\{ 
\begin{array}{lll}
\left(m_{j\ell_n}^{max}\right)^2,   & ~{\rm for}\ (2-R_{AB})^{-1} < R_{BC} < 1,   & {\rm case}\ (-,1), \\[4mm] 
\left(m_{j\ell(eq)}^{max}\right)^2, & ~{\rm for}\ R_{AB}< R_{BC}<(2-R_{AB})^{-1}, & {\rm case}\ (-,2),\\[4mm] 
\left(m_{j\ell(eq)}^{max}\right)^2, & ~{\rm for}\ 0< R_{BC}<R_{AB},               & {\rm case}\ (-,3);
\end{array}%
\right .
\label{jllodef}
\\ [4mm]
\left( m_{j\ell(hi)}^{max}\right)^2 &=& \left\{ 
\begin{array}{lll}
\left(m_{j\ell_f}^{max}\right)^2, & ~{\rm for}\ (2-R_{AB})^{-1} < R_{BC} < 1,   & {\rm case}\ (-,1), \\[4mm] 
\left(m_{j\ell_f}^{max}\right)^2, & ~{\rm for}\ R_{AB}< R_{BC}<(2-R_{AB})^{-1}, & {\rm case}\ (-,2), \\[4mm] 
\left(m_{j\ell_n}^{max}\right)^2, & ~{\rm for}\ 0< R_{BC}<R_{AB},               & {\rm case}\ (-,3);
\end{array}%
\right .
\label{jlhidef}
\eea
where
\beq
\left(m_{j\ell(eq)}^{max}\right)^2 = m_D^2\, (1-R_{CD})\, (1-R_{AB})\, (2-R_{AB})^{-1} \label{mjleqmax}
\eeq
and $m_{j\ell_n}^{max}$ and $m_{j\ell_f}^{max}$ were already defined in
(\ref{mjlnmax}) and (\ref{mjlfmax}), correspondingly.
With this approach, the original set of 4 endpoints
in eqs.~(\ref{lldef}-\ref{jlldef}) is replaced by
\begin{equation}
m_{\ell\ell}^{max}, m_{j\ell\ell}^{max}, m_{j\ell(lo)}^{max}, m_{j\ell(hi)}^{max}.
\label{4meas}
\end{equation}

In contrast to this conventional approach in the literature, 
we shall adopt a very different attitude towards resolving the 
problem of the near-far lepton ambiguity. We will do the simplest possible thing,
namely, we shall do nothing. We shall never ask the question 
``which lepton was $\ell_n$ and which one was $\ell_f$?''. 
We shall also {\em not} use the ordering (\ref{mjllodef},\ref{mjlhidef}).
Instead, we shall simply take the two $m_{j\ell}$ entries in each 
event, and always treat them in a symmetric fashion. 
For example, any observable invariant mass distribution that we will 
build out of the two measured quantities $m_{j\ell_n}$ and $m_{j\ell_f}$ 
should be invariant under the symmetry
\beq
m_{j\ell_n} \leftrightarrow m_{j\ell_f}\ .
\label{nearfarsymmetry}
\eeq
The advantages of our approach may not be immediately obvious at this point, 
but will become clear in the process of our mass determination analysis in 
Sec.~\ref{sec:analysis} below.

\subsection{Insufficient number of measurements.}
\label{sec:lindep}

The second problem associated with the original set of four 
measurements (\ref{lldef}-\ref{jlldef}),
as well as the alternative set (\ref{4meas}), is that
the measured endpoints may not all be independent from each other.
Indeed, there are certain regions of parameter space
where one finds the following correlation \cite{Gjelsten:2004ki}
\begin{equation}
\left( m_{j\ell\ell}^{max}\right)^2 = \left( m_{j\ell(hi)}^{max}\right)^2 + \left( m_{\ell\ell}^{max}\right)^2.
\label{mjllcorrelation}
\end{equation}
In this case, the four measurements (\ref{4meas}) are clearly 
insufficient to pin down all four independent input parameters 
$m_D$, $m_C$, $m_B$ and $m_A$. Therefore, one has to measure 
an additional independent endpoint. To this end, it has been 
suggested to consider the constrained distribution 
$m_{j\ell\ell(\theta>\frac{\pi}{2})}$, which exhibits a useful 
{\em lower} kinematic endpoint $m_{j\ell\ell(\theta>\frac{\pi}{2})}^{min}$ 
\cite{Allanach:2000kt}
\begin{eqnarray}
&& \left( m_{j\ell\ell(\theta>\frac{\pi}{2})}^{min}\right)^2 = 
\frac{1}{4}m_D^2 \Biggl\{ (1-R_{AB})(1-R_{BC})(1+R_{CD})  \label{jllthetadef} 
\\ \nonumber
&+& 2\, (1-R_{AC})(1-R_{CD})
-(1-R_{CD})\sqrt{(1+R_{AB})^2 (1+R_{BC})^2-16 R_{AC}}\Biggr\} .
\end{eqnarray}
The distribution $m_{j\ell\ell(\theta>\frac{\pi}{2})}$ is nothing but the
usual $m_{j\ell\ell}$ distribution over a subset of the original events,
subject to the additional dilepton mass constraint 
\begin{equation}
\frac{m_{\ell\ell}^{max}}{\sqrt{2}} < m_{\ell\ell} < m_{\ell\ell}^{max}\, .
\end{equation}
In the rest frame of particle $B$, this cut implies the following 
restriction on the opening angle $\theta$ between the two leptons \cite{Nojiri:2000wq}
\begin{equation}
\theta > \frac{\pi}{2}\, ,
\end{equation}
thus justifying the notation for $m_{j\ell\ell(\theta>\frac{\pi}{2})}$.

The advantage of the ``threshold'' endpoint measurement (\ref{jllthetadef})
is that it is always independent of the other four measurements
in (\ref{4meas}). As a result, it would appear that the enlarged set of five kinematic 
endpoint measurements
\begin{equation}
m_{\ell\ell}^{max}, m_{j\ell\ell}^{max}, m_{j\ell(lo)}^{max}, 
m_{j\ell(hi)}^{max}, m_{j\ell\ell(\theta>\frac{\pi}{2})}^{min}
\label{5meas}
\end{equation}
should be in principle sufficient to determine all four unknown masses
(see, however, Ref.~\cite{Burns:2009zi} and Section \ref{sec:regdep} below).

Unfortunately, the ``threshold'' (\ref{jllthetadef})
also suffers from certain disadvantages, which are mostly of experimental
nature. It is generally expected that the experimental precision on the
determination of the {\em lower} kinematic endpoint (\ref{jllthetadef}) 
will be rather inferior compared to the precision on the other four 
{\em upper} kinematic endpoints (\ref{4meas}) \cite{Gjelsten:2004ki}. 
There are several generic reasons for such a pessimistic attitude.
First, the region in the $m_{j\ell\ell(\theta>\frac{\pi}{2})}$
distribution near its lower endpoint (\ref{jllthetadef}) is rather
sparsely populated, resulting in a shallow edge and 
sizable statistical errors. To make matters worse,
the $m_{j\ell\ell(\theta>\frac{\pi}{2})}$ distribution near its
lower edge is a convex function \cite{Lester:2006yw},
which makes it even more difficult to tell where the 
signal ends and the tails from various sources begin \cite{Gjelsten:2004ki}.
Finally, the low mass region of almost any invariant mass distribution 
in SUSY is generally associated with larger SM (as well as SUSY 
combinatorial) backgrounds compared to its high mass counterpart. 

Overall we find all these disadvantages sufficiently 
convincing so that we will drop the measurement (\ref{jllthetadef})
altogether and will never use it in the course of our analysis in
Sec.~\ref{sec:analysis} below. We will be justified in doing so, since
the linear dependence problem (\ref{mjllcorrelation}), 
which has plagued previous studies and was the prime motivation 
for introducing the $m_{j\ell\ell(\theta>\frac{\pi}{2})}^{min}$
measurement in the first place, will have no effect on our analysis.
In fact, we will not be using the endpoint measurement 
$m_{j\ell(hi)}^{max}$ (for the reasons given in the previous 
subsection \ref{sec:near-far}) {\em and} we will not be using
the endpoint measurement $m_{j\ell\ell}^{max}$ (for the reasons given 
in the following subsection \ref{sec:regdep}). Once these two 
problematic measurements are removed from consideration,
the linear dependence problem (\ref{mjllcorrelation}) does not arise, and the 
``threshold'' measurement (\ref{jllthetadef}) is not central 
to the analysis any more.

\subsection{Parameter space region ambiguity}
\label{sec:regdep}

The third problem with the conventional set of measurements (\ref{5meas})
is immediately obvious from the defining equations 
(\ref{jlldef}), (\ref{jllodef}) and (\ref{jlhidef})
for the kinematic endpoints $m_{j\ell\ell}^{max}$, 
$m_{j\ell(lo)}^{max}$, and $m_{j\ell(hi)}^{max}$, correspondingly.
One can see that the relevant expressions are piecewise-defined functions,
i.e. they depend on the values of the independent variables 
$m_A$, $m_B$, $m_C$ and $m_D$. For example, 
there are four different cases for $m_{j\ell\ell}^{max}$, 
and three different cases for the pair of 
$(m_{j\ell(lo)}^{max},m_{j\ell(hi)}^{max})$. Altogether, these give 
rise to 9 different cases\footnote{The remaining 3 cases are 
always unphysical \cite{Gjelsten:2004ki}.} 
which must be separately considered \cite{Gjelsten:2004ki,Burns:2009zi}. 
Of course, this represents a problem, since the masses are a priori unknown, 
and it is not clear which case is the relevant one.
Barring any model-dependent assumptions, one is forced to 
consider all possibilities, obtain a solution for the spectrum,
and only at the very end, test whether the solution falls within
the parameter space applicable for the case at hand. 
This procedure may often result in several alternative solutions
\cite{Gjelsten:2004ki,ArkaniHamed:2005px,Gjelsten:2005sv,Gjelsten:2006as,MP1,KM1,Burns:2009zi}.
In fact, Ref.~\cite{Burns:2009zi} recently proved that 
there exists a sizable parameter space region in which 
even the full set of measurements (\ref{5meas}) would 
always yield two alternative solutions, 
even under ideal experimental conditions.
The problem is further exacerbated by the inevitable 
experimental errors on the measurements (\ref{5meas}),
which would allow for an even larger number of ``fake'' or 
``duplicate'' solutions \cite{Gjelsten:2005sv,Gjelsten:2006as,Burns:2009zi}.

Having identified the root of the duplication problem 
as the piecewise definition of the mathematical formulas
in (\ref{jlldef},\ref{jllodef},\ref{jlhidef}), 
our solution to the problem will be again very simple and conservative.
We will simply avoid using any kinematic endpoints which
are given in terms of piecewise-defined expressions.
This requirement automatically eliminates from consideration
the three conventional endpoints $m_{j\ell\ell}^{max}$, 
$m_{j\ell(lo)}^{max}$, and $m_{j\ell(hi)}^{max}$. 
Since we already gave up on $m_{j\ell\ell(\theta>\frac{\pi}{2})}^{min}$ 
in the previous subsection, this leaves $m_{\ell\ell}^{max}$
as {\em the only} measurement out of the conventional
set (\ref{5meas}) that we shall use in our analysis.
This is perhaps the most drastic difference between 
our approach and all previous studies in the literature.

\subsection{Posing the problem}
\label{sec:outline}

In the previous three subsections we discussed
each of the three generic theoretical\footnote{In addition, 
there are problems which are of experimental nature, 
e.g. identifying the correct jet and the correct 
lepton pair resulting from the decay chain in Fig.~\ref{fig:chain}.
There exists a set of standard experimental techniques 
which are aimed at overcoming these problems, e.g.
the opposite flavor subtraction for the two leptons and 
the mixed event subtraction for the jet \cite{Ball:2007zza}. 
Wrong $\ell\ell$ and $j\ell$ pairings can also be identified and 
a posteriori removed whenever an invariant mass entry for
$m_{\ell\ell}$, $m_{j\ell}$ or $m_{j\ell\ell}$ exceeds
the corresponding kinematic endpoint $m^{max}_{\ell\ell}$, 
$m^{max}_{j\ell(hi)}$ or $m^{max}_{j\ell\ell}$. 
In what follows we shall assume that those preliminary 
steps have already been done and the samples we are 
dealing with have already been appropriately subtracted 
to remove the combinatorial background.} 
problems with the previous applications of the 
kinematic endpoint method for mass determination. We are now 
ready to explicitly formulate our main goal in this paper.
We aim to design a method for measuring the masses of 
the particles in the decay chain of Fig.~\ref{fig:chain},
which is based on kinematic endpoint information, and
satisfies the following requirements:
\begin{itemize}
\item It does not make use of any kinematic endpoints 
whose interpretation is ambiguous, i.e. whose expressions 
in terms of the physical masses are piecewise-defined functions.
\item It does not make use of any {\em lower} kinematic 
endpoints such as the ``threshold''
$m_{j\ell\ell(\theta>\frac{\pi}{2})}^{min}$, due to the
experimental challenges with such measurements.
\item It relies solely on 1-dimensional distributions, 
unlike the methods recently advertised in 
\cite{MP1,KM1,Luc,Costanzo:2009mq,Burns:2009zi},
which utilize 2-dimensional correlation plots.
While the latter do provide a wealth 
of valuable information, they also typically require more 
data in order to obtain good enough statistics for drawing
any robust conclusions from them. In contrast, the one-dimensional 
distributions should be available rather early on, and
with sufficient statistics for endpoint measurements.
\end{itemize}
As already alluded to in the previous subsections,
the first two requirements already eliminate four
out of the five conventional inputs (\ref{5meas}).
Obviously, we will need to find a way to replace those with an
alternative set of kinematic endpoint measurements which 
nevertheless satisfy the above requirements. 
In Section~\ref{sec:variables} we introduce and 
investigate a new set of invariant mass variables 
whose upper endpoints can be useful for our analysis.
Then in Section~\ref{sec:analysis} we outline our basic method,
which makes use of some of these new variables.
We illustrate our discussion in Section~\ref{sec:examples}
with two numerical examples: the LM1 and LM6 CMS study points.
Section~\ref{sec:conclusions} is reserved for our conclusions.
In Appendix~\ref{app:shapes} we supply the analytic expressions
for the shapes of the 1-dimensional invariant mass distributions 
used in our main analysis in Sec.~\ref{sec:theory}. Those results
can be useful in improving the precision on the extraction 
of the kinematical endpoints.

\section{New variables}
\label{sec:variables}

In this section we propose a new set of invariant mass (squared) variables. 
As already explained in the Introduction, our variables should 
be composed of $m^2_{j\ell_n}$ and $m^2_{j\ell_f}$ in a symmetric way,
in accordance with (\ref{nearfarsymmetry}). Consequently, any
plotting manipulations or mathematical operations involving 
$m^2_{j\ell_n}$ and $m^2_{j\ell_f}$ should obey the symmetry implied by
eq.~(\ref{nearfarsymmetry}).

\subsection{The union $m^2_{j\ell_n} \cup m^2_{j\ell_f}$}
\label{sec:union}

We begin with the simplest case, where we postpone applying any 
mathematical operations to $m^2_{j\ell_n}$ and $m^2_{j\ell_f}$, and
instead simply plot them. The requirement of eq.~(\ref{nearfarsymmetry})
implies that the only possibility is to place both of them together 
on the same plot, in essence forming the union 
\beq
m^2_{jl(u)} \equiv m^2_{j\ell_n} \cup m^2_{j\ell_f}
\label{mjludef}
\eeq
of the individual $m^2_{j\ell_n}$ and $m^2_{j\ell_f}$ distributions.
Since each individual distribution is smooth and has a kinematic endpoint,
the same two kinematic endpoints should be visible on the 
combined distribution $m^2_{jl(u)}$ as well\footnote{For specific 
numerical examples, refer to Sec.~\ref{sec:examples}.}. 
We shall denote the larger of the two endpoints with 
\beq
\left(M^{max}_{jl(u)}\right)^2 
\equiv \max\left\{ \left(m_{j\ell_n}^{max}\right)^2, \left(m_{j\ell_f}^{max}\right)^2 \right\}
\label{mjluhi}
\eeq
and the smaller of the two endpoints with 
\beq
\left(m^{max}_{jl(u)}\right)^2 
\equiv \min\left\{ \left(m_{j\ell_n}^{max}\right)^2, \left(m_{j\ell_f}^{max}\right)^2 \right\}\ .
\label{mjlulo}
\eeq
The newly introduced quantities $M^{max}_{jl(u)}$ and $m^{max}_{jl(u)}$
are nothing but the usual kinematic endpoints $m_{j\ell_n}^{max}$
and $m_{j\ell_f}^{max}$, given by (\ref{mjlnmax}) and (\ref{mjlfmax}),
correspondingly. Of course, at this point we do not know 
which is which, and we have an apparent two-fold ambiguity: 
we can have either
\beq
M^{max}_{jl(u)}=m_{j\ell_n}^{max}, \qquad m^{max}_{jl(u)}=m_{j\ell_f}^{max}\ ,
\qquad {\rm if}\ R_{AB}\ge R_{BC},
\label{Mnear}
\eeq
or 
\beq
M^{max}_{jl(u)}=m_{j\ell_f}^{max}, \qquad m^{max}_{jl(u)}=m_{j\ell_n}^{max}\ ,
\qquad {\rm if}\ R_{AB}\le R_{BC}.
\label{Mfar}
\eeq
Notice that both (\ref{mjluhi}) and (\ref{mjlulo}) are officially {\em upper}
kinematic endpoints, and thus satisfy our basic requirements.

The benefits of our alternative treatment (\ref{mjludef}) in response 
to the near-far lepton ambiguity problem of Sec.~\ref{sec:near-far}, 
are now starting to emerge. With the conventional ordering (\ref{mjllodef},\ref{mjlhidef}) 
one has to deal with a {\em three-fold} ambiguity in the interpretation of the endpoints
$m^{max}_{j\ell(lo)}$ and $m^{max}_{j\ell(hi)}$, as seen in eqs.~(\ref{jllodef},\ref{jlhidef}).
Instead, the simple union (\ref{mjludef}) leads only to the {\em two-fold} ambiguity
of eqs.~(\ref{Mnear},\ref{Mfar}). More importantly, the analysis of Sec.~\ref{sec:theory}
below will reveal that in spite of the remaining two-fold ambiguity
in eqs.~(\ref{Mnear},\ref{Mfar}), one can nevertheless {\em uniquely}
determine all three of the masses $m_D$, $m_C$ and $m_A$!
We consider this to be one of the important results of this paper.

\subsection{The product $m_{j\ell_n} \times m_{j\ell_f}$}

In the remainder of this section, we shall construct new
invariant mass squared variables out of the two entries $m^2_{j\ell_n}$ and $m^2_{j\ell_f}$,
simply by applying various mathematical operations on them in a symmetric fashion.
We begin with the product
\beq
m^2_{j\ell(p)} \equiv m_{j\ell_n} m_{j\ell_f}
\label{mjlpdef}
\eeq
whose endpoint is given by 
\beq
\left(m_{j\ell(p)}^{max}\right)^2 \equiv \left\{ 
\begin{array}{lll}
\frac{1}{2}\, m_D^2 (1-R_{CD})\sqrt{1-R_{AB}},   & ~{\rm for}\ R_{BC} \le 0.5, \\[4mm] 
m_D^2 (1-R_{CD})\sqrt{R_{BC}(1-R_{BC})(1-R_{AB})}, & ~{\rm for}\ R_{BC} \ge 0.5.
\end{array}%
\right .
\label{mjlpmaxdef}
\eeq
Unfortunately, this endpoint also turns out to be piecewise-defined, thus
failing one of our basic requirements from the Introduction. Therefore 
we shall not use this endpoint in the course of our analysis.

\subsection{The sums $m^{2\alpha}_{j\ell_n} + m^{2\alpha}_{j\ell_f}$}

Another possibility is to consider various sums, for example 
$m^2_{j\ell_n} + m^2_{j\ell_f}$ or
$(m_{j\ell_n} + m_{j\ell_f})^2$, as originally proposed in \cite{Luc}. 
Here we generalize the discussion in \cite{Luc} and introduce 
a whole set of new variables, $m^2_{j\ell(s)}(\alpha)$,
labelled by the continuous parameter $\alpha$, which are defined as
\beq
m^2_{j\ell(s)}(\alpha) \equiv \left(m^{2\alpha}_{j\ell_n}+m^{2\alpha}_{j\ell_f}\right)^\frac{1}{\alpha}\ .
\label{mjlsdef}
\eeq
Since $\alpha$ is a continuous parameter, in principle there are 
infinitely many $m_{j\ell(s)}$ variables! Notice that the conventional
variables $m^2_{j\ell(lo)}$ and $m^2_{j\ell(hi)}$
from (\ref{mjllodef}) and (\ref{mjlhidef}) are also included 
in our set, and are simply given by
\bea
m^2_{j\ell(lo)} &\equiv& m^2_{j\ell(s)}(-\infty) \, ,  \label{lois-infty}  \\ [2mm]
m^2_{j\ell(hi)} &\equiv& m^2_{j\ell(s)}( \infty) \, .  \label{hiis+infty}  
\eea
We see that our new set (\ref{mjlsdef}) is a very broad generalization of 
the conventional definitions (\ref{mjllodef}) and (\ref{mjlhidef}), 
which just correspond to the two extreme cases $\alpha=\pm\infty$. 
Of course, the user is free to choose $\alpha$ at will, and 
any finite value of $\alpha$ will lead to a new variable $m^2_{j\ell(s)}(\alpha)$.

In order to make the new variables $m^2_{j\ell(s)}(\alpha)$
useful for mass spectrum studies, we need to provide the 
formulas for their kinematic endpoints $(m^{max}_{j\ell(s)}(\alpha))^2$.
These formulas are easy to derive, using the results from
\cite{Burns:2009zi}, and we present them in the next two
subsections, where it is convenient to consider 
separately the following two cases: $\alpha\ge 1$ (in Sec.~\ref{sec:alphamore1}) 
and $\alpha<1$, but $\alpha\ne 0$ (in Sec.~\ref{sec:alphaless1}).

\subsubsection{Kinematic endpoints of $m^2_{j\ell(s)}(\alpha)$ with $\alpha\ge 1$}
\label{sec:alphamore1}

When one chooses a value of $\alpha \ge 1$, the $m^2_{j\ell(s)}(\alpha)$
endpoint is given by the following expression
\beq
\left(m_{j\ell(s)}^{max}(\alpha\ge1)\right)^2 \equiv \left\{ 
\begin{array}{ll}
\left(m_{j\ell_f}^{max}\right)^2,   
           & ~R_{AB} \le 1 - \left(1-R_{BC}\right)\left(1-R^\alpha_{BC}\right)^{-\frac{1}{\alpha}}, \\[4mm] 
\left(m_{j\ell}^{max}(\alpha)\right)^2, 
           & ~R_{AB} \ge 1 - \left(1-R_{BC}\right)\left(1-R^\alpha_{BC}\right)^{-\frac{1}{\alpha}},
\end{array}%
\right .
\label{mjlsmax>1}
\eeq
where $m_{j\ell_f}^{max}$ was already defined in (\ref{mjlfmax}), and $m_{j\ell}^{max}(\alpha)$
is a newly defined, $\alpha$-dependent quantity
\beq
\left( m_{j\ell}^{max}(\alpha)\right)^2 \equiv
m_D^2 (1-R_{CD}) \Bigl[ R_{BC}^\alpha (1-R_{AB})^\alpha + (1-R_{BC})^\alpha\Bigr]^{\frac{1}{\alpha}}\ .
\label{mjlalphadef}
\eeq
As a cross-check, one can verify that in the limit $\alpha\to \infty$
the expression (\ref{mjlsmax>1}) reduces to (\ref{jlhidef}), in 
agreement with (\ref{hiis+infty}). In that case, the upper line in
(\ref{mjlsmax>1}) corresponds to options $(-,1)$ and $(-,2)$ in
(\ref{jlhidef}), where $m^{max}_{j\ell(hi)}=m^{max}_{j\ell_f}$,
while the lower line in
(\ref{mjlsmax>1}) corresponds to option $(-,3)$ in
(\ref{jlhidef}), where $m^{max}_{j\ell(hi)}=m^{max}_{j\ell_n}$.
Unfortunately, just like the product
endpoint (\ref{mjlpmaxdef}), the endpoint (\ref{mjlsmax>1}) 
is in general piecewise-defined, and does not meet our criteria.

However, there is one important exception, namely the case of
$\alpha=1$, in which we do get a singly defined function.
According to the general definition (\ref{mjlsdef}),
$m^2_{j\ell(s)}(\alpha=1)$ is simply the sum of the two $m^2_{j\ell}$ 
entries in each event:
\beq
m^2_{j\ell(s)}(\alpha=1) \equiv m^2_{j\ell_n}+m^2_{j\ell_f}\, .
\label{mjls1def}
\eeq
Using the identity 
\beq
m^2_{j\ell\ell} = m^2_{j\ell_n}+m^2_{j\ell_f} + m^2_{\ell\ell}\, ,
\eeq 
(\ref{mjls1def}) can be equivalently rewritten as
\beq
m^2_{j\ell(s)}(\alpha=1) \equiv m^2_{j\ell\ell}-m^2_{\ell\ell}\ .
\label{mjls1def2}
\eeq
To find the expression for its endpoint, one can set $\alpha=1$ 
in (\ref{mjlsmax>1}), and then realize that the logical 
condition for executing the upper line becomes $R_{AB}\le 0$,
which is impossible, since the mass ratios $R_{ij}$ in (\ref{RABdef}) 
are always positive definite. Therefore, the endpoint $m^{max}_{j\ell(s)}(\alpha=1)$ 
is always calculated according to the lower line in (\ref{mjlsmax>1}),
which results in \cite{Luc}
\beq
\left(m^{max}_{j\ell(s)}(1)\right)^2 \equiv m_D^2 (1-R_{CD}) (1-R_{AC})\ .
\label{mjls1max}
\eeq
Note that this endpoint is perfect for our purposes since the 
formula (\ref{mjls1max}) is always unique, i.e. it is independent 
of the parameter space region. The variable $m^2_{j\ell(s)}(\alpha=1)$ 
will thus play a crucial role in our analysis below.

\subsubsection{Kinematic endpoints of $m^2_{j\ell(s)}(\alpha)$ with $\alpha<1$ and $\alpha\ne 0$}
\label{sec:alphaless1}

Finally, in the case when $\alpha< 1$, but $\alpha\ne 0$,
the $m^2_{j\ell(s)}(\alpha)$ endpoint is given by the following expression
\beq
\left(m_{j\ell(s)}^{max}(\alpha< 1)\right)^2 \equiv \left\{ 
\begin{array}{ll}
\left(m_{j\ell}^{max}(\alpha)\right)^2,
           & R_{BC} \ge \left[1+\left(1-R_{AB}\right)^{\frac{\alpha}{\alpha-1}}\right]^{-1}, \\[4mm] 
m_D^2 (1-R_{CD}) \left[1+\left(1-R_{AB}\right)^{\frac{\alpha}{1-\alpha}} \right]^{\frac{1-\alpha}{\alpha}},   
           & R_{BC} \le \left[1+\left(1-R_{AB}\right)^{\frac{\alpha}{\alpha-1}}\right]^{-1}, 
\end{array}%
\right .
\label{mjlsmax<1}
\eeq
where $m_{j\ell}^{max}(\alpha)$ was already defined in (\ref{mjlalphadef}).
Again as a cross-check, one can verify that in the limit $\alpha\to -\infty$
the expression (\ref{mjlsmax<1}) reduces to (\ref{jllodef}), in 
agreement with (\ref{lois-infty}). In the $\alpha\to -\infty$ case, 
the upper line in (\ref{mjlsmax<1}) corresponds to option $(-,1)$ 
in (\ref{jllodef}), where $m^{max}_{j\ell(lo)}=m^{max}_{j\ell_n}$,
while the lower line in
(\ref{mjlsmax<1}) corresponds to options $(-,2)$ and $(-,3)$ in
(\ref{jllodef}), where $m^{max}_{j\ell(lo)}=m^{max}_{j\ell(eq)}$.
Unfortunately, the endpoint
function (\ref{mjlsmax<1}) is again piecewise-defined, and does 
not meet one of our basic criteria spelled out in the introduction.

In passing, we note that the special case of $\alpha=\frac{1}{2}$,
which involves the {\em linear} sum of the two masses
\beq
m^2_{j\ell(s)}(\alpha=\frac{1}{2}) \equiv \left(m_{j\ell_n}+m_{j\ell_f}\right)^2\, ,
\label{mjls0.5def}
\eeq
was previously explored in \cite{Heinemann,Luc}. In that case, from 
(\ref{mjlsmax<1}) we find for its endpoint
\beq
\left(m_{j\ell(s)}^{max}(\frac{1}{2})\right)^2 \equiv \left\{ 
\begin{array}{ll}
m_D^2 (1-R_{CD}) \left( \sqrt{R_{BC}(1-R_{AB})}+\sqrt{1-R_{BC}}\right)^2,
           & ~R_{BC} \ge \frac{1-R_{AB}}{2-R_{AB}}, \\[4mm] 
m_D^2 (1-R_{CD}) (2-R_{AB}),   
           & ~R_{BC} \le \frac{1-R_{AB}}{2-R_{AB}}.
\end{array}%
\right .
\eeq

\subsection{The difference $|m^2_{j\ell_n} - m^2_{j\ell_f}|$}

Finally, one can also consider a set of variables which involve the absolute value of 
differences between $m^2_{j\ell_n}$ and $m^2_{j\ell_f}$. In analogy with 
(\ref{mjlsdef}), we can define another infinite set of variables
\beq
m^2_{j\ell(d)}(\alpha) \equiv 
\left| m^{2\alpha}_{j\ell_n}-m^{2\alpha}_{j\ell_f}\right|^\frac{1}{\alpha}\ .
\label{mjlddef}
\eeq
Once again, the user is free to consider arbitrary values of $\alpha$.
However, this freedom is redundant, when it comes to the issue of the kinematic 
endpoints of the variables in (\ref{mjlddef}). It is not difficult to see that
the endpoints of $m^2_{j\ell(d)}(\alpha)$ are always given by 
\beq
\left( m^{max}_{j\ell(d)}(\alpha)\right)^2 \equiv \left(M^{max}_{jl(u)}\right)^2
\label{mjld1max}
\eeq 
and are in fact independent of $\alpha$! Therefore, for the purposes of our discussion, 
it is sufficient to consider just one particular value of $\alpha$. In the 
following we shall only use $\alpha=1$:
\beq
m^2_{j\ell(d)}(\alpha=1) \equiv 
\left| m^2_{j\ell_n}-m^2_{j\ell_f}\right|\ ,
\label{mjld1def}
\eeq
which is the analogue of $m^2_{j\ell(s)}(\alpha=1)$ defined in (\ref{mjls1def}).

The result (\ref{mjld1max}) implies that the endpoint of (\ref{mjld1def})
does not contain any new amount of information, which was not already present 
in the two kinematic endpoints $M^{max}_{jl(u)}$ and $m^{max}_{jl(u)}$ 
discussed in Sec.~\ref{sec:union}. Nevertheless, the independent 
measurement of $(m^{max}_{jl(d)}(1))^2$  can still be very useful, since it 
will mark the location of $(M^{max}_{jl(u)})^2$ on the $m^2_{jl(u)}$ 
distribution. Then one will be looking for the second endpoint $(m^{max}_{jl(u)})^2$ 
to the left, i.e.~in the region of smaller $m^2_{jl(u)}$ values.

This completes our discussion of the new invariant mass variables and 
their kinematic endpoints. For our basic proof-of-principle measurement 
technique presented in the next Section~\ref{sec:theory}, we shall use only 
three of them, namely $M^{max}_{j\ell(u)}$, $m^{max}_{j\ell(u)}$, 
and $m^{max}_{j\ell(s)}(\alpha=1)$. However, the remaining variables
are in principle just as good, their only disadvantage being that 
they failed our arbitrarily imposed condition at the beginning that 
the endpoint functions should all be region independent. Of course, 
one could, and in fact should, use all of the available kinematic endpoint 
information, which in a global fit analysis can only increase the 
experimental precision of the sparticle mass determination.

\section{Theoretical analysis}
\label{sec:analysis}

\subsection{Our method and the solution for the mass spectrum}
\label{sec:theory}

Our starting point is the set of four measurements
\beq
m^{max}_{\ell\ell}, M^{max}_{j\ell(u)}, m^{max}_{j\ell(u)}, m^{max}_{j\ell(s)}(\alpha=1)
\label{ourmeas}
\eeq
in place of the conventional set (\ref{5meas}). It is easy to verify that 
the measurements (\ref{ourmeas}) are always independent of each other, 
and thus never suffer from the linear dependence problem discussed in Section~\ref{sec:lindep}.

Given the set of four measurements (\ref{ourmeas}), it is easy to solve for the 
mass spectrum. To simplify the notation, we introduce the following shorthand notation
for the endpoints of the mass {\em squared} distributions
\beq
L\equiv \left(m^{max}_{\ell\ell}\right)^2, \quad 
M\equiv \left(M^{max}_{j\ell(u)}\right)^2, \quad
m\equiv \left(m^{max}_{j\ell(u)}\right)^2, \quad
S\equiv \left(m^{max}_{j\ell(s)}(\alpha=1)\right)^2
\label{shorthand}
\eeq
The solution for the mass spectrum is then given by
\bea
m^2_D &=& \frac{Mm(L+M+m-S)}{(M+m-S)^2};
\label{mDsol}
\\ [2mm] 
m^2_C &=& \frac{MmL}{(M+m-S)^2};
\label{mCsol}
\\ [2mm] 
m^2_B &=& \left\{ 
\begin{array}{ll}
\frac{ML(S-M)}{(M+m-S)^2}, & ~~~{\rm if}\ R_{AB} \ge R_{BC}, \\[4mm] 
\frac{mL(S-m)}{(M+m-S)^2}, & ~~~{\rm if}\ R_{AB} \le R_{BC}; \end{array}%
\right .  
\label{mBsol} 
\\ [2mm] 
m^2_A &=& \frac{L(S-m)(S-M)}{(M+m-S)^2}.
\label{mAsol}
\eea
It is easy to verify that the right-hand side expressions in these
equations are always positive definite, so that one can safely
take the square root and compute the linear masses $m_D$, $m_C$, $m_B$ and $m_A$.
Notice that in spite of the two-fold ambiguity (\ref{Mnear},\ref{Mfar}),
the solution for $m_D$, $m_C$ and $m_A$ is unique! Indeed, the expressions for
$m_D$, $m_C$ and $m_A$ are symmetric under the interchange $M\leftrightarrow m$.
The remaining two-fold ambiguity for $m_B$ is precisely the result of 
the ambiguous interpretation (\ref{Mnear},\ref{Mfar})
of the two $m^2_{j\ell(u)}$ endpoints, and is related to the symmetry 
under (\ref{nearfarsymmetry}), or equivalently, under the interchange
\beq
R_{AB}\leftrightarrow R_{BC}\ .
\label{yzsymmetry}
\eeq
In the next subsection we discuss several ways in which one can lift 
the remaining two-fold degeneracy for $m_B$ which is due to (\ref{yzsymmetry}). 

Notice the great simplicity of this method. The expressions for 
(\ref{mDsol}), (\ref{mCsol}) and (\ref{mAsol}) are region independent 
and therefore one does not have to go through the standard trial and error 
procedure involving the 9 parameter space regions $(N_{j\ell\ell},N_{j\ell})$ 
\cite{Gjelsten:2004ki,Burns:2009zi} associated with the various interpretations
of the endpoints $m^{max}_{j\ell\ell}$, $m^{max}_{j\ell(lo)}$ and $m^{max}_{j\ell(hi)}$.

\subsection{Disambiguation of the two solutions for $m_B$}
\label{sec:disambiguation}

The method outlined in Sec.~\ref{sec:theory} allowed 
us to find the true masses of particles $A$, $C$ and $D$, but
yields two separate possible solutions for the mass $m_B$ of particle $B$.
We shall now discuss several ways of lifting the 
remaining two-fold degeneracy for $m_B$. 

\subsubsection{Invariant mass endpoint method}
\label{sec:1dim}

One possibility is to use an additional measurement of an 
invariant mass endpoint. Indeed, as shown in Secs.~\ref{sec:introduction}
and \ref{sec:variables}, there are still quite a few
one-dimensional invariant mass distributions at our disposal, 
which we have not used so far. Those include the conventional
distributions of $m^2_{j\ell\ell}$, $m^2_{j\ell(lo)}$ and $m^2_{j\ell(hi)}$,
as well as the new distributions $m^2_{j\ell(p)}$,
$m^2_{j\ell(s)}(\alpha)$ and $m^2_{j\ell(d)}(1)$
which we introduced in Sec.~\ref{sec:variables}.
Which of them can be used for our purposes? 
Note that the duplication in (\ref{mBsol}) arose due 
to the symmetry (\ref{yzsymmetry}), so that
any kinematic endpoint which violates this symmetry 
will be able to distinguish between the two solutions. 

Let us begin with the conventional distributions
$m^2_{j\ell\ell}$, $m^2_{j\ell(lo)}$, $m^2_{j\ell(hi)}$
and $m^2_{j\ell\ell(\theta>\frac{\pi}{2})}$, whose endpoints
we did not use in our analysis so far. It is easy to check that
$m_{j\ell\ell}^{max}$, $m_{j\ell(hi)}^{max}$ and 
$m_{j\ell\ell(\theta>\frac{\pi}{2})}^{min}$
are invariant under the interchange (\ref{yzsymmetry}) 
and cannot be used for discrimination. However,
$m_{j\ell(lo)}^{max}$ is {\em not} symmetric under (\ref{yzsymmetry})
and can do the job. In fact, one can show that
the two duplicate solutions for $m_B$ 
always\footnote{The only exception is the trivial case of 
$R_{AB}=R_{BC}$, but then the two solutions for $m_B$ coincide,
and $m_B$ is again uniquely determined.}
give different predictions for $m_{j\ell(lo)}^{max}$.

\FIGURE[ht]{
\epsfig{file=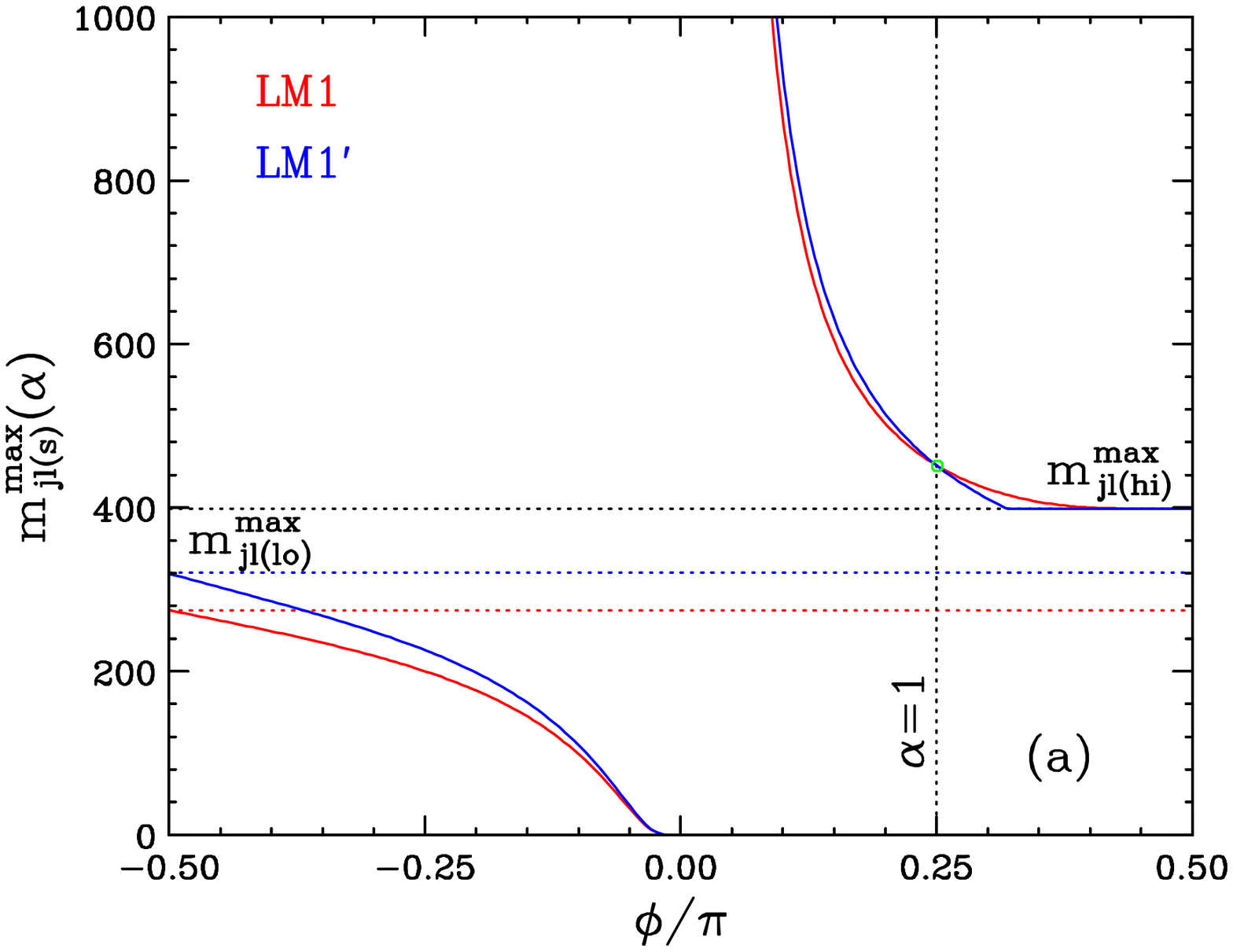,width=7.0cm}~~~
\epsfig{file=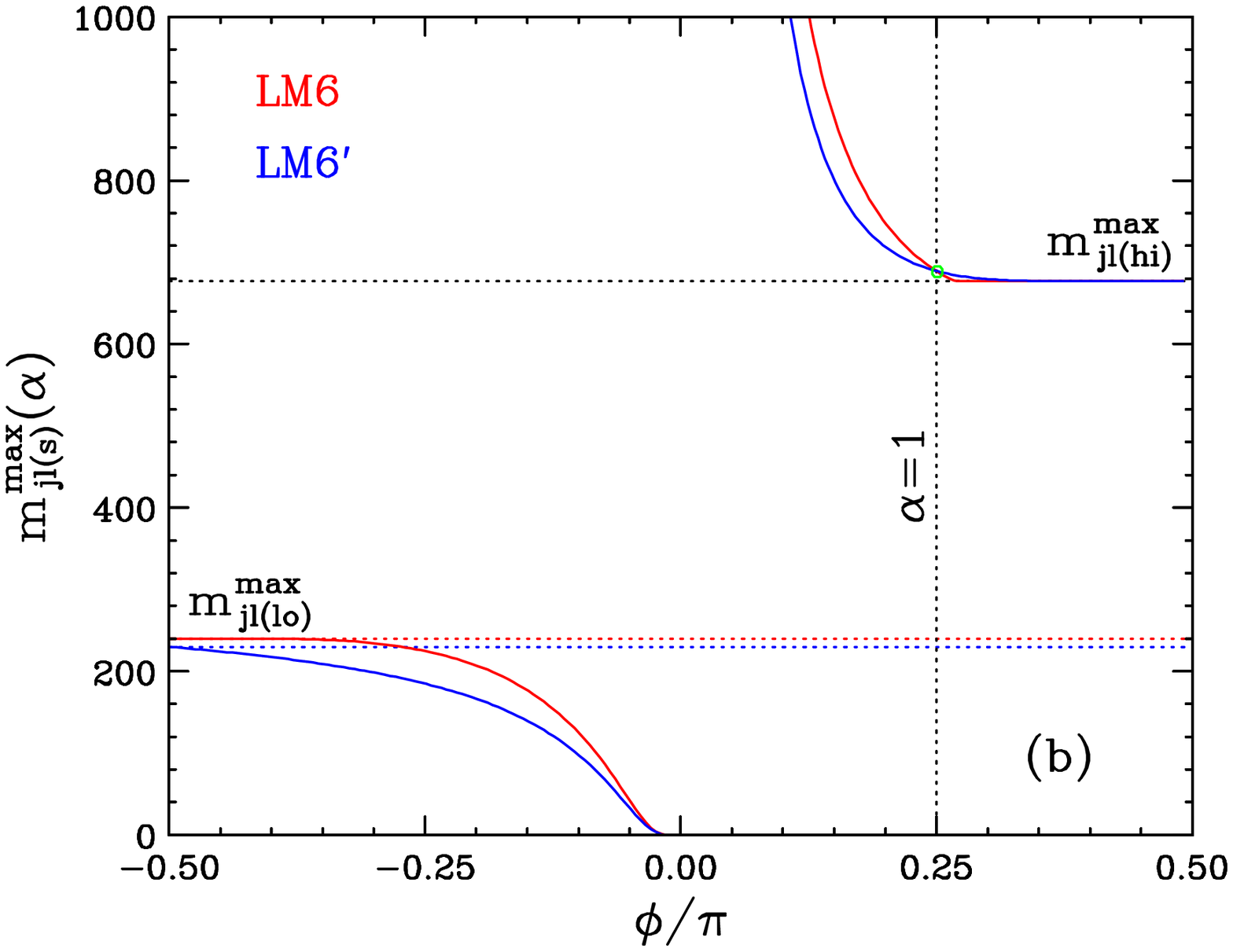,width=7.0cm}
\caption{\sl Comparison of the predictions 
for the kinematic endpoints $m^{max}_{j\ell(s)}(\alpha)$
of the real and fake solutions, as a function of $\phi\equiv\arctan\alpha$
(in units of $\pi$), for the two examples discussed
in detail in Section \ref{sec:examples}: (a) the LM1 CMS study point
and (b) the LM6 CMS study point. In each panel, the prediction of the
real (fake) solution is plotted in red (blue). The vertical dotted line 
indicates the
case of $\phi=\frac{\pi}{4}$ ($\alpha=1$), for which the two solutions 
give an identical answer, marked with a green dot. 
The horizontal dotted lines show the corresponding
asymptotic values $m_{j\ell(hi)}^{max}$ and $m_{j\ell(lo)}^{max}$,
obtained at $\alpha\to\pm \infty$ ($\phi\to\pm\frac{\pi}{2}$).
}
\label{fig:alpha}}

More importantly, many of our new variables from Sec.~\ref{sec:variables}
can provide an independent cross-check on the correct choice 
for the solution. For example, the kinematic endpoint (\ref{mjlpmaxdef})
of the product variable $m^2_{j\ell(p)}$, also violates the symmetry
(\ref{yzsymmetry}) and distinguishes among the two solutions.
The infinite set of variables $m^2_{j\ell(s)}(\alpha)$
can also be used, and for almost the whole range of $\alpha<1$.
To see this, in Fig.~\ref{fig:alpha} we compare the predictions 
for the kinematic endpoints $m^{max}_{j\ell(s)}(\alpha)$
of the real and fake solutions, for the two examples discussed
in detail in Section \ref{sec:examples}: (a) the LM1 CMS study point
and (b) the LM6 CMS study point. The corresponding mass spectra
are listed in Table~\ref{tab:spectra} below. For convenience, we plot 
versus the parameter 
\beq
\phi\equiv\arctan\alpha\, ,
\eeq
which allows us to map the whole definition domain $(-\infty,\infty)$
for $\alpha$ into the finite region $(-\frac{\pi}{2},\frac{\pi}{2})$ for $\phi$.
Fig.~\ref{fig:alpha} shows that for most of the allowed $\phi$ range, the
two solutions predict different values for the kinematic endpoints 
$m^{max}_{j\ell(s)}(\alpha)$. In fact, for $\phi< \frac{\pi}{4}$, 
the two predictions are always different, apart from the trivial case 
of $\phi=0$ ($\alpha=0$). Even for $\phi>\frac{\pi}{4}$, there 
still exists a range of $\phi$, for which, at least theoretically, 
a discrimination can be made.
The predictions are guaranteed to coincide only for $\phi=\frac{\pi}{4}$ ($\alpha=1$)
(as they should, see (\ref{ourmeas})), and for a certain range of the 
largest possible values of $\phi$.

\subsubsection{Invariant mass correlations}
\label{sec:2dim}

Another way to resolve the twofold ambiguity in
our solution (\ref{mBsol}) is to simply go back to the 
original measurements of $M^{max}_{jl(u)}$ and $m^{max}_{jl(u)}$
and already at that point try to decide which of the two measured 
$m_{jl(u)}$ endpoints is $m_{j\ell_n}^{max}$ and which one is 
$m_{j\ell_f}^{max}$. As already discussed in \cite{Luc,Costanzo:2009mq},
this identification is in principle possible, if one considers 
the correlations which are present in the two-dimensional
distribution $m^2_{jl(u)}$ versus $m^2_{ll}$. The basic idea
is illustrated in Fig.~\ref{fig:jlvsll}, where we show scatter plots of 
$m_{j\ell(u)}$ versus $m_{\ell\ell}$, for the two examples 
used in Fig.~\ref{fig:alpha} and discussed in detail later
in Section~\ref{sec:examples}. Fig.~\ref{fig:jlvsll}(a)
(Fig.~\ref{fig:jlvsll}(b)) shows the result for the real (fake) 
solution corresponding to the LM1 study point, while
Figs.~\ref{fig:jlvsll}(c) and \ref{fig:jlvsll}(d)
show the analogous results for the LM6 study point.
In each plot we used 10,000 entries, which roughly corresponds 
to $20\ {\rm fb}^{-1}$ ($200\ {\rm fb}^{-1}$) of data for the actual
LM1 (LM6) SUSY study point. Here and below we show the ideal case where
we neglect smearing effects due to the finite detector resolution,
finite particle widths and combinatorial backgrounds.
All of our plots are at the parton level (using our own Monte-Carlo 
phase space generator) and without any cuts.
Notice that in order to avoid dealing with the large 
numerical values of the squared masses, we use a 
quadratic power scale on both axes, which allows us to 
preserve the simple shapes of the scatter plots when plotting 
versus the linear masses themselves.

\FIGURE[ht]{
\epsfig{file=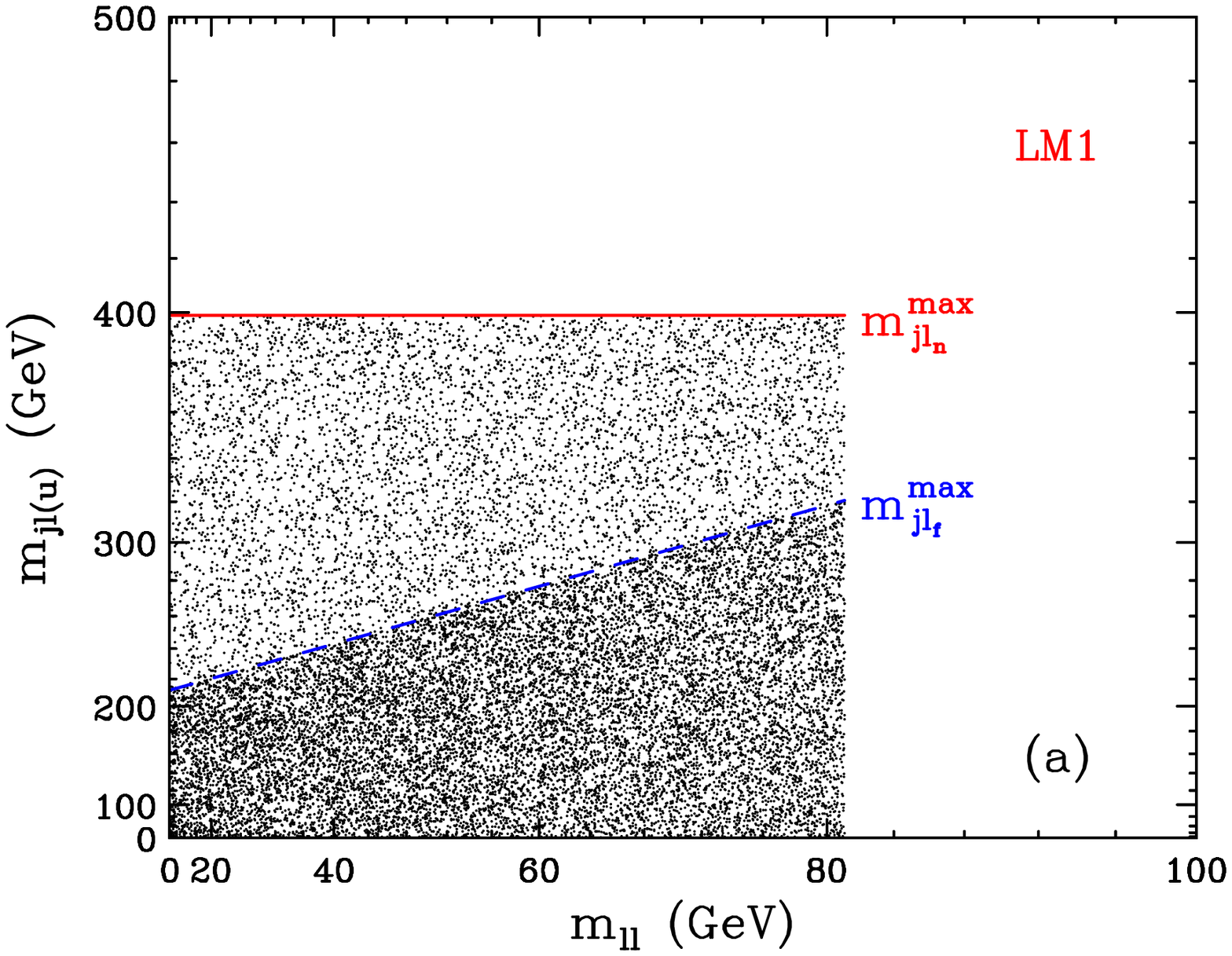,width=7.0cm}~~~
\epsfig{file=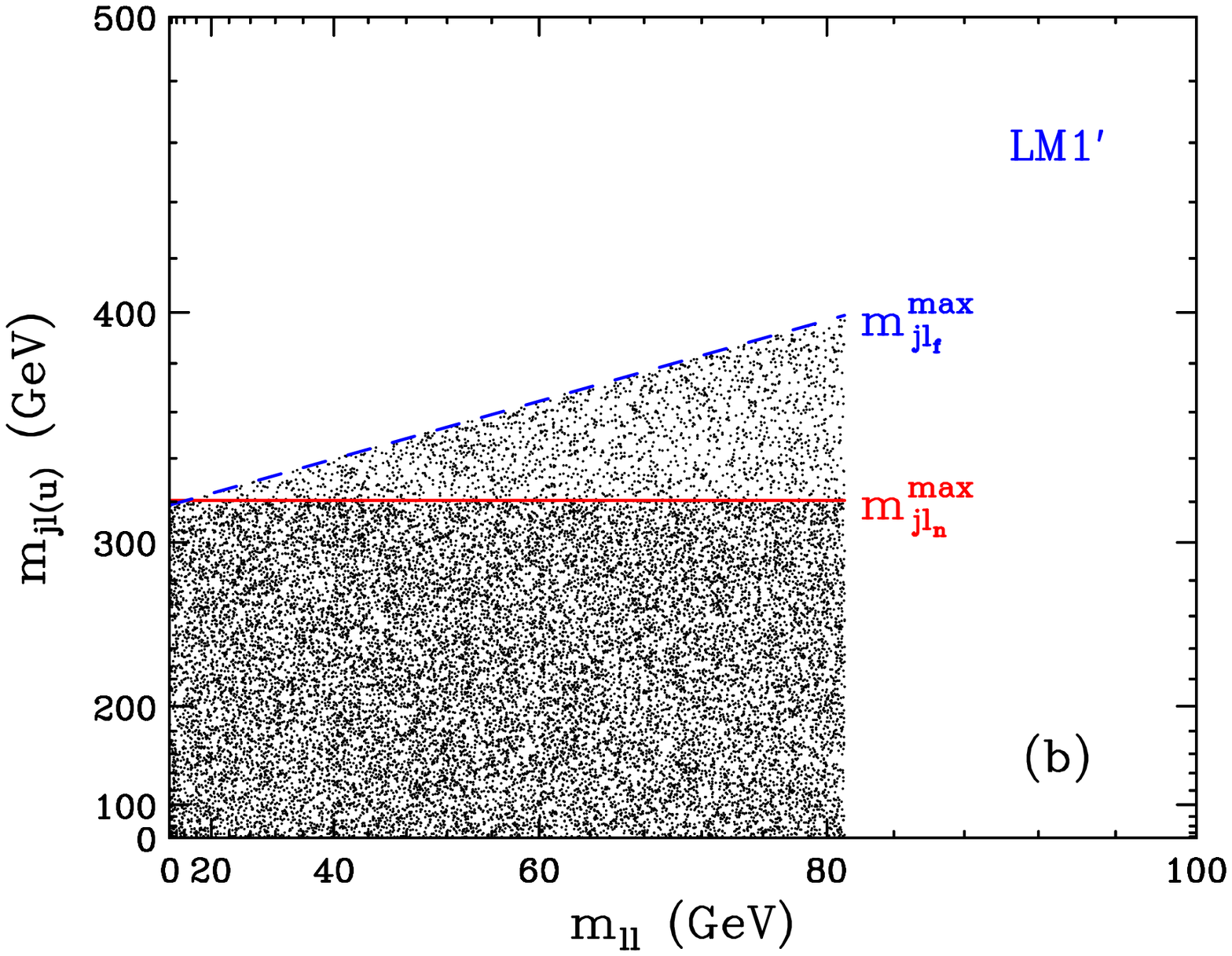,width=7.0cm}\\
\epsfig{file=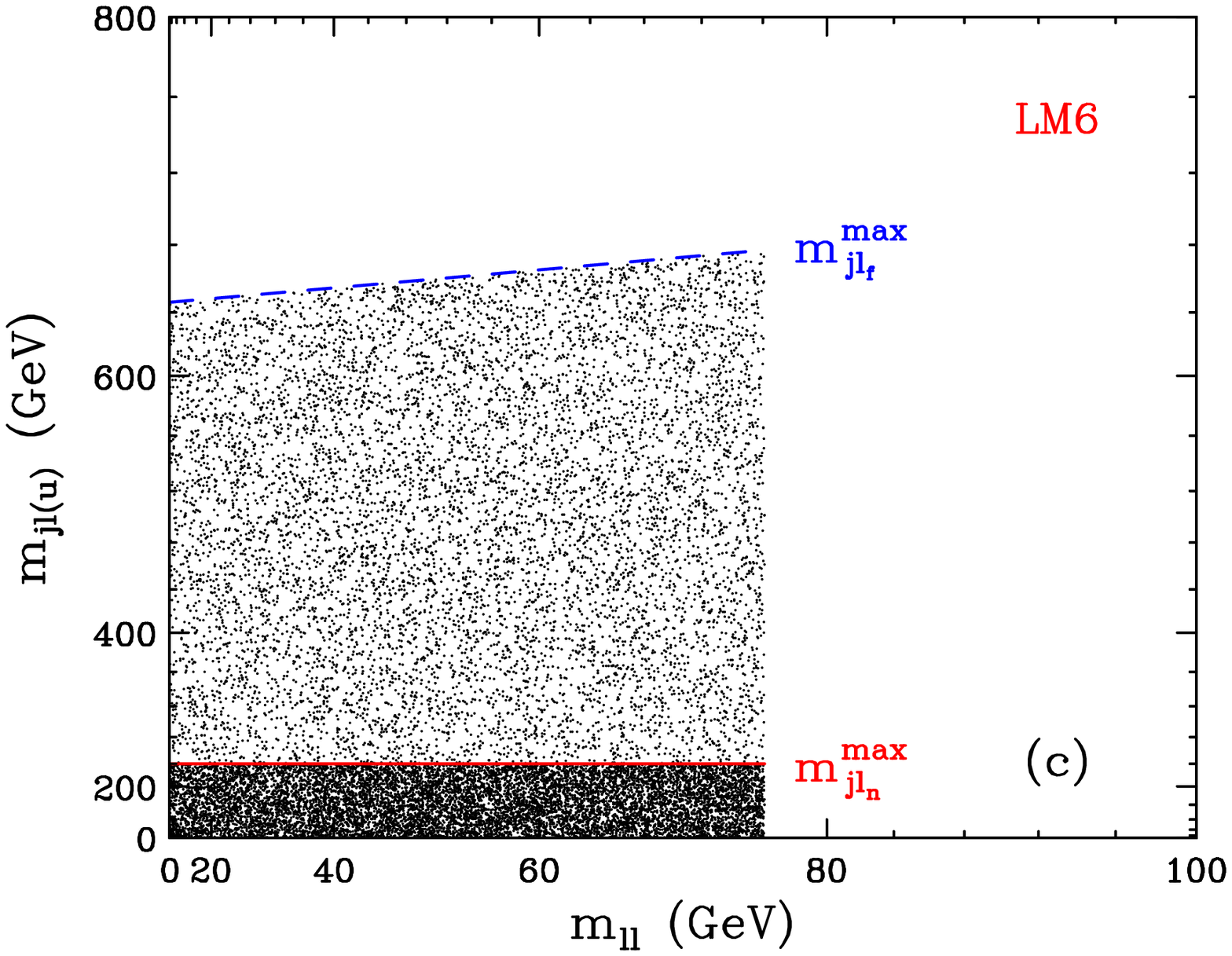,width=7.0cm}~~~
\epsfig{file=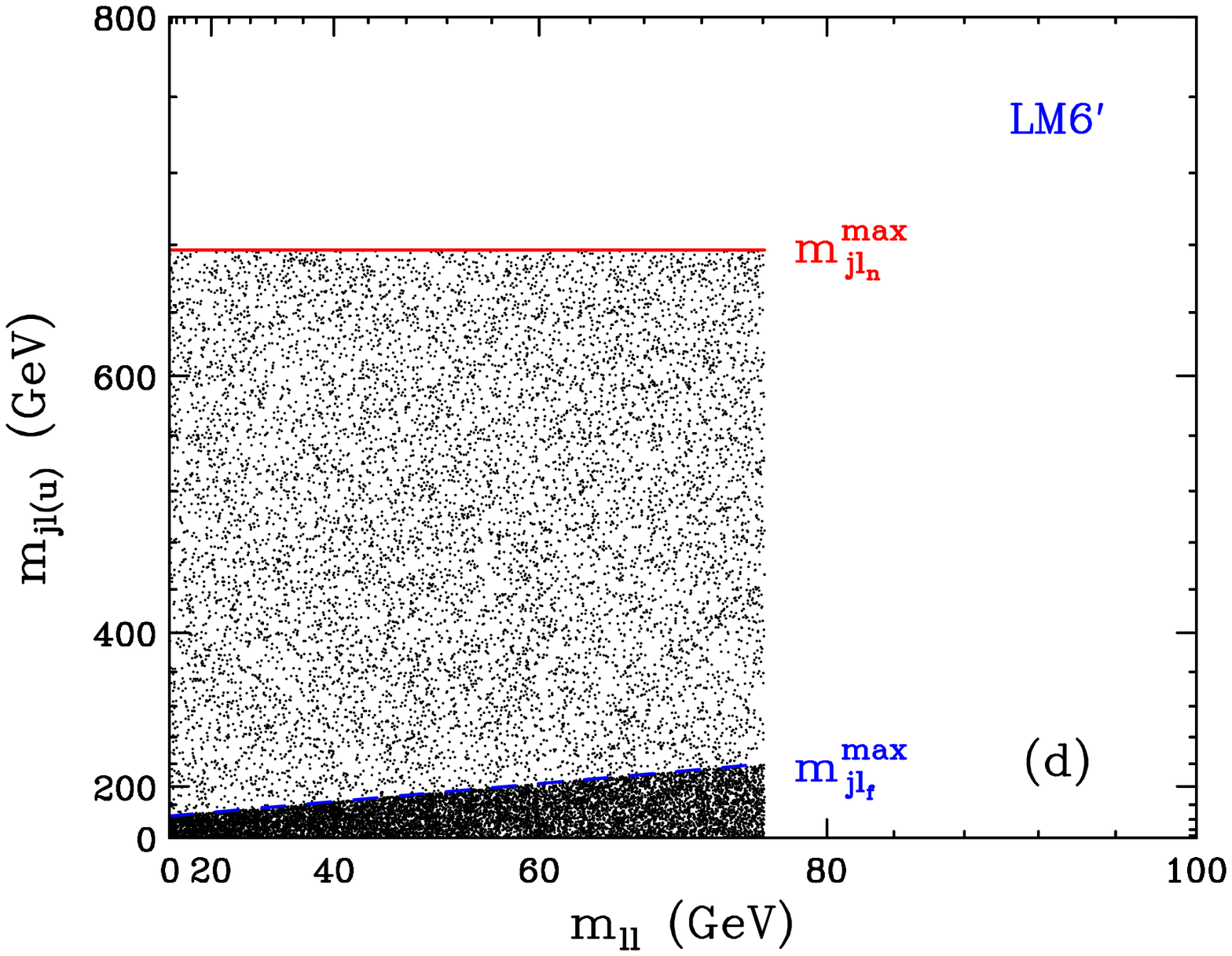,width=7.0cm}
\caption{\sl Predicted scatter plots of $m_{j\ell(u)}$ versus $m_{\ell\ell}$, 
for the case of the real and fake solutions for each of the two study points 
LM1 and LM6:
(a) the real solution LM1; (b) the fake solution LM1';
(c) the real solution LM6; and (d) the fake solution LM6'.
The red solid horizontal (blue dashed inclined) line indicates the 
conditional maximum $m^{max}_{j\ell_n}(m_{\ell\ell})$
($m^{max}_{j\ell_f}(m_{\ell\ell})$) given by eq.~(\ref{condmjln})
(eq.~(\ref{condmjlf})).
Each panel contains 10,000 entries. The results shown here are
idealized in the sense that we neglect smearing effects due to 
the finite detector resolution,
finite particle widths and combinatorial backgrounds.
Notice the use of quadratic power scale on the two axes,
which preserves the simple shapes of the scatter plots, 
even when plotted versus the linear masses $m_{j\ell(u)}$ and $m_{\ell\ell}$.
}
\label{fig:jlvsll}}

Fig.~\ref{fig:jlvsll} shows that the combined
distribution $m^2_{jl(u)}$ is simply composed of the two separate 
distributions $m^2_{j\ell_n}$ and $m^2_{j\ell_f}$, but they are correlated 
{\em differently} with the dilepton distribution $m^2_{\ell\ell}$.
In particular, let us concentrate on the conditional maxima
$m^{max}_{j\ell_n}(m_{\ell\ell})$ and 
$m^{max}_{j\ell_f}(m_{\ell\ell})$, i.e.
the maximum allowed values of $m_{j\ell_n}$ and 
$m_{j\ell_f}$, respectively, for a given fixed value of 
$m_{\ell\ell}$ \cite{Luc,Costanzo:2009mq}.
A close inspection of Fig.~\ref{fig:jlvsll}
shows that the values of $m^2_{j\ell_n}$
and $m^2_{\ell\ell}$ are uncorrelated, 
and as a result, the conditional maximum
$m^{max}_{j\ell_n}(m_{\ell\ell})$ does not depend on 
$m_{\ell\ell}$. In turn, this implies that the endpoint value 
$(m^{max}_{j\ell_n})^2$ given in (\ref{mjlnmax})
can be obtained for any $m^2_{\ell\ell}$:
\beq
n \equiv \left(m_{j\ell_n}^{max}\right)^2 = 
\left[m^{max}_{j\ell_n}(m_{\ell\ell})\right]^2 = m_D^2\, (1-R_{CD})\, (1-R_{BC}), 
\quad \forall \, m_{\ell\ell}\in \left[\, 0, m_{\ell\ell}^{max}\, \right]\ .
\label{condmjln}
\eeq
Because of (\ref{condmjln}), the shape
of the $m^2_{j\ell_n}$ versus $m^2_{\ell\ell}$ scatter plot is a 
simple rectangle \cite{Luc,Costanzo:2009mq}. This is confirmed
by the plots in Fig.~\ref{fig:jlvsll}, where the (red) horizontal 
solid line indicates the constant value (\ref{condmjln})
for the conditional maximum $m^{max}_{j\ell_n}(m_{\ell\ell})$.

In contrast, 
the values of $m^2_{j\ell_f}$ and $m^2_{\ell\ell}$ are correlated.
The conditional maximum $m^{max}_{j\ell_f}(m_{\ell\ell})$
{\em does} depend on the value of $m_{\ell\ell}$ as follows:
\beq
\left( m^{max}_{j\ell_f}(m_{\ell\ell}) \right)^2
= p + \frac{f-p}{L} m_{\ell\ell}^2\, ,
\label{condmjlf}
\eeq
where we introduce the shorthand notation used in \cite{Burns:2009zi}
\bea
f &\equiv & \left(m_{j\ell_f}^{max}\right)^2       = m_D^2\, (1-R_{CD})\, (1-R_{AB}), \label{fdef1} \\ [2mm]
p &\equiv & R_{BC}\, f                             = m_D^2\, (1-R_{CD})\, R_{BC}\,(1-R_{AB}). \label{pdef1}
\eea
The absolute maximum of $m^2_{j\ell_f}$, which is given by (\ref{mjlfmax}) and denoted here by $f$,
can only be obtained when $m^2_{\ell\ell}$ itself is at a maximum \cite{Luc,Costanzo:2009mq}:
\beq
f \equiv 
\left[m^{max}_{j\ell_f}(m_{\ell\ell}^{max})\right]^2\, .
\label{fcond}
\eeq
On the other hand, the conditional maximum $m^{max}_{j\ell_f}(m_{\ell\ell})$
obtains its minimum value at $m^2_{\ell\ell}=0$ and corresponds to
\cite{Luc,Costanzo:2009mq}
\beq
p \equiv \left[m^{max}_{j\ell_f}(0)\right]^2 \le f\, .
\label{pcond}
\eeq
Eqs.~(\ref{fcond},\ref{pcond}) imply that the shape of the $m^2_{j\ell_f}$ 
versus $m^2_{\ell\ell}$ scatter plot is a right-angle trapezoid. This is
confirmed by the plots in Fig.~\ref{fig:jlvsll}, where we mark with 
a (blue) dashed line the conditional maximum (\ref{condmjlf}).
With sufficient statistics, 
this difference in the kinematic boundaries may be observable, 
and would reveal the identity of $m^{max}_{j\ell_n}$ and 
$m^{max}_{j\ell_f}$ \cite{Luc,Costanzo:2009mq}. 
Once the individual $m^{max}_{j\ell_n}$ and 
$m^{max}_{j\ell_f}$ are known, the solution for the mass spectrum is unique
-- see e.g.~Appendix A in \cite{Burns:2009zi}.
Of course, in cases where $p\sim f$, namely $R_{BC}\sim 1$,
it may be difficult in practice to tell which of the two boundaries 
in the scatter plot is inclined and which one is horizontal\footnote{A 
separate problem, which arises in the case of $p\sim f$, will be discussed 
below in Sec.~\ref{sec:measurement}.}.
One example of this sort is offered by point LM6, which has 
$R_{BC}=0.91$ and leads to a rather flat $m^{max}_{j\ell_f}(m_{\ell\ell})$
function, as seen in Fig.~\ref{fig:jlvsll}(c).

An alternative and somewhat related method will be to investigate the 
shapes of the one-dimensional distributions themselves \cite{Georgia}. 
In Appendix~\ref{app:shapes} we provide the analytical expressions 
for the shapes of the four invariant mass distributions $m^2_{\ell\ell}$, 
$m^2_{j\ell(u)}$, $m^2_{j\ell(s)}(1)$ and $m^2_{j\ell(d)}(1)$ 
used in our basic analysis from Sec.~\ref{sec:theory}. 
Given what we have already seen in Fig.~\ref{fig:jlvsll}, it is
not surprising that the true and the fake solutions predict 
different shapes for the one-dimensional distributions 
as well. In the LM1 and LM6 examples considered below
in Sec.~\ref{sec:examples}, this difference is particularly
noticeable for the $m^2_{j\ell(u)}$ and $m^2_{j\ell(d)}(1)$ 
distributions (see Figs.~\ref{fig:LM1meas}(b), \ref{fig:LM1meas}(d),
\ref{fig:LM6meas}(b) and \ref{fig:LM6meas}(d)), 
and can be tested experimentally.

\subsubsection{$M_{T2}$ endpoint method}
\label{sec:MT2}

Let us note that if we identify particle $A$ with the LSP,
we have a rather peculiar situation, in which we {\em know}
the LSP mass $m_A$, and we are unsure about the NLSP mass $M_B$,
for which we have to choose among two alternatives. This goes against 
the common lore which considers the LSP mass (in this case $m_A$) 
to be the least constrained among the masses appearing in the decay 
chain in Fig.~\ref{fig:chain}. For example, the method of the Cambridge $M_{T2}$ 
variable \cite{Lester:1999tx,Barr:2003rg} treats the LSP mass 
as a continuous unknown parameter. At this point of our analysis 
we already know the LSP mass, and we can use this knowledge to our advantage.
For example, if we can collect a sufficient number of events 
of $B$ pair-production, we can apply the idea of $M_{T2}$ for
the $B\to A$ decay as in the original $M_{T2}$ proposal \cite{Lester:1999tx}. 
When we use for the trial LSP mass the known true value of 
$m_A$ given by (\ref{mAsol}), the kinematic endpoint of the $M_{T2}$ 
distribution will reveal the correct value of the mass
$m_B$ of the parent particle $B$, thus selecting the true
solution in (\ref{mBsol}).

As emphasized in Ref.~\cite{Burns:2008va}, the $M_{T2}$ 
endpoint method does not necessarily rely on
$A$ being the LSP (i.e. the very last particle in the decay chain)
or $B$ being the ``grandparent'' (i.e. the
very first particle in the decay chain). For example, suppose that
$A$ decays further. In that case, one simply needs to apply the more general
``subsystem'' variable $M_{T2}^{(n,p,c)}$ \cite{Burns:2008va} with 
$A$ being the ``child'' particle: $c=A$. Similarly, the two $B$
particles do not have to be the two grandparents 
initiating the decay chains: it is sufficient to consider 
$M_{T2}^{(n,p,c)}$ with $p=B$ and arbitrary $n$ \cite{Burns:2008va}.
Finally, for the purposes of selecting the correct solution
in (\ref{mBsol}) it is also possible to apply the subsystem  
variable $M_{T2}^{(n,p,c)}$ in a different way, 
where $B$ is the child, and the parent is either $D$ or $C$. 
In this case, we know the parent mass, which is respectively 
given by (\ref{mDsol}) or (\ref{mCsol}), and we are asking the question, 
which of the two test masses in (\ref{mBsol}) gives the correct
answer for the $M_{T2}$ endpoint.

\section{Numerical examples}
\label{sec:examples}

We shall now illustrate the ideas of the previous section with
two specific numerical examples: the LM1 and LM6 SUSY study points
in CMS \cite{Ball:2007zza}. The mass spectra at LM1 and LM6 are 
listed in Table~\ref{tab:spectra}. Point LM1 is similar to
benchmark point A (A') in Ref.~\cite{Battaglia:2001zp} (Ref.~\cite{Battaglia:2003ab})
and to benchmark point SPS1a in Ref.~\cite{Allanach:2002nj}.
Point LM6 is similar to 
benchmark point C (C') in Ref.~\cite{Battaglia:2001zp} (Ref.~\cite{Battaglia:2003ab}). 
The table also lists the corresponding duplicate solutions LM1' and LM6',
which are obtained by interchanging $R_{BC} \leftrightarrow R_{AB}$, or equivalently, by
replacing the mass of $B$ via
\beq
m_B \rightarrow m'_B = \frac{m_A m_C}{m_B}\, .
\label{mBreplacement}
\eeq

%
\TABULAR[ht]{|l||r|r||r|r|}{
\hline
Variable             &  LM1           &  LM1'       &  LM6         &   LM6'        \\  [1mm] \hline \hline
$m_A$ (GeV)          &  \multicolumn{2}{c||}{94.9}  & \multicolumn{2}{c|}{158.15}  \\ [1mm] \hline
$m_B$ (GeV)          &  118.9         &  143.35     & 291.0        &  165.65       \\ [1mm] \hline
$m_C$ (GeV)          &  \multicolumn{2}{c||}{179.6} & \multicolumn{2}{c|}{304.8}   \\ [1mm] \hline
$m_D$ (GeV)          &  \multicolumn{2}{c||}{561.6} & \multicolumn{2}{c|}{861.9}   \\ [1mm] \hline 
$R_{AB}$             &  0.6370        &    0.4383   & 0.2954       & 0.9115        \\ [1mm] \hline
$R_{BC}$             &  0.4383        &    0.6370   & 0.9115       & 0.2954        \\ [1mm] \hline
$R_{CD}$             &  \multicolumn{2}{c||}{0.1023}& \multicolumn{2}{c|}{0.1251}  \\ [1mm] \hline
\hline
$m_{\ell\ell}^{max}$ (GeV)     
                     & \multicolumn{2}{c||}{81.10} 
                                                    & \multicolumn{2}{c|}{76.12} \\ [1mm] \hline
$M_{j\ell(u)}^{max}$ (GeV)    
                     & \multicolumn{2}{c||}{398.8} 
                                                    & \multicolumn{2}{c|}{676.8} \\ [1mm] \hline
$m_{j\ell(u)}^{max}$ (GeV)    
                     & \multicolumn{2}{c||}{320.6} 
                                                    & \multicolumn{2}{c|}{239.8} \\ [1mm] \hline
$m_{j\ell(s)}^{max}(\alpha=1)$ (GeV)  
                     & \multicolumn{2}{c||}{451.8} 
                                                    & \multicolumn{2}{c|}{689.2} \\ [1mm] \hline
\hline
$m_{j\ell\ell}^{max}$ (GeV)   
                     & \multicolumn{2}{c||}{451.8} 
                                                    & \multicolumn{2}{c|}{689.2} \\ [1mm] \hline
$m_{j\ell\ell(\theta>\frac{\pi}{2})}^{min}$ (GeV)   
                     & \multicolumn{2}{c||}{215.2} 
                                                    & \multicolumn{2}{c|}{176.4} \\ [1mm] \hline
$m_{j\ell(hi)}^{max}$ (GeV)  
                     & \multicolumn{2}{c||}{398.8} 
                                                    & \multicolumn{2}{c|}{676.8} \\ [1mm] \hline
$m_{j\ell(s)}^{max}(\alpha=2)$ (GeV)  
                     & 406.6   & 398.8   &  676.8   & 677.0 \\ [1mm] \hline
$m_{j\ell(s)}^{max}(\alpha=1.5)$ (GeV)  
                     & 417.9   & 402.5   & 676.8    & 678.4 \\ [1mm] \hline
$m_{j\ell(s)}^{max}(\alpha=0.5)$ (GeV)  
                     & 611.0  &  638.9   & 886.0    & 807.1     \\ [1mm] \hline
$m_{j\ell(s)}^{max}(\alpha=-0.5)$ (GeV)  
                     & 142.9  &  159.7   & 174.9    & 138.0     \\ [1mm] \hline
$m_{j\ell(s)}^{max}(\alpha=-1)$ (GeV)  
                     & 200.1  &  225.9   & 224.8    & 184.8     \\ [1mm] \hline
$m_{j\ell(lo)}^{max}$ (GeV)  
                     & 274.6  &  319.1   & 239.8    & 229.9     \\ [1mm] \hline
$m_{j\ell(p)}^{max}$ (GeV)     
                     &  292.0 &  319.4   & 393.7    & 310.9     \\ [1mm] \hline\hline
$m_{j\ell_n}^{max}$ (GeV)
                     & 398.8          & 320.6    & 239.8   &  676.8     \\ [1mm] \hline
$m_{j\ell_f}^{max}$ (GeV) 
                     & 320.6          & 398.8    & 676.8   &  239.8    \\ [1mm] \hline
}{\label{tab:spectra} The relevant part of the SUSY mass spectrum for the LM1 and LM6 study points.
The corresponding duplicated solutions LM1' and LM6' are obtained by 
interchanging $R_{BC} \leftrightarrow R_{AB}$ as in (\ref{yzsymmetry}).
In the table we also list the corresponding values for various
invariant mass endpoints. The first four of those represent our basic 
set of measurements (\ref{ourmeas}) discussed in detail in Section~\ref{sec:measurement}, 
while the last two ($m_{j\ell_n}^{max}$ and $m_{j\ell_f}^{max}$) are not directly 
observable. The remaining invariant mass endpoints are 
considered in Section~\ref{sec:discrimination}.
In the case of $m^{max}_{j\ell(s)}(\alpha)$,
we show several representative values for $\alpha$. For the complete
$\alpha$ variation, refer to Fig.~\ref{fig:alpha}. Recall that 
$m_{j\ell(s)}^{max}(+\infty)=m_{j\ell(hi)}^{max}$ and
$m_{j\ell(s)}^{max}(-\infty)=m_{j\ell(lo)}^{max}$.
}

It is interesting to note that LM1 and LM6 represent both sides
of the ambiguity (\ref{yzsymmetry}): at LM1, we have $R_{AB}>R_{BC}$
and correspondingly, $m^{max}_{j\ell_n}> m^{max}_{j\ell_f}$
and (\ref{Mnear}) applies. On the other hand, at LM6 we have $R_{AB}<R_{BC}$
and $m^{max}_{j\ell_n} < m^{max}_{j\ell_f}$, so that (\ref{Mfar}) applies.
Another interesting difference is that at LM1 particle $B$ is the
{\em right-handed} slepton $\tilde \ell_R$, while at LM6
the role of particle $B$ is played\footnote{Although the 
right-handed slepton $\tilde \ell_R$ is also kinematically accessible at 
point LM6, the wino-like neutralino $\tilde\chi^0_2$ decays 
much more often to $\tilde \ell_L$ as opposed to $\tilde \ell_R$.} 
by the {\em left-handed} slepton $\tilde \ell_L$. 
Of course, to the extent that we are 
interested in kinematical features, this difference is not relevant,
and particle $B$ of the LM6 spectrum may very well have been the 
right-handed slepton instead.

\subsection{Mass measurements at points LM1 and LM6}
\label{sec:measurement}

Given the mass spectra in Table~\ref{tab:spectra}, it is straightforward 
to construct and investigate the relevant invariant mass distributions. 
For the purposes of illustration, we shall ignore spin correlations, 
referring the readers interested in those effects to 
Refs.~\cite{Smillie:2005ar,Athanasiou:2006ef,Burns:2008cp}.
We are justified to do so for several reasons. First, our method 
relies on the measurement of kinematic endpoints, whose location is 
unaffected by the presence of spin correlations. Second, in the case
of supersymmetry (which is really what we have in mind here), 
particle $B$ is a scalar, which automatically washes out any 
spin effects in the $m^2_{\ell\ell}$ and $m^2_{j\ell_f}$
distributions. Furthermore, if particles $D$ and their antiparticles $\bar{D}$ 
are produced in equal numbers, as would be the case if the dominant
production is from $gg$ and/or $q\bar{q}$ initial state, 
any spin correlations in the $m^2_{j\ell_n}$ distribution are also washed out.
Under those circumstances, therefore, the pure phase space distributions 
shown here are in fact the correct answer.

\FIGURE[t]{
\epsfig{file=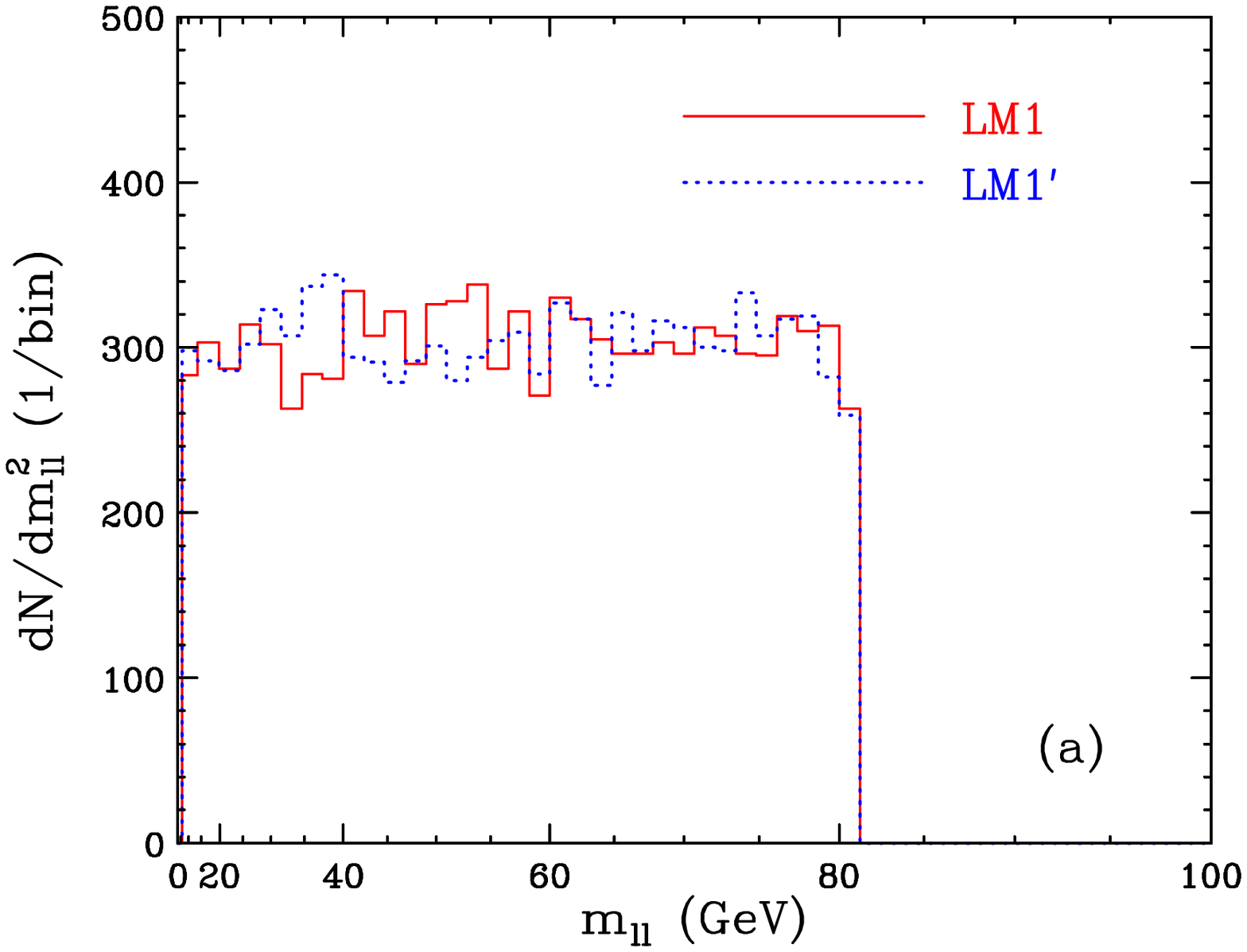,width=7.0cm}~~~
\epsfig{file=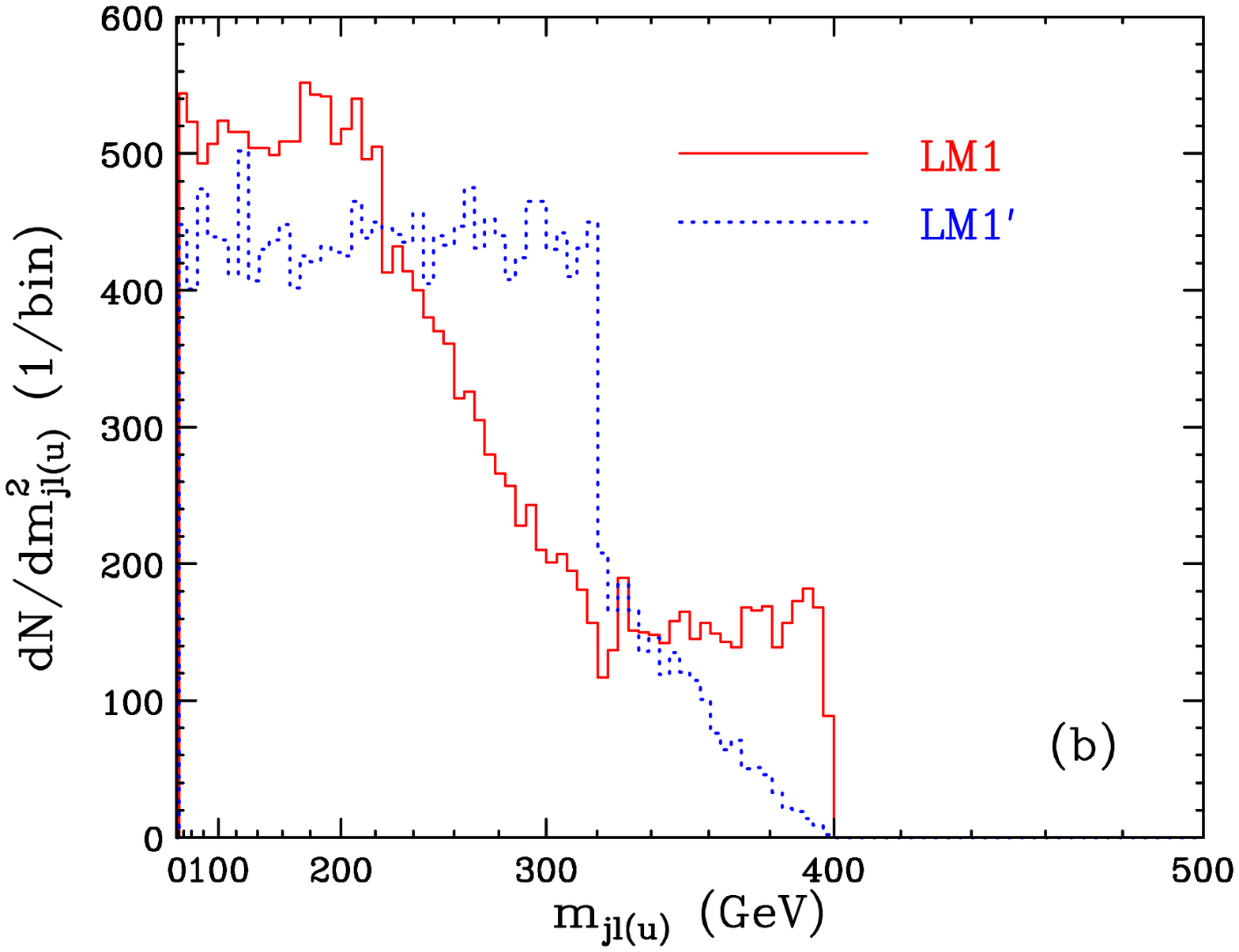,width=7.0cm}\\ [2mm]
\epsfig{file=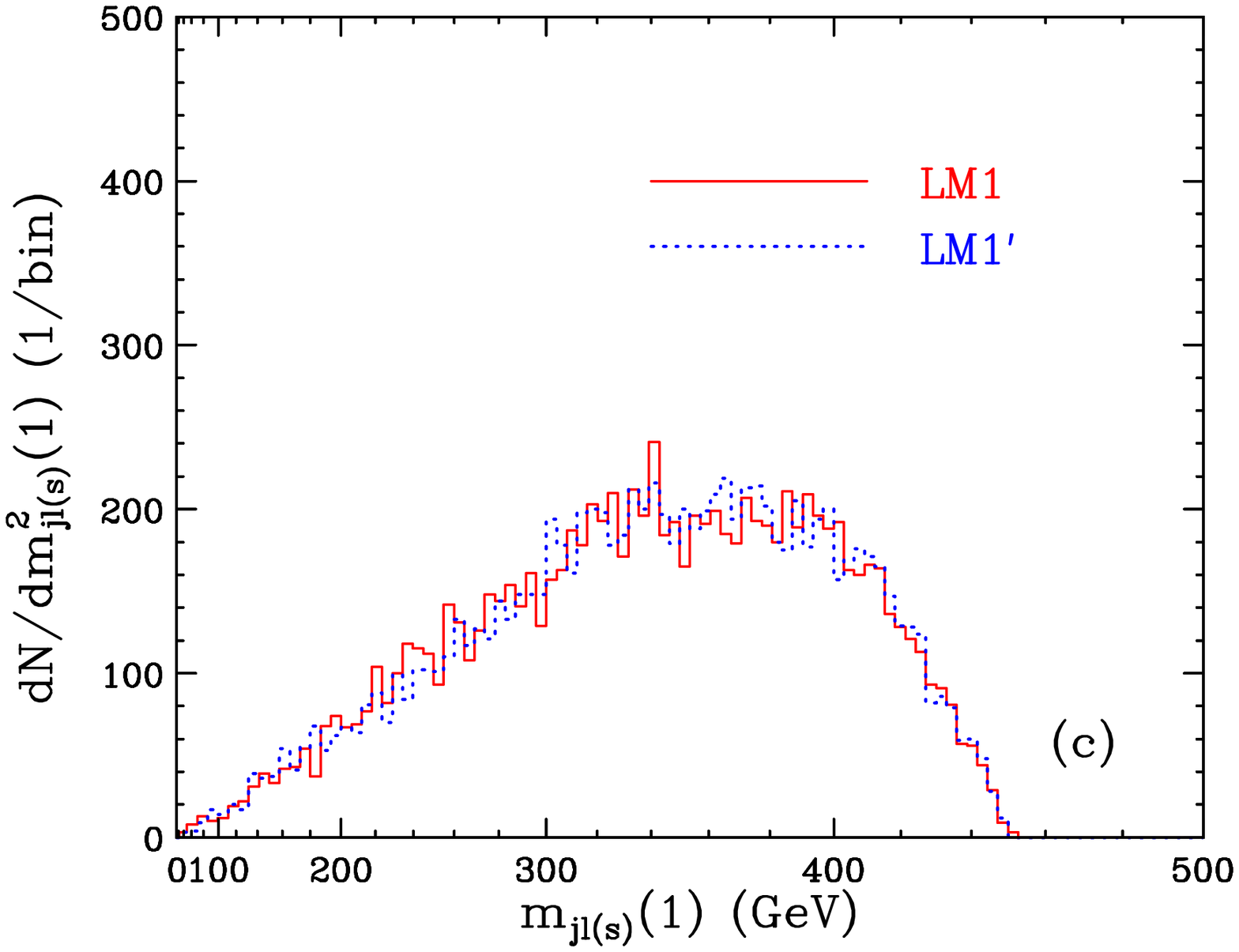,width=7.0cm}~~~
\epsfig{file=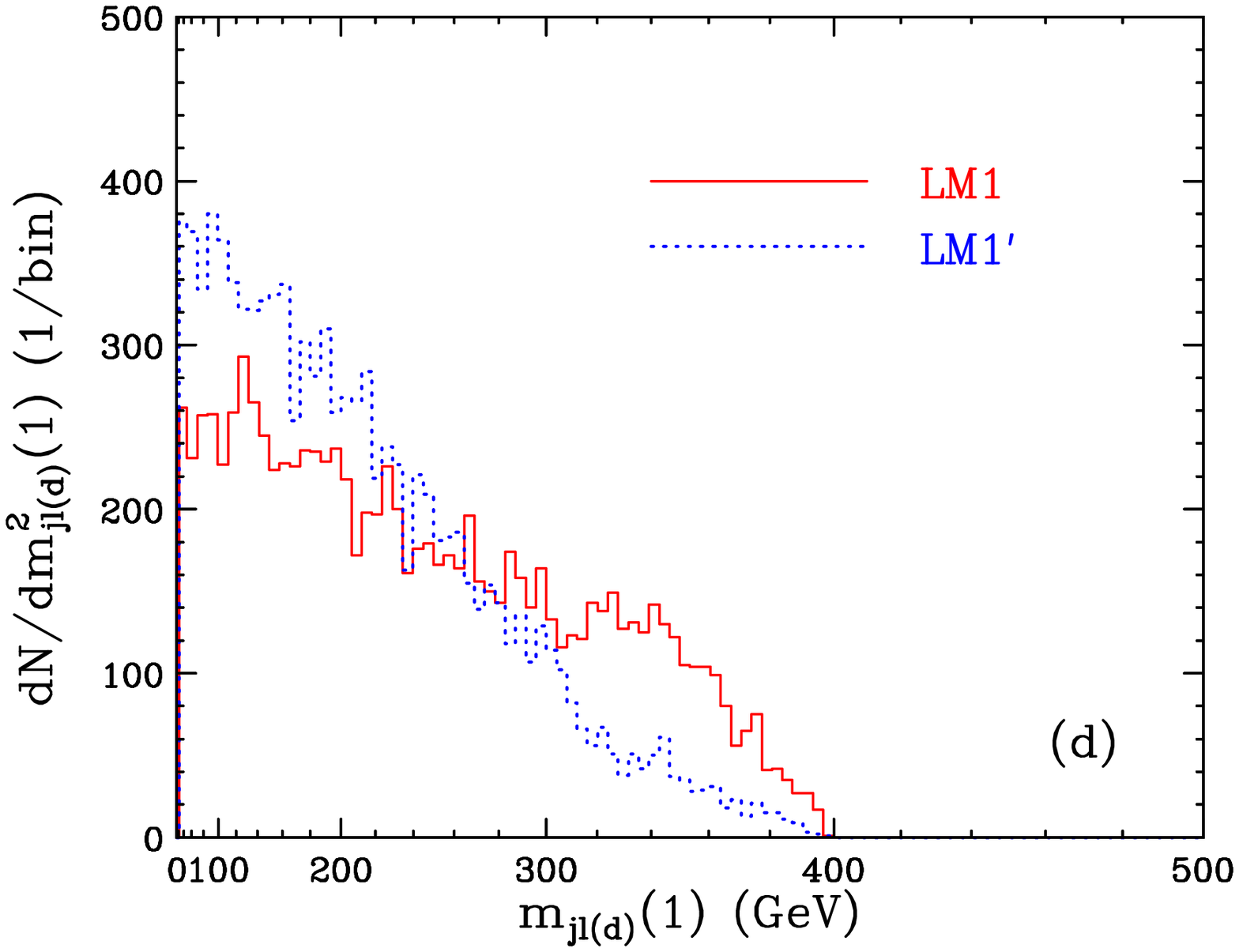,width=7.0cm}
\caption{\sl One-dimensional invariant mass distributions for the
case of LM1 (red solid lines) and LM1' (blue dotted lines) spectra. The kinematic endpoints 
(\ref{ourmeas}) used in our analysis in Section~\ref{sec:theory}
can be observed from these distributions as follows:
$m^{max}_{\ell\ell}$ is the upper kinematic endpoint 
of the $m_{\ell\ell}$ distribution in panel (a);
$M^{max}_{j\ell(u)}$ is the absolute upper kinematic 
endpoint seen in both the combined $m_{j\ell(u)}$ 
distribution in panel (b), or the difference
distribution $m_{j\ell(d)}(1)$ in panel (d);
$m^{max}_{j\ell(u)}$ is the intermediate kinematic endpoint seen 
in panel (b); and 
$m^{max}_{j\ell(s)}(\alpha=1)$ is the upper kinematic endpoint 
of the $m_{j\ell(s)}(\alpha=1)$ distribution in panel (c). }
\label{fig:LM1meas}}

\FIGURE[t]{
\epsfig{file=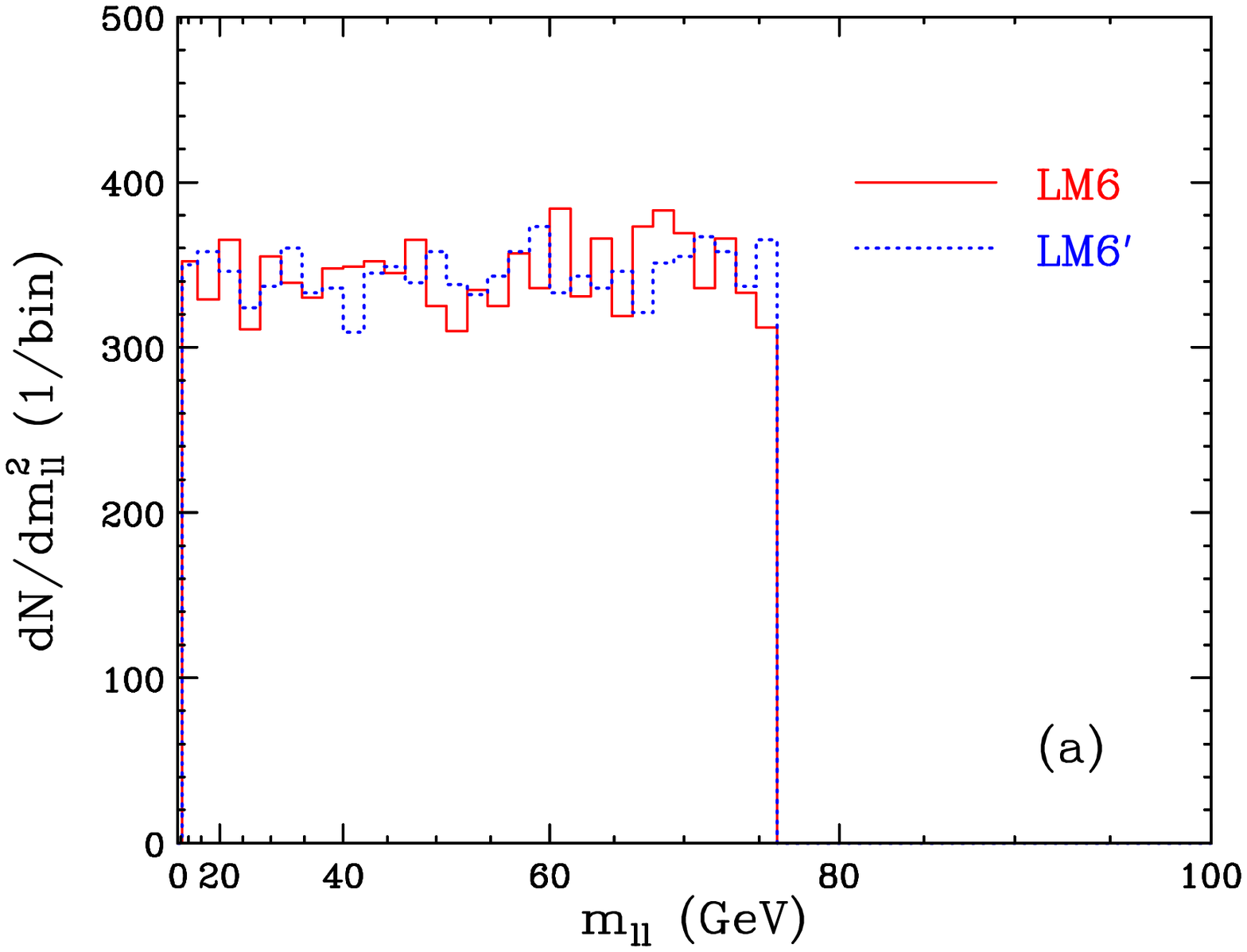,width=7.0cm}~~~
\epsfig{file=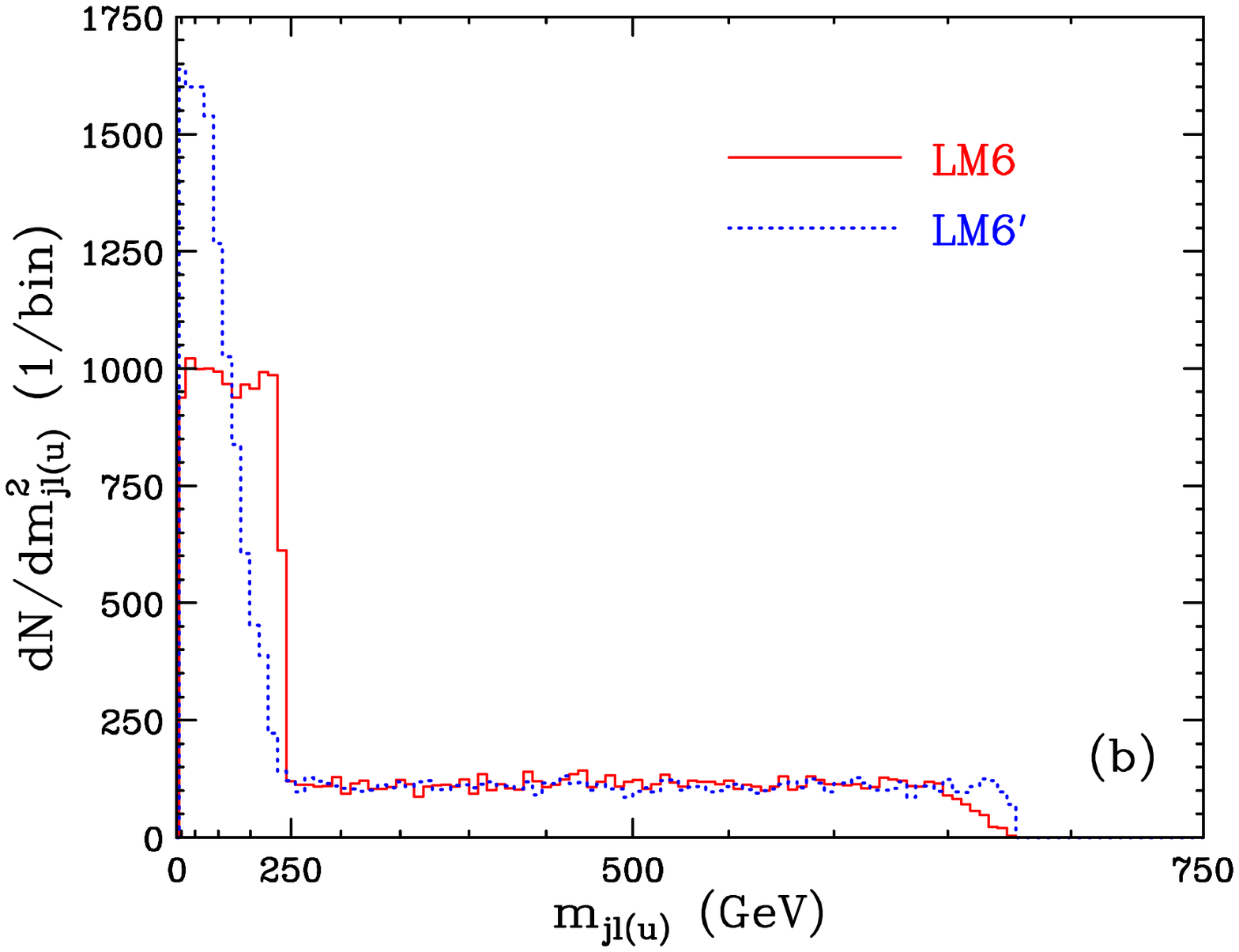,width=7.0cm}\\ [2mm]
\epsfig{file=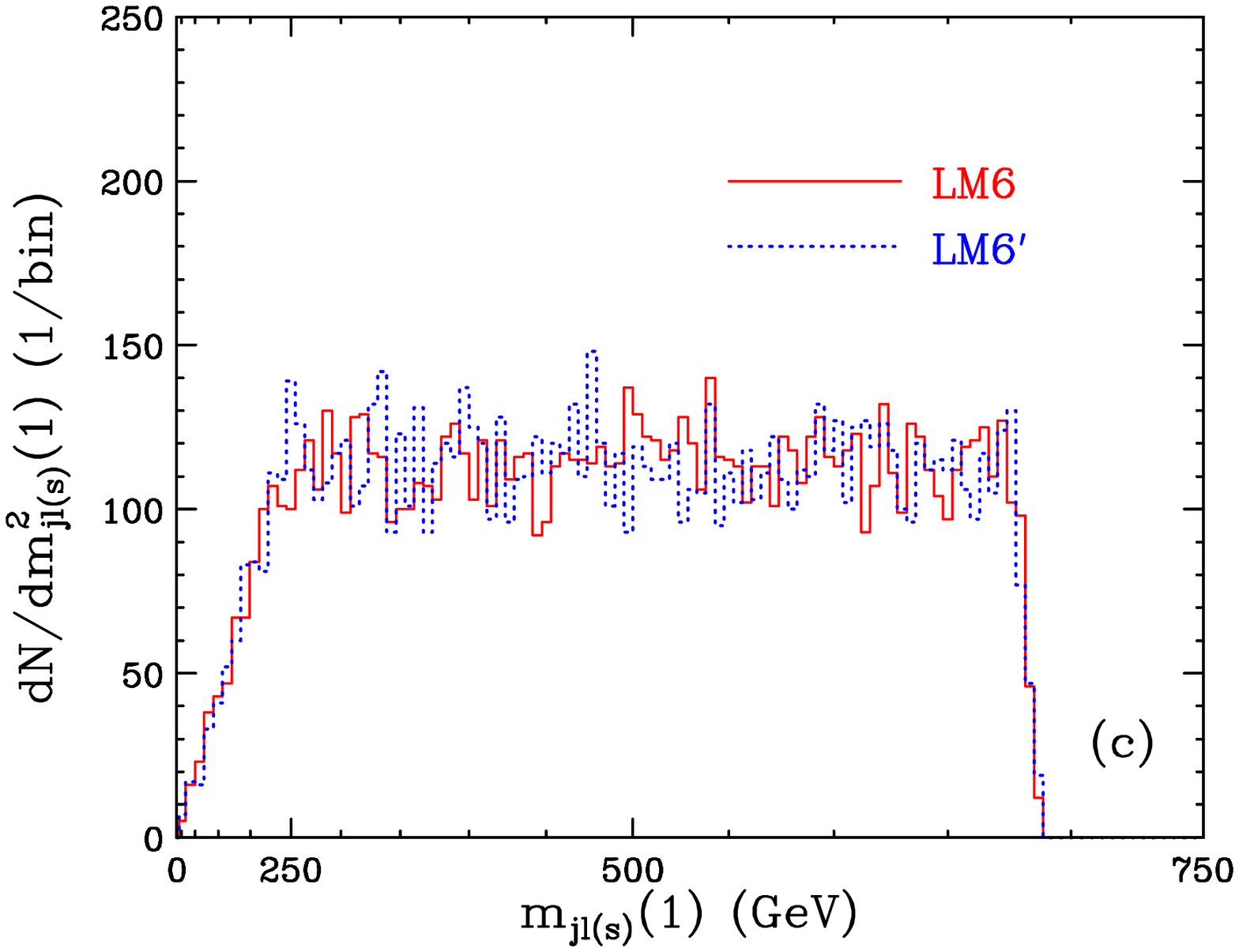,width=7.0cm}~~~
\epsfig{file=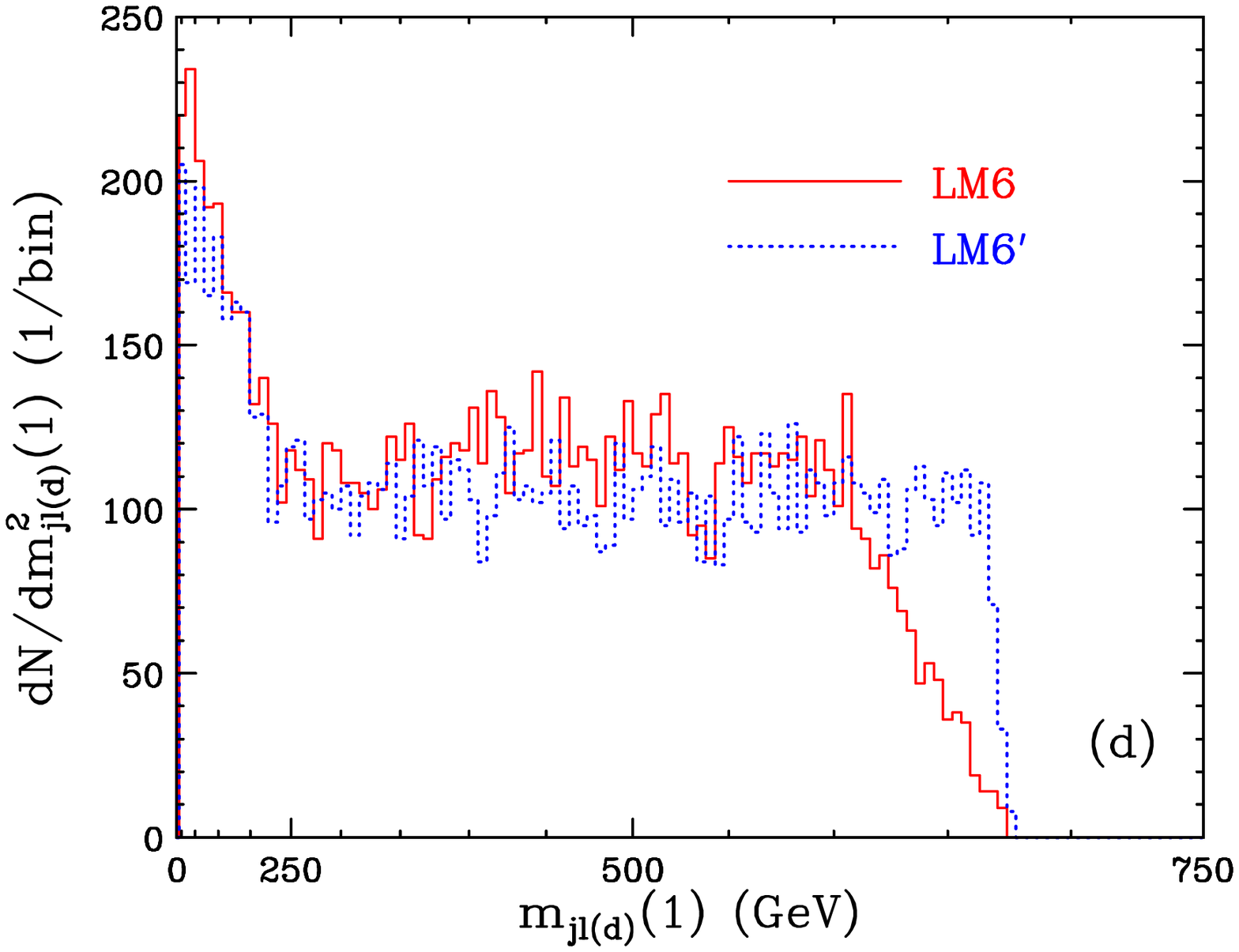,width=7.0cm}
\caption{\sl The same as Fig.~\ref{fig:LM1meas}, but for 
the LM6 mass spectrum (red solid lines) and the 
LM6' mass spectrum (blue dotted lines). }
\label{fig:LM6meas}}

We begin our discussion with the four invariant mass distributions
$m^2_{\ell\ell}$, $m^2_{j\ell(u)}$, $m^2_{j\ell(s)}(\alpha=1)$ and
$m^2_{j\ell(d)}(\alpha=1)$, which form the basis of our method 
outlined in Sec.~\ref{sec:theory}. Fig.~\ref{fig:LM1meas}
(Fig.~\ref{fig:LM6meas}) shows those four distributions
for the case of study point LM1 (LM6). In each panel, the 
red (solid) histogram corresponds to the nominal spectrum (LM1 or LM6),
while the blue (dotted) histogram corresponds to the ``fake'' solution
(LM1' or LM6'), which is obtained through the replacement 
(\ref{mBreplacement}). For all figures in this section, we use 
the same 4 samples of 10,000 events each, which were already
used to make Fig.~\ref{fig:jlvsll}. Notice our somewhat unconventional 
way of filling and then plotting the histograms in this section.
First, we show differential distributions in the corresponding 
mass squared, i.e. $dN/dm^2$. This is done in order to preserve the
connection to the analytical results in Appendix~\ref{app:shapes},
which are written the same way. More importantly, the 
shapes of the one-dimensional histograms are much 
simpler in the case of $dN/dm^2$ as opposed to $dN/dm$
\cite{Smillie:2005ar,Athanasiou:2006ef,Burns:2008cp}.
In the next step, however, we choose to plot the thus obtained 
histogram versus the mass itself rather than the mass squared.
This allows one to read off immediately the corresponding 
endpoint and compare directly to the values listed in 
Table~\ref{tab:spectra}. It also keeps the $x$-axis range 
within a manageable range. However, since the histograms were 
binned on a mass squared scale, if we were to use a linear scale on the 
$x$-axis, we would get bins with varying size. This 
would be rather inconvenient and more importantly, would distort the 
nice simple shapes of the $dN/dm^2$ distributions. 
Therefore, we use a quadratic scale on the $x$-axis,
which preserves the nice shapes and leads to a constant 
bin size on each plot. 

Figs.~\ref{fig:LM1meas} and \ref{fig:LM6meas}
illustrate how each one of the measurements (\ref{ourmeas}) 
can be obtained. For example, $m^{max}_{\ell\ell}$ is the 
classic upper kinematic endpoint of the $m_{\ell\ell}$ distributions
in Figs.~\ref{fig:LM1meas}(a) and \ref{fig:LM6meas}(a).
This endpoint is very sharp and should be easily observable.
$M^{max}_{j\ell(u)}$ is the absolute upper kinematic 
endpoint seen in the combined $m_{j\ell(u)}$ 
distribution in Figs.~\ref{fig:LM1meas}(b)
and \ref{fig:LM6meas}(b). Notice that the same endpoint 
can independently also be observed as the absolute upper kinematic 
limit of the difference distributions $m_{j\ell(d)}(1)$ 
shown in Figs.~\ref{fig:LM1meas}(d) and \ref{fig:LM6meas}(d). 
The fact that there are two independent ways of getting to
the endpoint $M^{max}_{j\ell(u)}$ should allow for a reasonable 
accuracy of its measurement. Upon closer inspection of the 
combined $m_{j\ell(u)}$ distribution in Figs.~\ref{fig:LM1meas}(b)
and \ref{fig:LM6meas}(b), we also notice the 
intermediate kinematic endpoint $m^{max}_{j\ell(u)}$ seen 
around 320 GeV in Fig.~\ref{fig:LM1meas}(b)
and around 240 GeV in Fig.~\ref{fig:LM6meas}(b).
Finally, $m^{max}_{j\ell(s)}(\alpha=1)$ is the upper kinematic endpoint 
of the $m_{j\ell(s)}(\alpha=1)$ distribution shown in
Figs.~\ref{fig:LM1meas}(c) and \ref{fig:LM6meas}(c).
It is also rather well defined, and should be 
well measured in the real data.

At this point we would like to comment on one potential problem 
which is not immediately obvious, but nevertheless has been encountered 
in practical applications of the invariant mass technique for
SUSY mass determinations \cite{Georgia}. It has been noted that
in the case of $p\sim f$ (see eqs.~(\ref{fdef1},\ref{pdef1})), 
the numerical fit for the mass spectrum becomes rather unstable.
Given our analytical results in Sec.~\ref{sec:theory}, 
we are now able to trace the root of the problem.
Notice that $p\sim f$ implies that $R_{BC}\sim 1$.
In this limit, from eqs.~(\ref{lldef}), (\ref{mjlnmax}), (\ref{mjlfmax}) 
and (\ref{mjls1max}) we find
\beq
\lim_{R_{BC}\to 1} \left(L\right) = 0, 
\qquad \lim_{R_{BC}\to 1} \left(n\right) = 0, 
\qquad \lim_{R_{BC}\to 1} \left(M+m-S\right) = 0.
\eeq
This means that the functions (\ref{mDsol}-\ref{mAsol})
giving the solution for the mass spectrum will all behave as $\frac{0^2}{0^2}$,
and, given the statistical fluctuations in an actual analysis,
will have very poor convergence properties. We note that this
problem is not limited to our preferred set of measurements 
(\ref{ourmeas}) and is rather generic, but has been missed
in most previous studies simply because the case of $R_{BC}\sim 1$ 
was rarely considered. 

Figs.~\ref{fig:LM1meas} and \ref{fig:LM6meas} reveal that,
as expected, the real (red solid lines) and fake (blue dotted lines) 
solutions always give identical results for our basic set 
of four endpoint measurements (\ref{ourmeas}).
This is by design, and in order to discriminate among 
the real and the fake solution, we need additional experimental 
input, as discussed in Section~\ref{sec:disambiguation}.
Before we proceed with the disambiguation analysis in the next subsection, 
we should stress once again that the real and fake solutions 
agree on 75\% of the relevant mass spectrum, i.e. they
give the same values for the masses of particles $D$, $C$ and $A$
(see Table~\ref{tab:spectra}).
The only question mark at this point is, what is the mass of 
particle $B$. This issue is addressed in the following subsection.

\subsection{Eliminating the fake solution for $m_B$}
\label{sec:discrimination}

As already discussed in Section~\ref{sec:disambiguation},
there are several handles which could discriminate among the
two alternative values of $m_B$ in the real and the fake solution.
One possibility, reviewed in Sec.~\ref{sec:MT2}, is to use 
additional independent measurements of $M_{T2}$ kinematic 
endpoints. We shall not pursue this direction here, 
referring the interested readers to Ref.~\cite{Burns:2008va}
for details. Another possibility, discussed in
Sec.~\ref{sec:2dim} and demonstrated explicitly 
with Fig.~\ref{fig:jlvsll}, is to use the different
correlations in the 2-dimensional invariant mass distributions
$(m^2_{\ell\ell},m^2_{j\ell_n})$ and $(m^2_{\ell\ell},m^2_{j\ell_f})$.
The near-far lepton ambiguity is avoided by studying the
scatter plot of $(m^2_{\ell\ell},m^2_{j\ell(u)})$, 
shown in Fig.~\ref{fig:jlvsll}, which should be in principle
sufficient to discriminate among the two alternatives.

\FIGURE[t]{
\epsfig{file=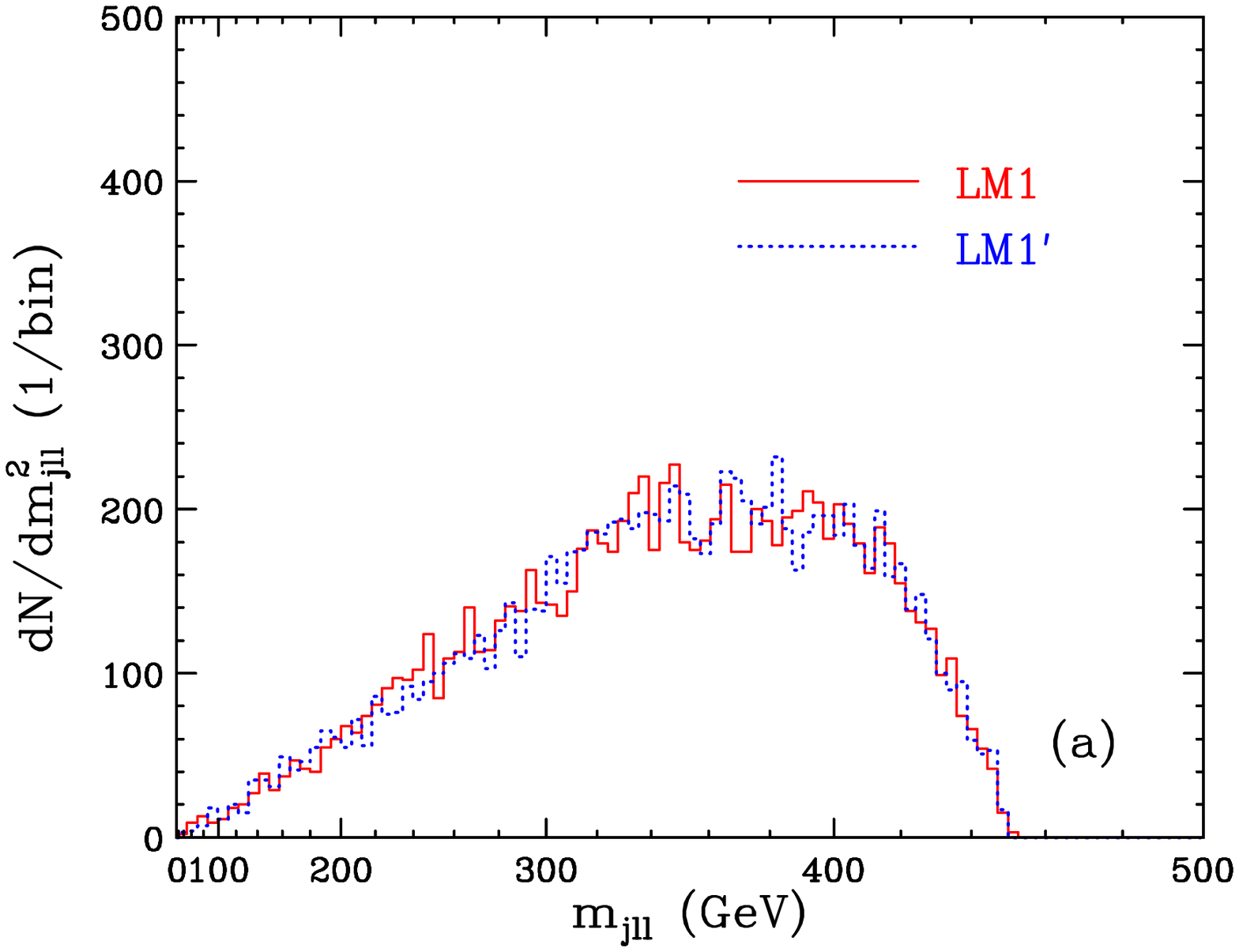,width=7.0cm}~~~
\epsfig{file=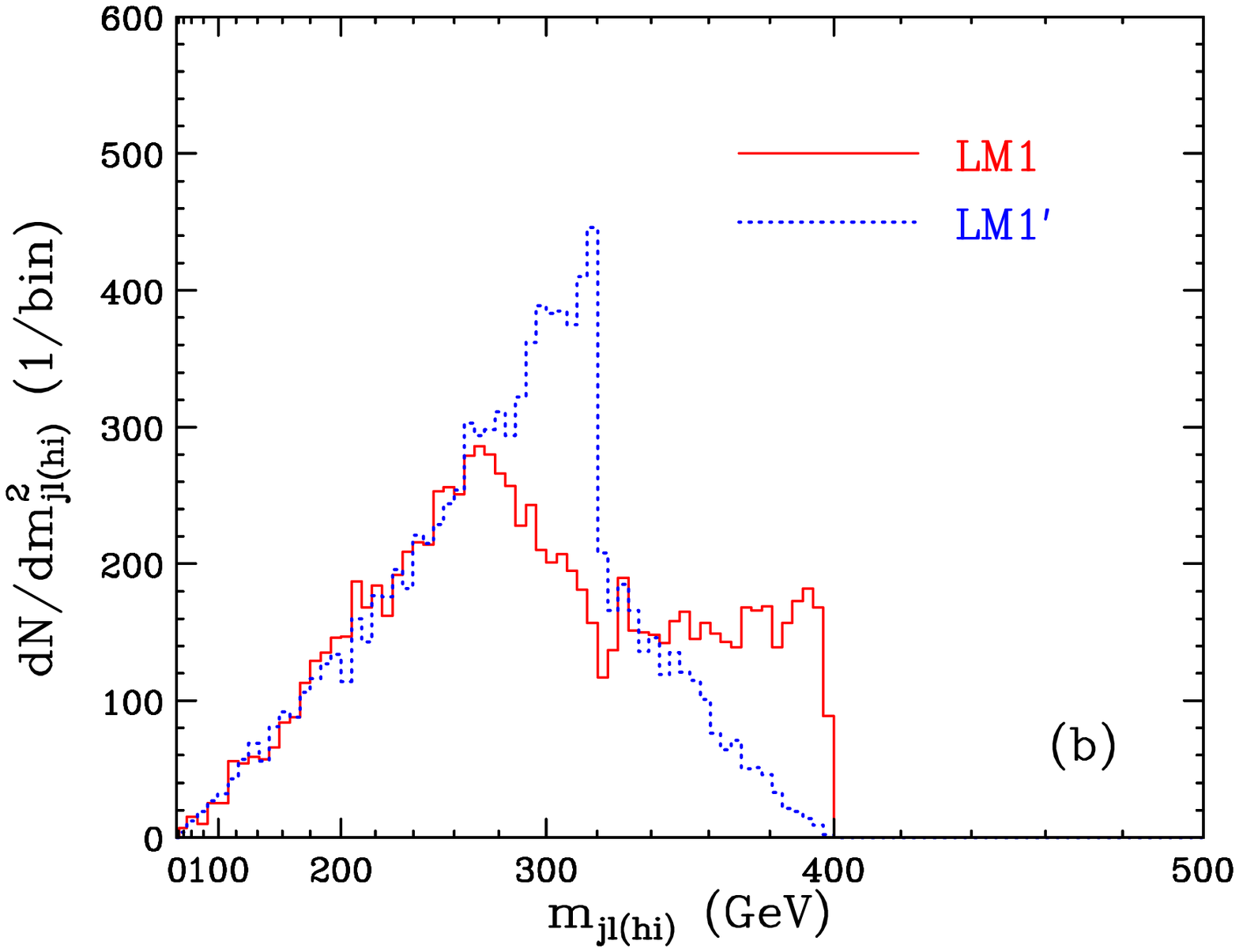,width=7.0cm}\\ [2mm]
\epsfig{file=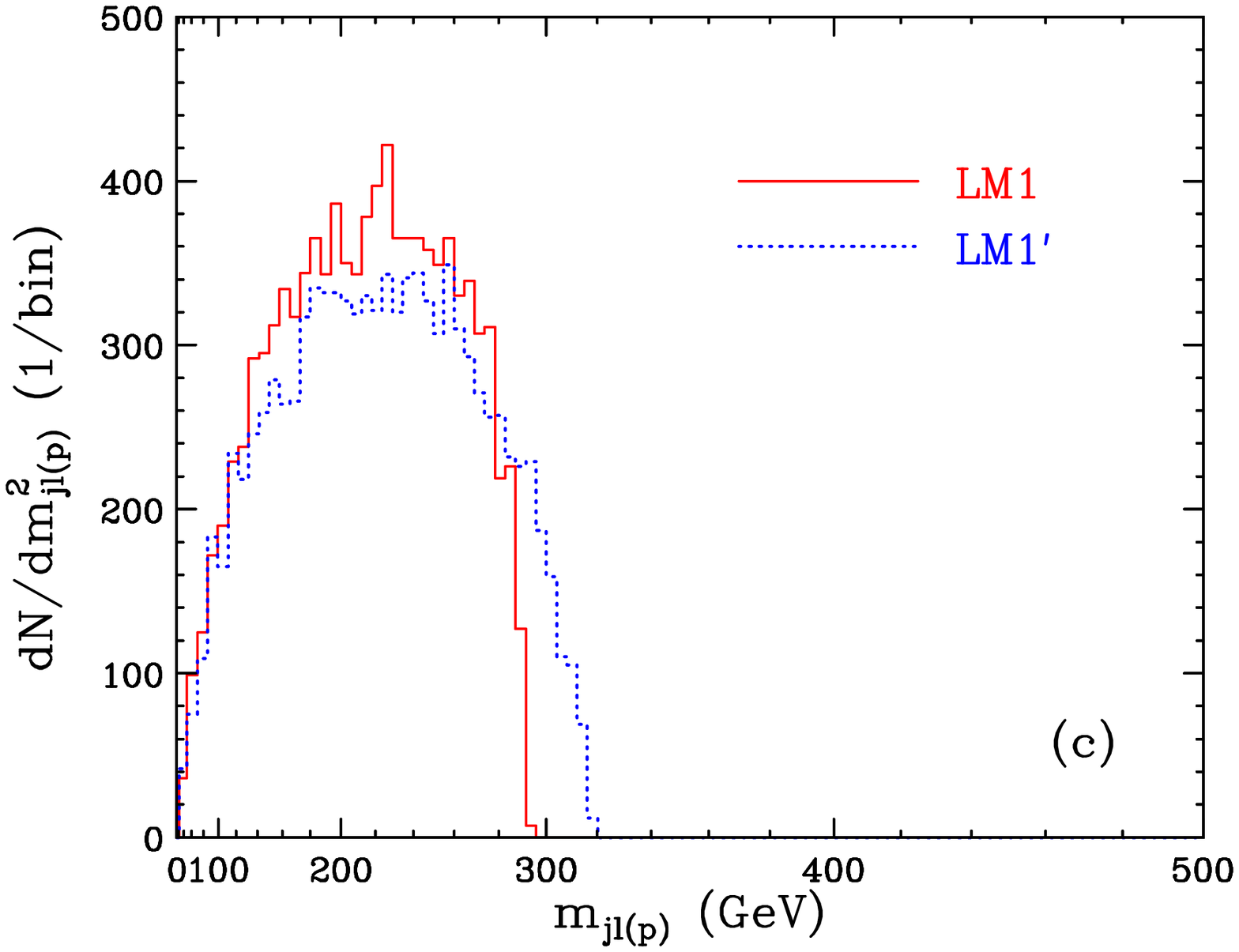,width=7.0cm}~~~
\epsfig{file=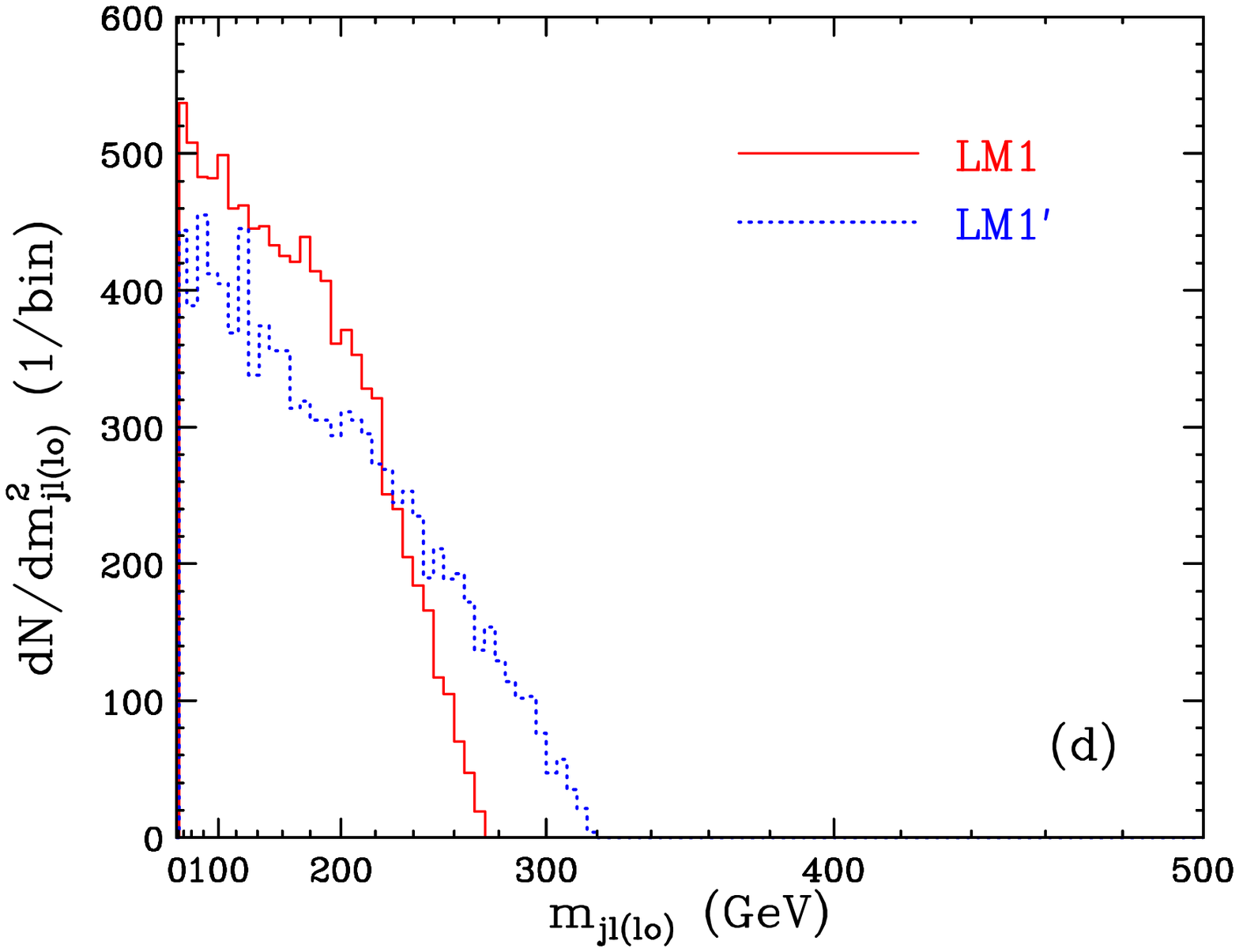,width=7.0cm}\\ [2mm]
\epsfig{file=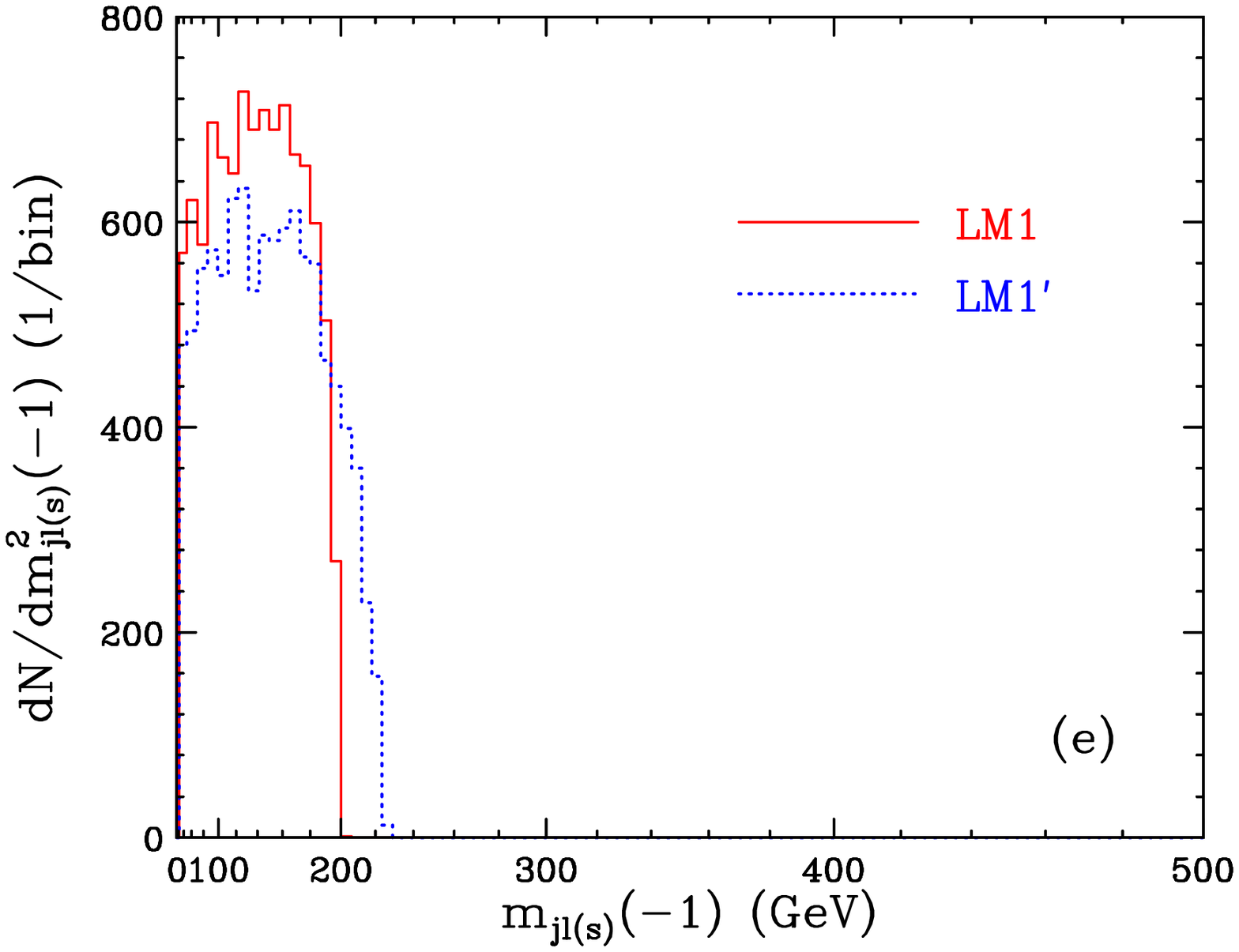,width=7.0cm}~~~
\epsfig{file=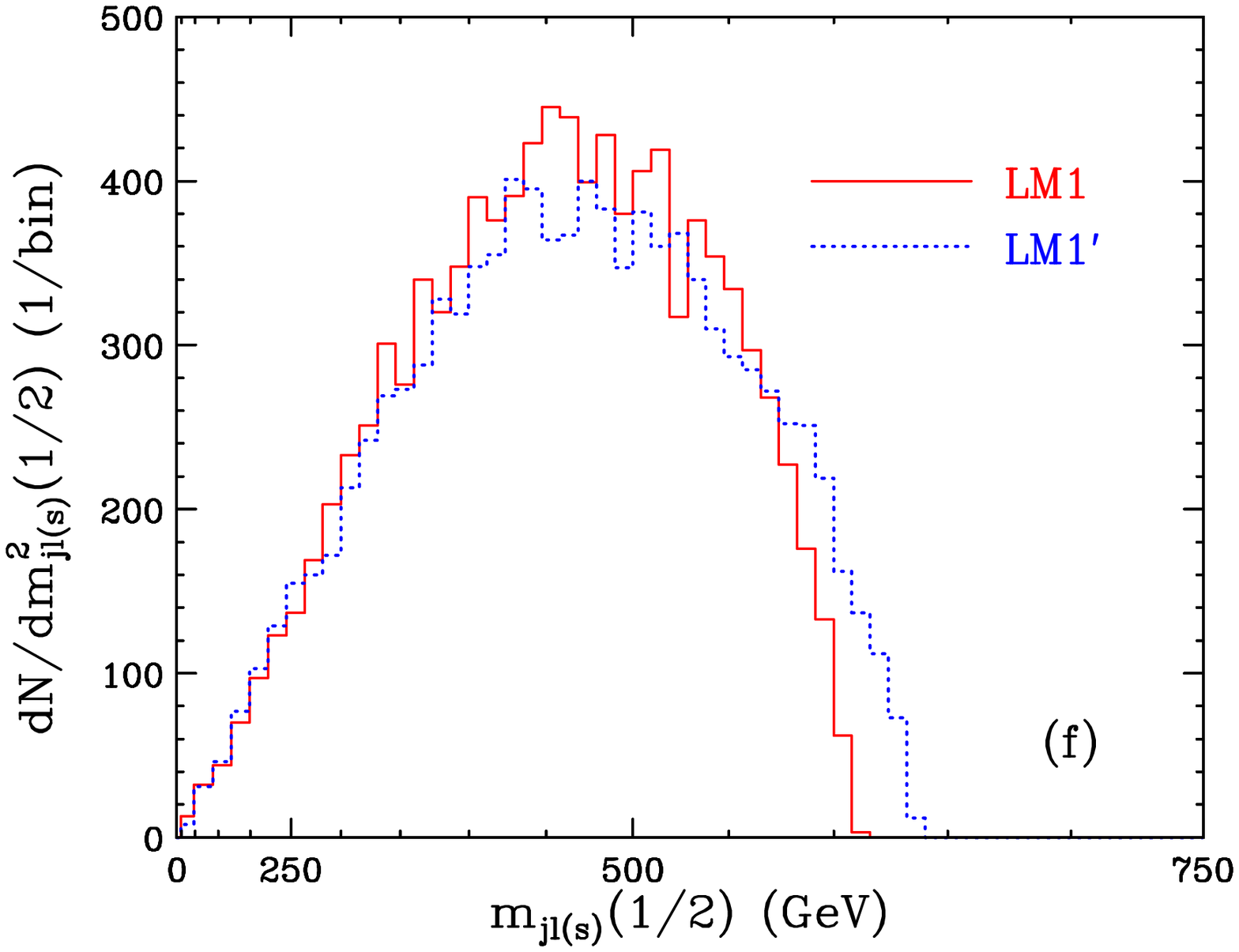,width=7.0cm}
\caption{\sl Some other one-dimensional invariant mass 
distributions of interest, for the 
case of the LM1 mass spectrum (red solid lines) and LM1' 
mass spectrum (blue dotted lines): 
(a) $m^2_{j\ell\ell}$ distribution;
(b) $m^2_{j\ell(hi)}$ distribution;
(c) $m^2_{j\ell(p)}$ distribution;
(d) $m^2_{j\ell(lo)}$ distribution;
(e) $m^2_{j\ell(s)}(\alpha=-1)$ distribution;
(f) $m^2_{j\ell(s)}(\alpha=\frac{1}{2})$ distribution.
All distributions are then plotted versus the corresponding 
mass, on a quadratic scale for the $x$-axis.
}
\label{fig:LM1disc}}

\FIGURE[t]{
\epsfig{file=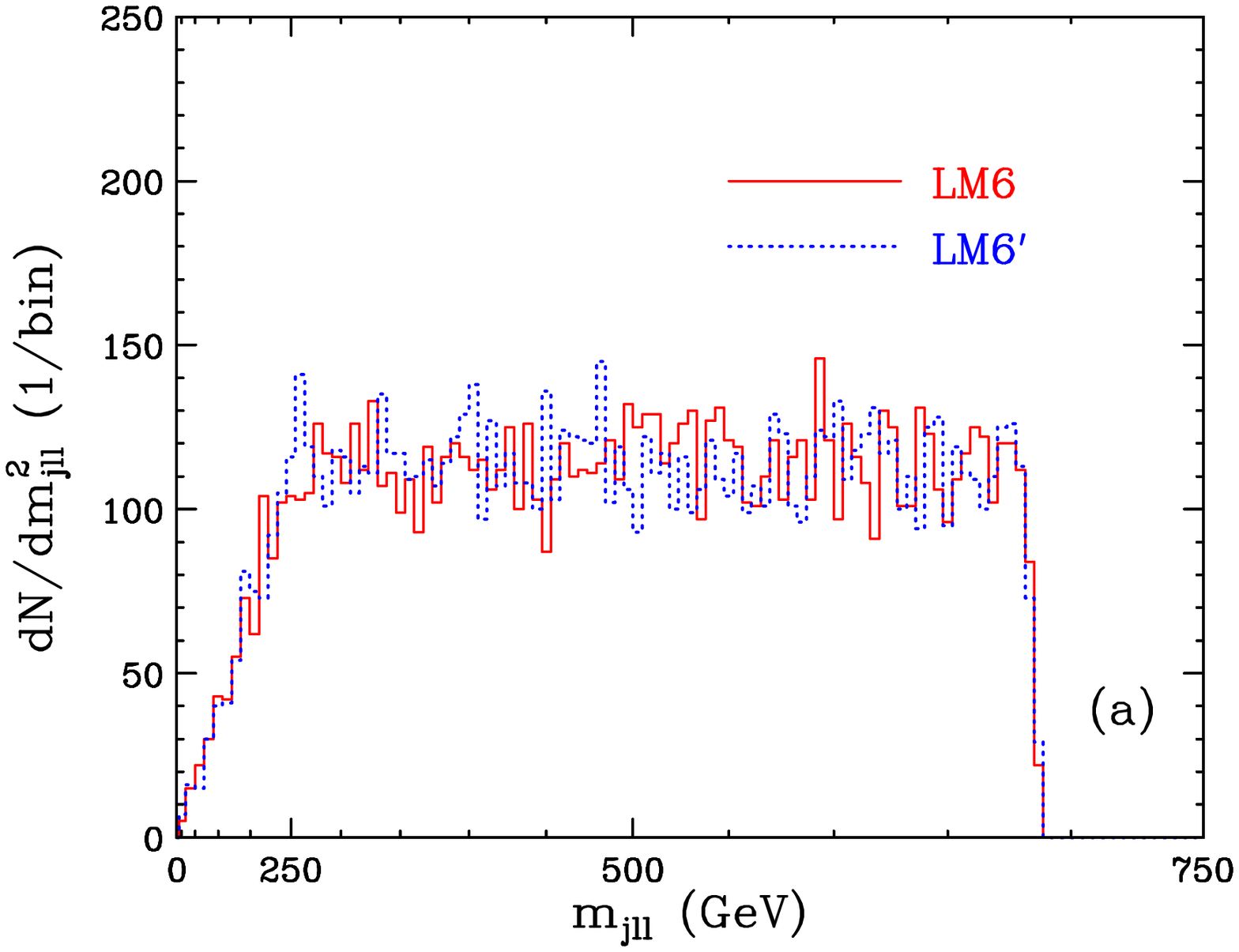,width=7.0cm}~~~
\epsfig{file=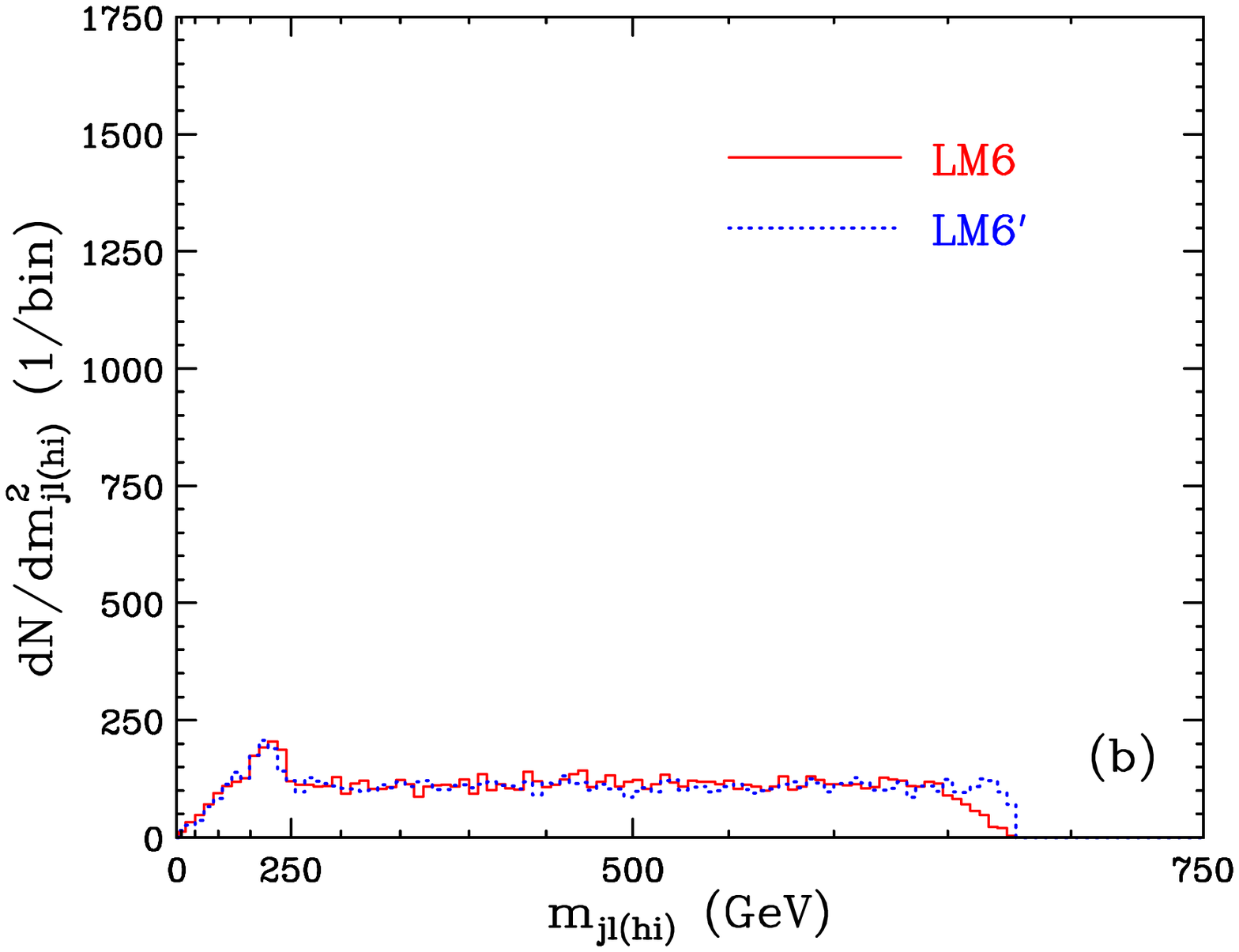,width=7.0cm}\\ [2mm]
\epsfig{file=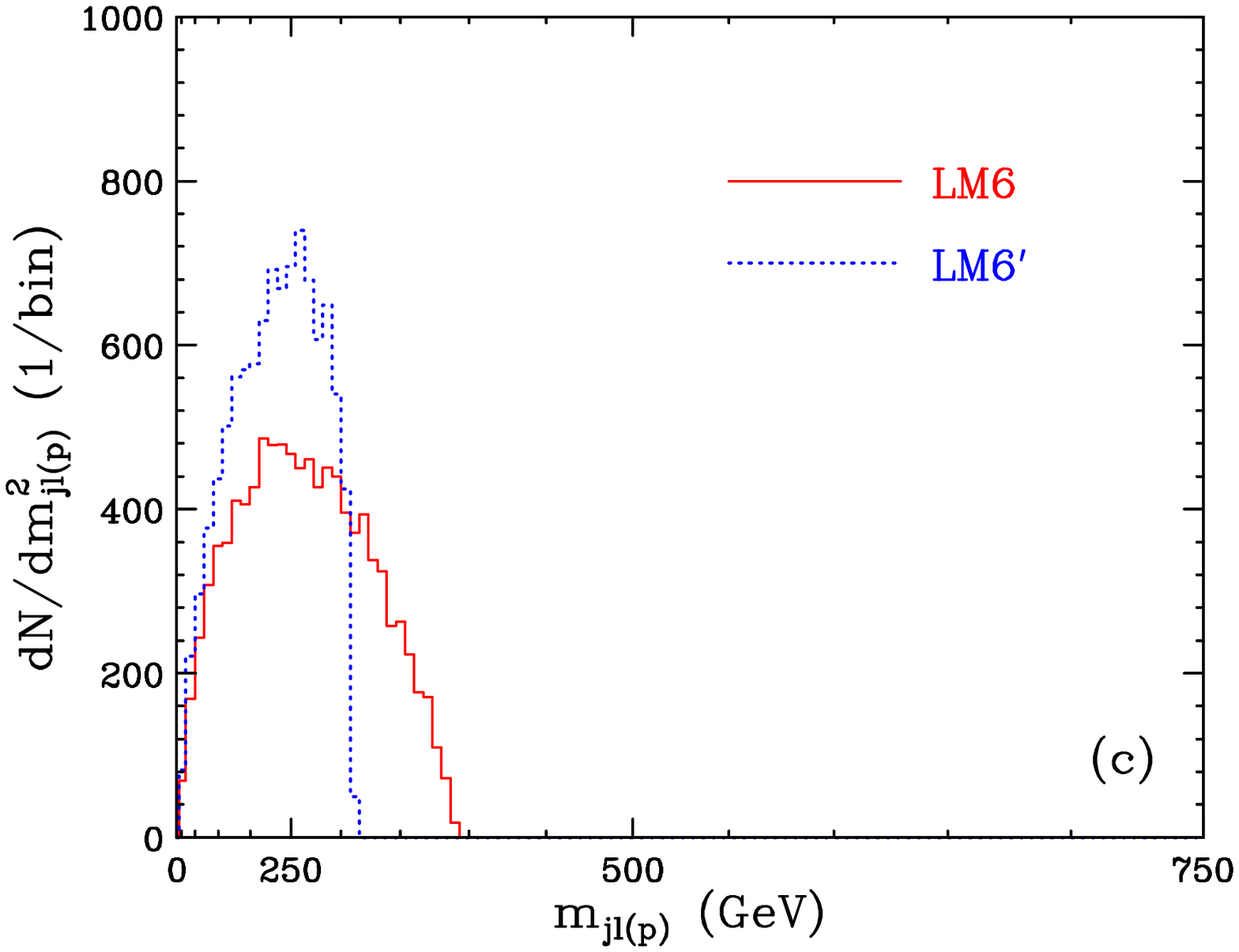,width=7.0cm}~~~
\epsfig{file=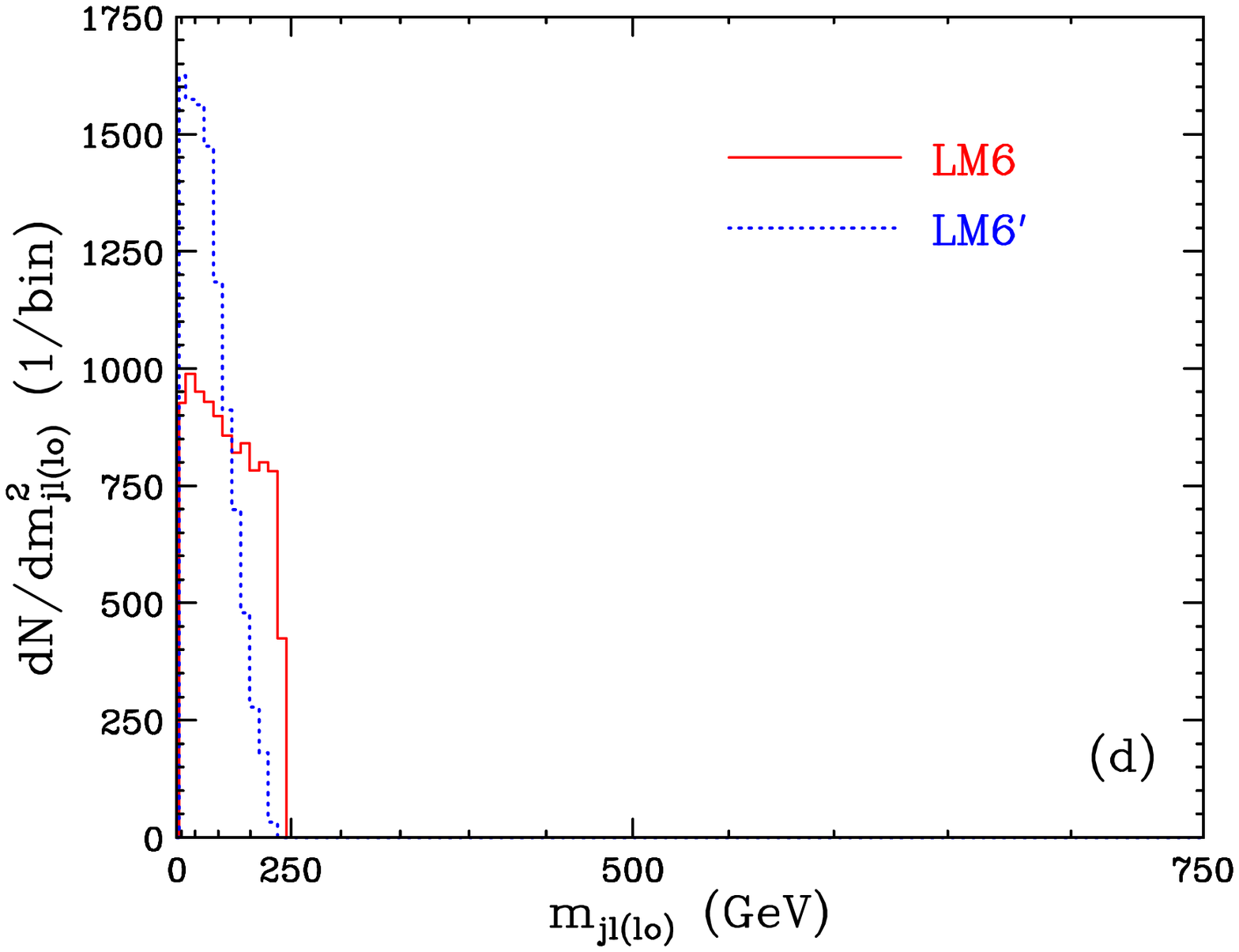,width=7.0cm}\\ [2mm]
\epsfig{file=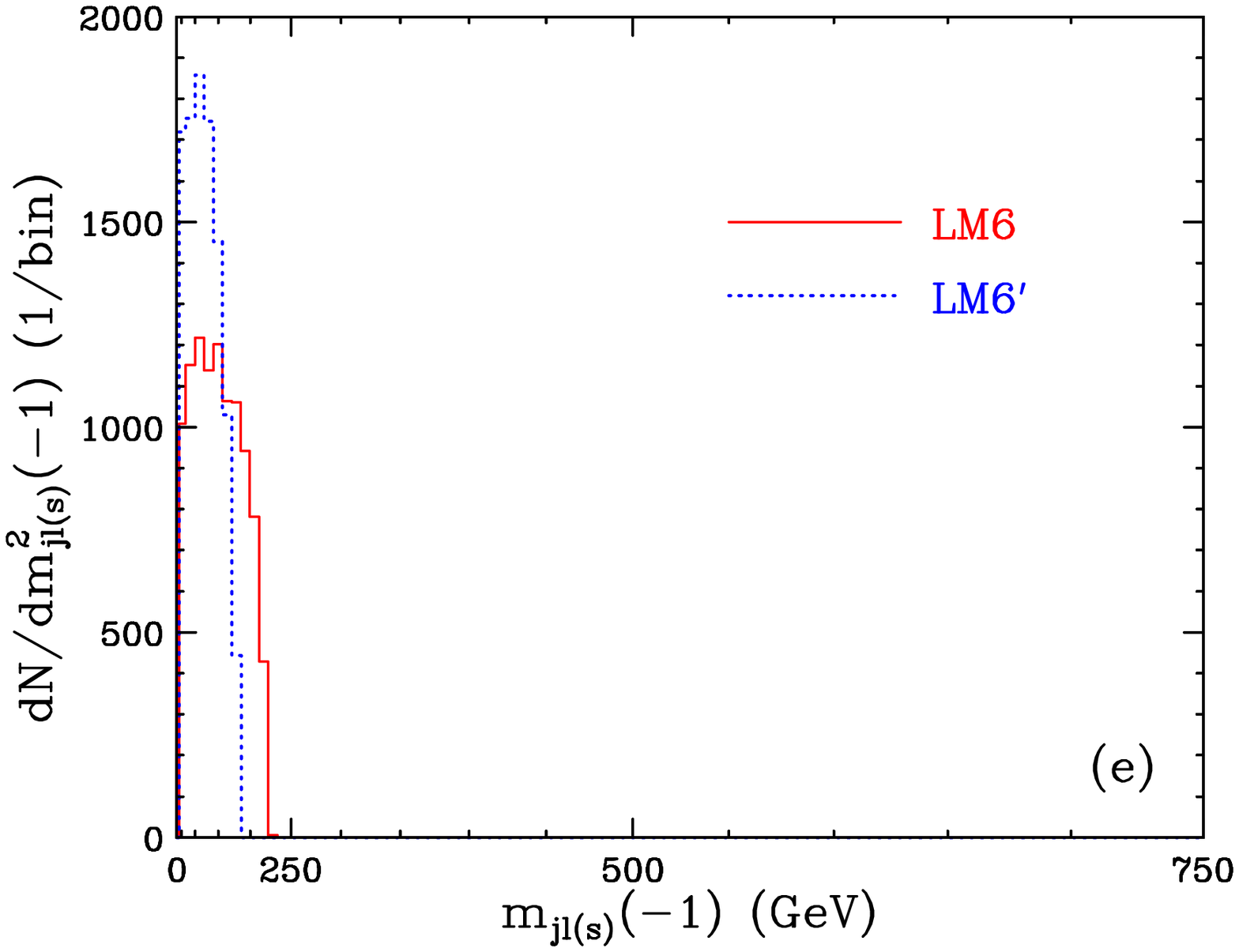,width=7.0cm}~~~
\epsfig{file=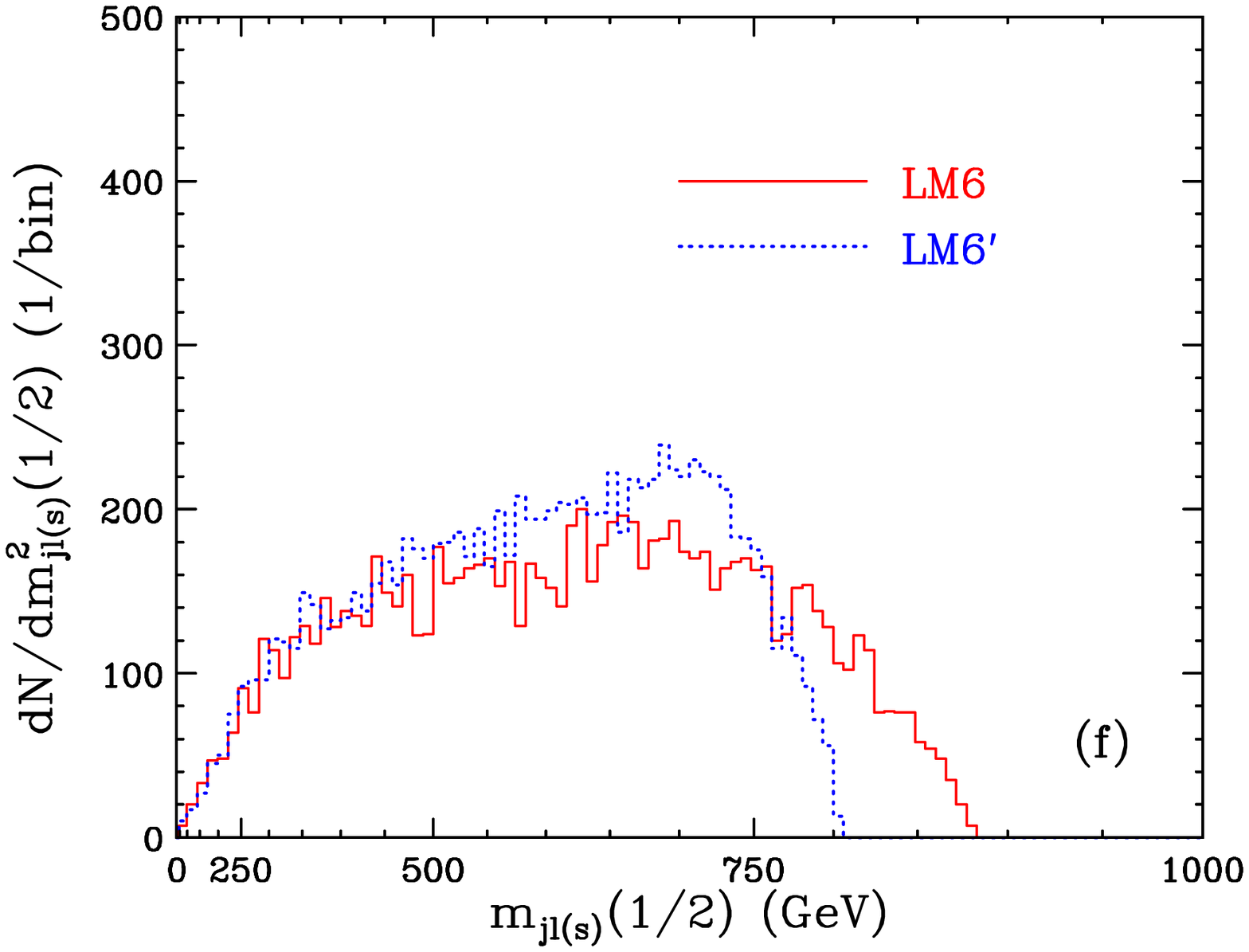,width=7.0cm}
\caption{\sl The same as Fig.~\ref{fig:LM1disc}, but for 
the LM6 mass spectrum (red solid lines) and the 
LM6' mass spectrum (blue dotted lines). }
\label{fig:LM6disc}}

In keeping with the main theme of this paper, 
in this subsection we shall concentrate on the third 
possibility, already suggested in Sec.~\ref{sec:1dim}.
We shall simply explore additional invariant mass endpoint
measurements, which would hopefully discriminate among
the two solutions for $m_B$.
Figs.~\ref{fig:LM1disc} and \ref{fig:LM6disc} 
show several invariant mass distributions which 
have already been mentioned at one point or another
in the course of our previous discussion.
Fig.~\ref{fig:LM1disc} shows the following 6 distributions:
(a) $m^2_{j\ell\ell}$;
(b) $m^2_{j\ell(hi)}$;
(c) $m^2_{j\ell(p)}$;
(d) $m^2_{j\ell(lo)}$;
(e) $m^2_{j\ell(s)}(\alpha=-1)$ and
(f) $m^2_{j\ell(s)}(\alpha=\frac{1}{2})$,
for the LM1 mass spectrum (red solid lines) and its LM1' 
counterpart (blue dotted lines). Fig.~\ref{fig:LM6disc}
shows the same 6 distributions, but for the 
LM6 and LM6' mass spectra. In both figures, we follow the 
same plotting conventions as in Figs.~\ref{fig:LM1meas} and \ref{fig:LM6meas}:
we form the mass squared distribution $dN/dm^2$, and then
plot versus the corresponding linear mass $m$ using a quadratic
scale on the $x$-axis. Notice that the sum of the 
$m^2_{j\ell(hi)}$ distribution in Fig.~\ref{fig:LM1disc}(b) (Fig.~\ref{fig:LM6disc}(b)) and the
$m^2_{j\ell(lo)}$ distribution in Fig.~\ref{fig:LM1disc}(d) (Fig.~\ref{fig:LM6disc}(d)) 
precisely equals the combined distribution
$m^2_{j\ell(u)}$ in Fig.~\ref{fig:LM1meas}(b) (Fig.~\ref{fig:LM6meas}(b)).
In order to be able to see this by the naked eye, we have kept the 
same $x$ and $y$ ranges on the corresponding plots.

As seen in Figs.~\ref{fig:LM1disc} and \ref{fig:LM6disc},
not all of the remaining invariant mass distributions 
are able to discriminate among the two $m_B$ solutions. 
As explained in Sec.~\ref{sec:1dim}, the suitable distributions
are those whose endpoints violate the symmetry (\ref{yzsymmetry}), 
which caused the $m_B$ ambiguity in the first place. 
For example, Figs.~\ref{fig:LM1disc}(a) and \ref{fig:LM6disc}(a)
show that the endpoint of the $m^2_{j\ell\ell}$ distribution
is the same for the real and the fake solution. This 
is to be expected, since the defining expression (\ref{jlldef}) 
for $m^{max}_{j\ell\ell}$ is symmetric under (\ref{yzsymmetry}).
Figs.~\ref{fig:LM1disc}(a) and \ref{fig:LM6disc}(a) also 
show that even the {\em shapes} of the $m^2_{j\ell\ell}$ distributions
for the real and fake solution are very similar.
In spite of this, the observation of the $m^2_{j\ell\ell}$ endpoint
can still be very useful, e.g. in reducing the experimental 
error on the mass determination.

Similar comments apply to the $m^2_{j\ell(hi)}$ distributions
shown in Figs.~\ref{fig:LM1disc}(b) and \ref{fig:LM6disc}(b). 
Here again the endpoint is a symmetric function of $R_{AB}$
and $R_{BC}$, and the real and fake solutions predict identical 
endpoints. However, while the endpoints are the same, this time
the shapes are not. The shape difference is more pronounced 
in the case of LM1 shown in Fig.~\ref{fig:LM1disc}(b), and less
visible in the case of LM6 shown in Fig.~\ref{fig:LM6disc}(b).

The remaining four distributions shown in 
Figs.~\ref{fig:LM1disc}(c-f) and \ref{fig:LM6disc}(c-f)
already have different endpoints and can thus be used 
for discrimination among the real and fake solution for $m_B$. 
All of the endpoints in Figs.~\ref{fig:LM1disc}(c-f) and \ref{fig:LM6disc}(c-f)
are relatively sharp and should be measured rather well.
One should not forget that in Figs.~\ref{fig:LM1disc} 
and \ref{fig:LM6disc} we show $m^2_{j\ell(s)}(\alpha)$ 
distributions for only three representative values of $\alpha$: 
$\alpha=-\infty$ in panels (d),
$\alpha=-1$ in panels (e), and 
$\alpha=0.5$ in panels (f).
As seen in Fig.~\ref{fig:alpha}, there are infinitely many
other choices for $\alpha$, which would still exhibit different 
endpoints for the real and fake $m_B$ solutions. 
Our conclusion is that through a suitable combination of 
additional endpoint measurements one would be able to tell 
apart the real solution for $m_B$ from its fake cousin.

\section{Summary and conclusions}
\label{sec:conclusions}

In this paper we revisited the classic technique for SUSY 
mass determinations via invariant mass endpoints.
We set out to redesign the standard algorithm for
performing these studies, by pursuing two main objectives
(see Section~\ref{sec:outline}):
\begin{itemize}
\item {\em Improving on the experimental precision of the
SUSY mass determination.} For example, we required that our 
analysis be based exclusively on {\em upper} 
invariant mass endpoints, which are expected to be 
measured with a greater precision than the corresponding
lower endpoints (a.k.a. thresholds). Consequently, we did 
not make use of the ``threshold'' measurement
$m_{j\ell\ell(\theta>\frac{\pi}{2})}^{min}$, which has
been an integral part of most SUSY studies since 
Ref.~\cite{Allanach:2000kt}. In the same vein, we also 
demanded that we should not rely on any
features observed in a two- or a three-dimensional 
invariant mass distribution --- such measurements 
are expected to be less precise than the (upper) endpoints
extracted from simple one-dimensional histograms. 
\item {\em Avoiding any parameter space region ambiguities.}
It is well known that some of the invariant mass endpoints 
used in the conventional analyses are piecewise-defined functions.
This feature may sometimes lead to multiple solutions 
for the SUSY mass spectrum in the ``LHC inverse problem''
\cite{Gjelsten:2004ki,ArkaniHamed:2005px,Gjelsten:2005sv,Gjelsten:2006as,Burns:2009zi}.
In order to safeguard against this possibility, we conservatively 
demanded from the outset that none of our endpoint measurements
be given by piecewise defined functions. This rather strict
requirement rules out three of the standard endpoint measurements
$m_{j\ell\ell}^{max}$, $m_{j\ell(lo)}^{max}$, and $m_{j\ell(hi)}^{max}$.
\end{itemize}

In order to meet these objectives, in Section~\ref{sec:variables}
we proposed a set of new invariant mass variables whose upper
kinematic endpoints can be alternatively used for SUSY mass
reconstruction studies. Then in Section~\ref{sec:analysis}
we outlined a simple analysis which was based on the particular 
set of four invariant mass variables (\ref{ourmeas}), all of which satisfy 
our requirements. In Section~\ref{sec:theory} we provided simple 
analytical formulas for the SUSY mass spectrum in terms of the 
four measured endpoints in eq.~(\ref{ourmeas}). Our solutions
revealed a surprise: in spite of the two-fold ambiguity 
(\ref{Mnear},\ref{Mfar}) in the interpretation of two of our 
endpoints $M^{max}_{j\ell(u)}$ and $m^{max}_{j\ell(u)}$,
the answer for three ($m_D$, $m_C$ and $m_A$) out of the 
four SUSY masses is unique! The fourth mass ($m_B$) is 
also known, up to the two-fold ambiguity (\ref{mBreplacement}),
which can be easily resolved by a variety of methods 
discussed and illustrated 
in Sections~\ref{sec:disambiguation} and \ref{sec:discrimination}.
In Section~\ref{sec:examples} we applied our technique to
two specific examples --- the LM1 and LM6 CMS study points.

Our method contains a number of elements which help in achieving
our two main objectives. For example, the precision 
of the SUSY mass determination is expected to improve, due
to the following factors:
\begin{enumerate}
\item {\em Precise knowledge of the whole shape of the invariant 
mass distribution.} In Appendix~\ref{app:shapes} we list 
the analytical expressions for {\em all} differential invariant mass
distributions used in our basic analysis from Section~\ref{sec:theory}: 
$m^2_{\ell\ell}$, $m^2_{j\ell(u)}$ and $m^2_{j\ell(s)}(1)$.
We also provide the corresponding expression for
the $m^2_{j\ell(d)}(1)$ distribution, whose upper endpoint 
offers an independent measurement of $M^{max}_{j\ell(u)}$
(see eq.~(\ref{mjld1max})). Finally, we also list the formula for
the differential distribution of $m^2_{jl(p)}$, whose endpoint can 
be used for selecting the correct $m_B$ solution, as shown in
Figs.~\ref{fig:LM1disc}(c) and \ref{fig:LM6disc}(c).
The knowledge of the shape of the whole distribution is indispensable
and greatly improves the accuracy of the endpoint extraction. 
In the absence of any analytical results like those in
Appendix~\ref{app:shapes}, one would be forced to use simple
linear extrapolations, which would lead to a significant systematic error.
\item {\em The number of available measurements tremendously exceeds 
the number of unknown mass parameters.} In principle, in order to 
extract 4 mass parameters, one needs a set of 4 measurements, 
for which we chose (\ref{ourmeas}). On the other hand, 
Section~\ref{sec:variables} contains a number of 
additional variables, whose endpoints will also be measured,
and possibly even better than our basic set (\ref{ourmeas}).
The addition of these extra measurements cannot hurt, and 
can only improve the overall accuracy of the SUSY mass 
determinations. 
\item {\em Improved precision on the endpoint measurements}.
Clearly, not all invariant mass variables will have their 
endpoints measured with exactly the same precision -- some endpoints 
will be measured better than others. This difference can be due to
many factors, e.g. the slope of the distribution near the endpoint,
the shape (convex versus concave) of the distribution near the endpoint,
the actual location of the endpoint, the level of SM and SUSY combinatorial 
background near the endpoint, etc. 
Our analysis in Sec.~\ref{sec:theory} was based on a specific set 
of 4 endpoint measurements (\ref{ourmeas}), which were chosen due
to the simplicity in their theoretical interpretation. However, 
these may not necessarily be {\em the best measured} endpoints.
In fact one can already anticipate from 
Figs.~\ref{fig:LM1disc} and \ref{fig:LM6disc}
that the endpoints of some of the 
$m_{j\ell\ell}$, $m_{j\ell(lo)}$, $m_{j\ell(p)}$
and $m_{j\ell(s)}(\alpha\ne 1)$ distributions
might be measured even better.
For example, the distributions in Figs.~\ref{fig:LM1disc}(c),
\ref{fig:LM1disc}(e) and \ref{fig:LM1disc}(f) are all steeper 
near their endpoints, compared to the distribution in
Fig.~\ref{fig:LM1meas}(c) that we used.
By the same token, one might expect that the endpoints in 
Figs.~\ref{fig:LM6disc}(a), \ref{fig:LM6disc}(d) and \ref{fig:LM6disc}(e)
will be measured more precisely than the upper endpoints of 
Figs.~\ref{fig:LM6meas}(b) and \ref{fig:LM6meas}(d).
\item {\em Controlled selection of an optimum set of measurements.} 
Notice that the variable $m^2_{j\ell(s)}(\alpha)$ defined in
eq.~(\ref{mjlsdef}) depends on a {\em continuous} parameter $\alpha$
whose value can be dialed up by the experimenter at will.
This has several advantages. For example, as we have seen in 
Fig.~\ref{fig:alpha}, the discriminating power of 
$m^2_{j\ell(s)}(\alpha)$ in rejecting the wrong solution in
(\ref{mBsol}) depends on the value of $\alpha$. Having obtained
a preliminary information about the two competing solutions, 
one can then choose the optimum value (or a range of values) 
for $\alpha$ for a subsequent study. Similarly, after the 
initial solution for the mass spectrum has been obtained, 
one can analyze by Monte Carlo the shapes of the $m^2_{j\ell(s)}(\alpha)$
distributions as a function of $\alpha$ and select for further study
specific values of $\alpha$ for which the corresponding endpoints 
$m^{max}_{j\ell(s)}(\alpha)$ are expected to be measured with 
a much better experimental precision. 
\end{enumerate}

In meeting our second objective, our method 
shows a certain improvement on the theoretical side as well:
\begin{enumerate}
\item {\em Reduced sensitivity to the parameter space region.}
All of the new variables introduced in Sec.~\ref{sec:variables}
exhibit milder sensitivity to the parameter space region, 
in comparison to the conventional endpoint $m^{max}_{j\ell\ell}$. 
As can be seen from the formulas in
Sec.~\ref{sec:variables}, the endpoint for each of our variables 
is given by at most two different expressions, as opposed to four 
in the case of $m^{max}_{j\ell\ell}$. A notable exception
is the variable $m_{j\ell(s)}(1)$, whose endpoint is actually uniquely 
predicted, and is independent of the parameter space region. 
We therefore strongly encourage the use of $m_{j\ell(s)}(1)$
in future analyses of SUSY mass determinations.
\item {\em Uniqueness of the solution.}
It is worth emphasizing that with only the 4 measurements of
eq.~(\ref{ourmeas}) we can already uniquely determine three out of the four 
masses involved in the problem. Then, the addition of a fifth measurement,
as discussed in Secs.~\ref{sec:1dim} and \ref{sec:discrimination},
is sufficient to pin down all four of the SUSY masses.
In contrast, with the conventional approach, one also starts with 
four measurements as in (\ref{4meas}), but in the worst case scenario
this results in infinitely many solutions, due to the linear dependence problem
(\ref{mjllcorrelation}) discussed in Section~\ref{sec:lindep}. 
Adding a fifth measurement as in (\ref{5meas}) helps, 
but once again, the worst case scenario leads to two alternative 
solutions \cite{Burns:2009zi}.
In order to resolve the remaining duplication, and thus guarantee uniqueness 
of the solution under any circumstances, one needs at least 6 measurements.
\end{enumerate}

In conclusion, our main accomplishment in this paper was to expand the
experimenter's arsenal with several new tools 
which can be used for SUSY mass determinations via kinematic endpoints.
We believe that the variables suggested in Section~\ref{sec:variables} 
and the shapes of their distributions listed in Appendix~\ref{app:shapes}
will eventually find their way into the actual experimental analyses 
after the discovery of SUSY (or any other new physics exhibiting the 
decay chain of Fig.~\ref{fig:chain}). 

\acknowledgments
We thank M.~Burns for collaboration in the early stages of this project.
This work is supported in part by a US Department of Energy 
grant DE-FG02-97ER41029. 

\appendix
\section{Appendix: \ Analytical expressions for the shapes of the invariant mass distributions}
\label{app:shapes}
\allowdisplaybreaks
\renewcommand{\theequation}{A.\arabic{equation}}
\setcounter{equation}{0}

In this appendix we will provide the analytical expressions for the 
shapes of the invariant mass distributions $m^2_{\ell\ell}$, 
$m^2_{j\ell(u)}\equiv m^2_{j\ell_n}\cup m^2_{j\ell_f}$,
$m^2_{j\ell(s)}(1)\equiv m^2_{j\ell_n}+m^2_{j\ell_f}$,
$m^2_{j\ell(d)}(1)\equiv |m^2_{j\ell_n}-m^2_{j\ell_f}|$, and
$m^2_{jl(p)}$.
To simplify the expressions, we introduce the shorthand notation for the 
corresponding endpoints, which was already introduced in 
(\ref{shorthand}), (\ref{condmjln}), (\ref{fdef1}) and (\ref{pdef1}):
\bea
L &\equiv & \left(m_{\ell\ell}^{max}\right)^2      = m_D^2\, R_{CD}\, (1-R_{BC})\, (1-R_{AB}), \\ [2mm]
n &\equiv & \left(m_{j\ell_n}^{max}\right)^2       = m_D^2\, (1-R_{CD})\, (1-R_{BC}),  \label{ndef} \\ [2mm]
f &\equiv & \left(m_{j\ell_f}^{max}\right)^2       = m_D^2\, (1-R_{CD})\, (1-R_{AB}),     \label{fdef}\\ [2mm]
p &\equiv & R_{BC}\, f                             = m_D^2\, (1-R_{CD})\, R_{BC}\,(1-R_{AB}). \label{pdef}
\eea
In this appendix, we shall ignore spin correlations and consider
only pure phase space decays. General results including 
spin correlations for $m^2_{\ell\ell}$, $m^2_{j\ell_n}$
and $m^2_{j\ell_f}$ exist and can be found in \cite{Burns:2008cp}.
We shall unit-normalize the $m^2_{\ell\ell}$, $m^2_{j\ell(s)}$, 
$m^2_{j\ell(d)}$ and $m^2_{j\ell(p)}$ distributions,
to which each event contributes a single entry. In contrast, the 
union distribution $m^2_{j\ell(u)}$ has two entries per event, so it
will be normalized to 2 instead. It is also convenient to
write the distributions in terms of masses squared instead of linear masses.
Of course, the two are trivially related by
\beq
\frac{dN}{d m} = 2 m \frac{dN}{d m^2}\ .
\eeq

\subsection{Dilepton mass distribution $m^2_{\ell\ell}$}

The differential dilepton invariant mass distribution is given by
\beq
\frac{dN}{d m^2_{\ell\ell}} = \frac{1}{L}\ ,
\eeq
which is unit-normalized:
\beq
\int_0^L dm^2_{\ell\ell} \left(\frac{dN}{d m^2_{\ell\ell}}\right) = 1\ .
\eeq

\subsection{Combined jet-lepton mass distribution $m^2_{j\ell(u)}$}

The differential distribution for $u\equiv m^2_{j\ell(u)}$ is given by
\beq
\frac{dN}{d u} 
= \theta\left(n-u\right)\, \theta\left(u  \right)\, \frac{1}{n} 
+ \theta\left(p-u\right)\, \theta\left(u  \right)\, \frac{\ln (f/p)}{f-p}   
+ \theta\left(f-u\right)\, \theta\left(u-p\right)\, \frac{\ln (f/u)}{f-p}  \, ,
\label{dNdu}
\eeq
where $\theta(x)$ is the usual Heaviside step function
\beq
\theta(x) \equiv \left\{ 
\begin{array}{ll}
1, & ~x\ge 0, \\[4mm] 
0, & ~x<  0.
\end{array}%
\right .
\eeq
It is easy to verify the normalization condition
\beq
\int_0^{M} du \left( \frac{dN}{d u} \right) = 2,
\eeq
where $M\equiv (M^{max}_{j\ell(u)})^2$ was already defined in (\ref{shorthand}).

\FIGURE[ht]{
\epsfig{file=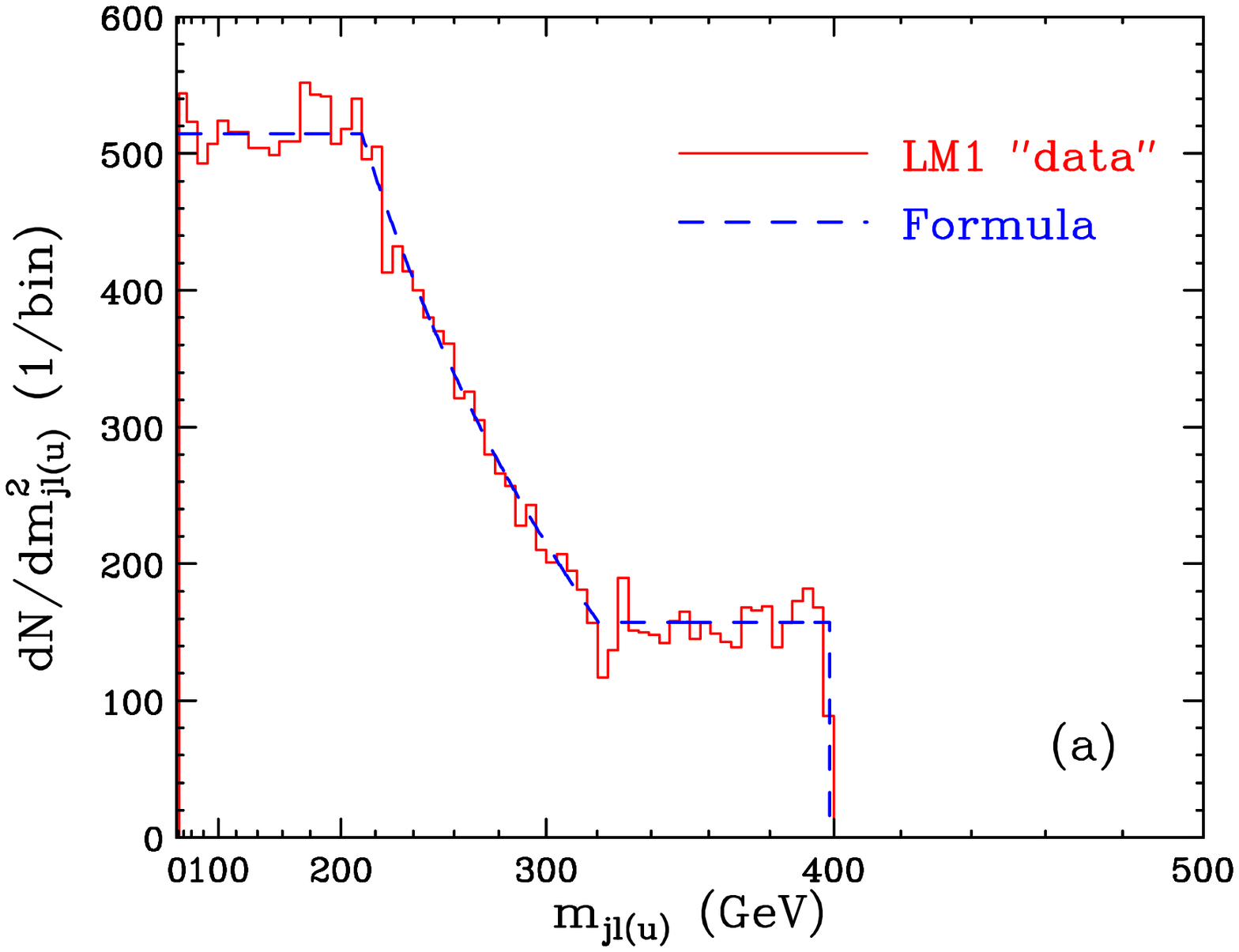,width=7.0cm}~~~
\epsfig{file=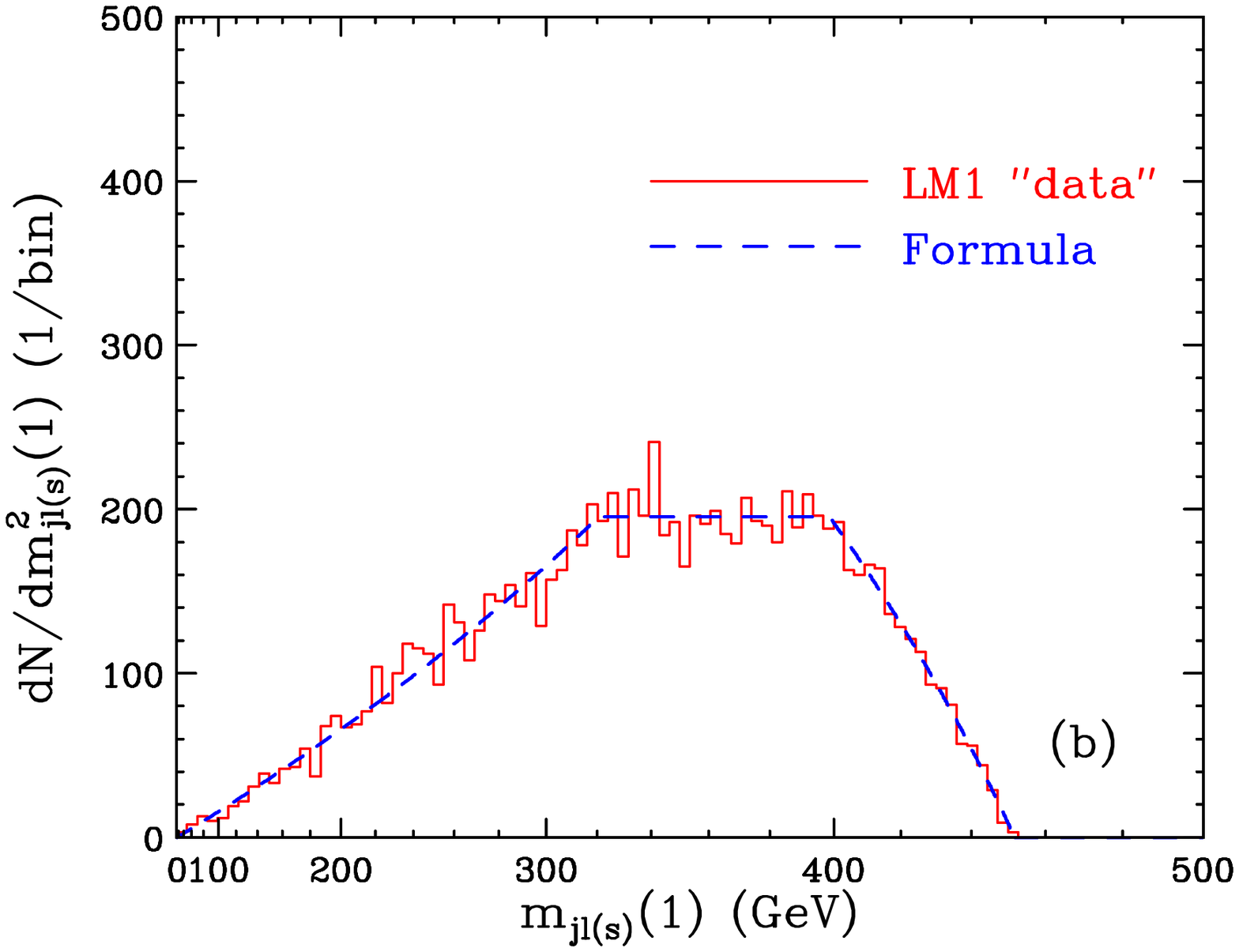,width=7.0cm}\\
\epsfig{file=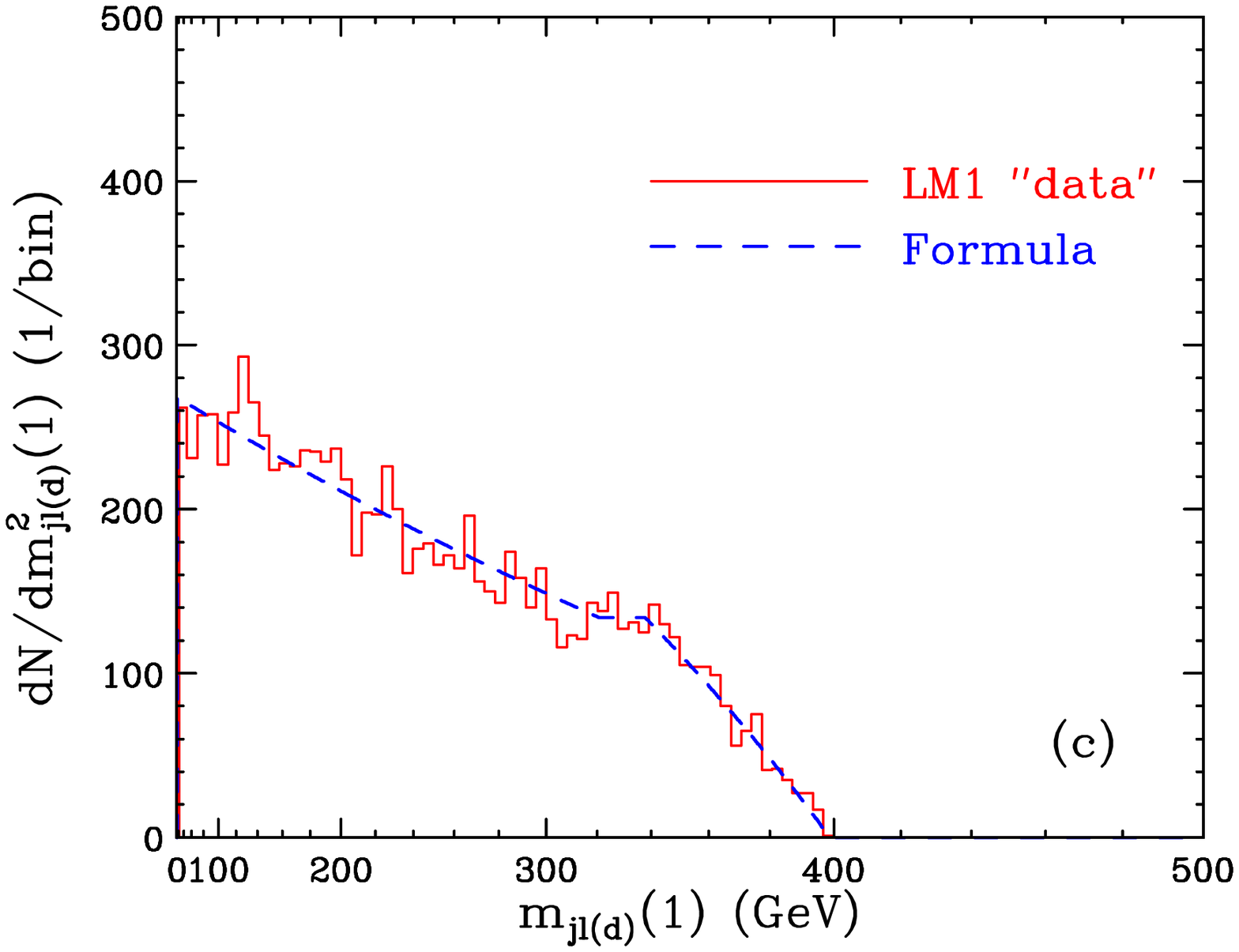,width=7.0cm}~~~
\epsfig{file=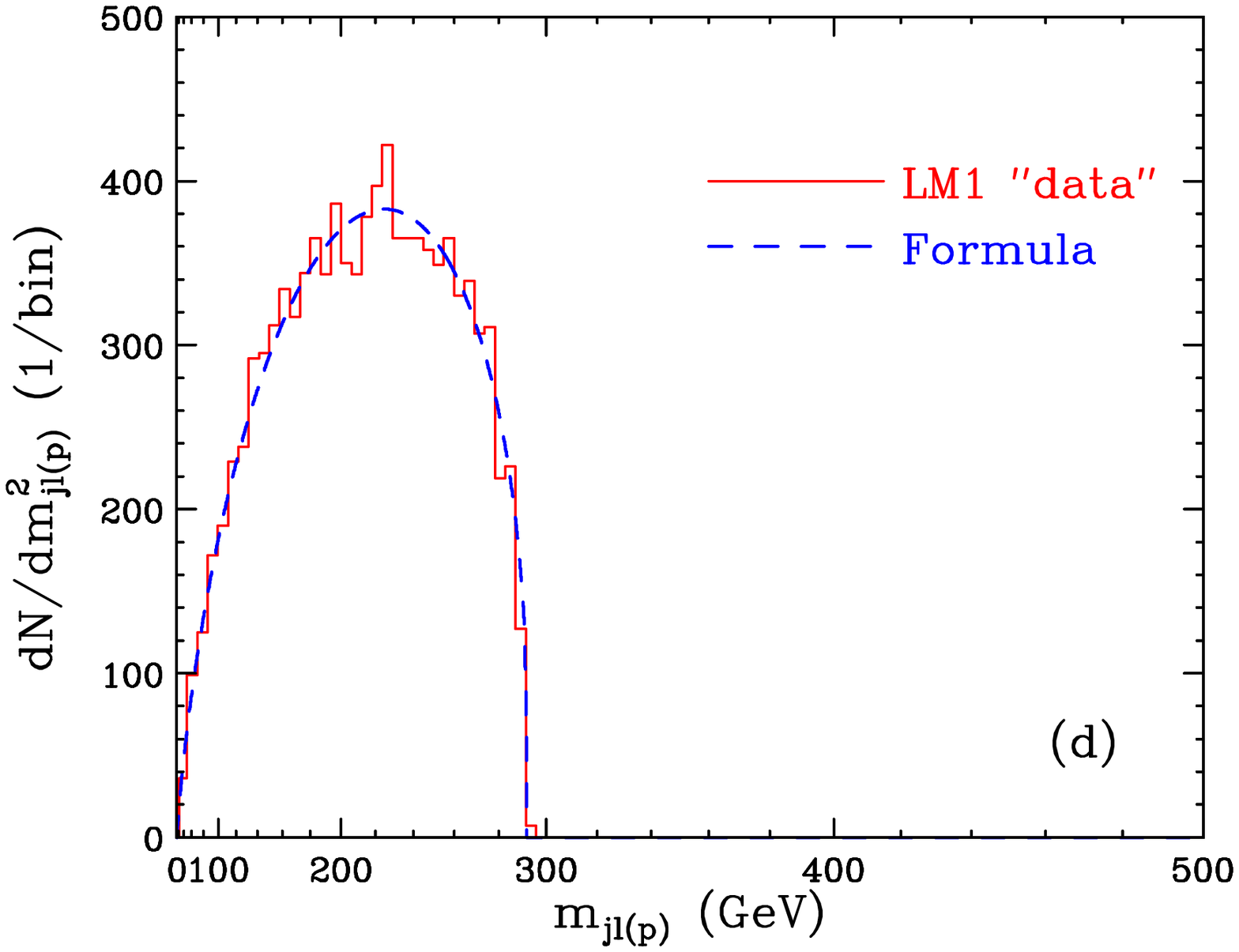,width=7.0cm}
\caption{\sl Comparison of the numerically obtained 
differential invariant mass distributions for study point LM1
(red solid lines) with the analytical results 
presented in this appendix (blue dashed lines):
(a) the distribution of the combined jet-lepton mass $u\equiv m^2_{j\ell(u)}$ 
from Fig.~\ref{fig:LM1meas}(b) versus the analytical prediction of eq.~(\ref{dNdu});
(b) the distribution of the sum $\sigma\equiv m^2_{j\ell(s)}(\alpha=1)$
from Fig.~\ref{fig:LM1meas}(c) versus the analytical prediction of eq.~(\ref{dNdsigma});
(c) the distribution of the difference $\Delta \equiv m^2_{j\ell(d)}(\alpha=1)$ 
from Fig.~\ref{fig:LM1meas}(d) versus the analytical prediction of eqs.~(\ref{dNdDelta1}-\ref{dNdDelta5});
(d) the distribution of the product $\rho \equiv m^2_{jl(p)}$
from Fig.~\ref{fig:LM1disc}(c) versus the analytical prediction of eqs.~(\ref{dNdrho1}-\ref{dNdrho2}).
}
\label{fig:formLM1}}

In Fig.~\ref{fig:formLM1}(a) we cross-check the prediction of eq.~(\ref{dNdu})
(blue dashed line) with the numerically obtained $m^2_{j\ell(u)}$ 
distribution in Fig.~\ref{fig:LM1meas}(b) (red solid line), 
for the case of study point LM1. We see that within the statistical errors, 
our formula is in perfect agreement with the numerical result.

\subsection{Distribution of the sum $m^2_{j\ell(s)}(\alpha=1)$}

The differential distribution for $\sigma\equiv m^2_{j\ell(s)}(\alpha=1)$ is given by
\bea
\frac{dN}{d \sigma} 
= \frac{1}{f-p} &\Biggl\{& \theta(m-\sigma)\,\theta(\sigma) 
\ln\left(\frac{fn}{fn-\sigma(f-p)}\right) \nonumber  \\ [2mm]
&+& \theta(M-\sigma)\,\theta(\sigma-m) 
\ln\left(\frac{M}{M-(f-p)}\right) \nonumber  \\ [2mm]
&+& \theta(n+p-\sigma)\,\theta(\sigma-M)
\ln\left(\frac{fn-\sigma(f-p)}{p(n+p-f)}\right) \Biggr\}\, ,
\label{dNdsigma}
\eea
where $m\equiv (m^{max}_{j\ell(u)})^2$ was defined in (\ref{shorthand}),
and $n$, $f$ and $p$ were defined in (\ref{ndef}-\ref{pdef}).
The normalization condition for (\ref{dNdsigma}) reads
\beq
\int_0^{S} d\sigma \left( \frac{dN}{d \sigma} \right) = 1\, ,
\eeq
where $S$ is defined in (\ref{shorthand}).

As a cross-check, Fig.~\ref{fig:formLM1}(b) shows that our analytical formula in
eq.~(\ref{dNdsigma}) agrees with the numerical result from Fig.~\ref{fig:LM1meas}(c)
for the LM1 study point.

\subsection{Distribution of the difference $m^2_{j\ell(d)}(\alpha=1)$}

The differential distribution for the difference
$\Delta \equiv m^2_{j\ell(d)}(\alpha=1)$ depends on the 
values of $R_{BC}$ and $R_{AB}$. To simplify the notation, 
we define an antisymmetric function
\beq
L(x,y) = - L(y,x) \equiv \ln\left( \frac{nf + x(f-p)}{nf+y(f-p)}\right)\, ,
\eeq
which we heavily use in writing down the result for the differential 
$\Delta$ distribution. Notice that there are various equivalent ways to 
write down these formulas, due to the transitivity property
\beq
L(x,y) + L(y,z) = L(x,z)\ .
\eeq

For $\Delta \equiv m^2_{j\ell(d)}(\alpha=1)$  one needs to consider
five separate cases:\\

\noindent
If $\frac{2}{3-R_{AB}}\le R_{BC} < 1$, then
\bea
\frac{dN}{d \Delta} 
= \frac{1}{f-p} &\Biggl\{& \theta(n-\Delta)\,\theta(\Delta) 
\Bigl[ L(0,-n)+L(-\Delta,-n) \Bigr] \nonumber \\ [2mm]
&+& \theta(p-n-\Delta)\,\theta(\Delta-n)\, L(0,-n) \nonumber \\ [2mm]
&+& \theta(f-\Delta)\,\theta(\Delta-(p-n))\, L(f,\Delta) \Biggr\}\ .
\label{dNdDelta1}
\eea
\noindent
If $\frac{1}{2-R_{AB}}\le R_{BC} < \frac{2}{3-R_{AB}}$, then
\bea
\frac{dN}{d \Delta} 
= \frac{1}{f-p} &\Biggl\{& \theta(p-n-\Delta)\,\theta(\Delta) 
\Bigl[ L(0,-n)+L(-\Delta,-n) \Bigr] \nonumber \\ [2mm]
&+& \theta(n-\Delta)\,\theta(\Delta-(p-n))\, \Bigl[ L(f,\Delta)+L(-\Delta,-n) \Bigr] \nonumber \\ [2mm]
&+& \theta(f-\Delta)\,\theta(\Delta-n)\, L(f,\Delta) \Biggr\}\ .
\label{dNdDelta2}
\eea
\noindent
If $R_{AB}\le R_{BC} < \frac{1}{2-R_{AB}}$, then
\bea
\frac{dN}{d \Delta} 
= \frac{1}{f-p} &\Biggl\{& \theta(n-p-\Delta)\,\theta(\Delta) 
\Bigl[ L(f,\Delta)+L(f,0) \Bigr] \nonumber \\ [2mm]
&+& \theta(n-\Delta)\,\theta(\Delta-(n-p))\, \Bigl[ L(f,\Delta)+L(-\Delta,-n) \Bigr] \nonumber \\ [2mm]
&+& \theta(f-\Delta)\,\theta(\Delta-n)\, L(f,\Delta) \Biggr\}\ .
\label{dNdDelta3}
\eea
\noindent
If $\frac{R_{AB}}{2-R_{AB}}\le R_{BC} < R_{AB}$, then
\bea
\frac{dN}{d \Delta} 
= \frac{1}{f-p} &\Biggl\{& \theta(n-p-\Delta)\,\theta(\Delta) 
\Bigl[ L(f,\Delta)+L(f,0) \Bigr] \nonumber \\ [2mm]
&+& \theta(f-\Delta)\,\theta(\Delta-(n-p))\, \Bigl[ L(f,\Delta)+L(-\Delta,-n) \Bigr] \nonumber \\ [2mm]
&+& \theta(n-\Delta)\,\theta(\Delta-f)\, L(-\Delta,-n) \Biggr\}\ .
\label{dNdDelta4}
\eea
\noindent
If $0 \le R_{BC} < \frac{R_{AB}}{2-R_{AB}}$, then
\bea
\frac{dN}{d \Delta} 
= \frac{1}{f-p} &\Biggl\{& \theta(f-\Delta)\,\theta(\Delta) 
\Bigl[ L(f,\Delta)+L(f,0) \Bigr] \nonumber \\ [2mm]
&+& \theta(n-p-\Delta)\,\theta(\Delta-f)\,  L(f,0) \nonumber \\ [2mm]
&+& \theta(n-\Delta)\,\theta(\Delta-(n-p))\, L(-\Delta,-n) \Biggr\}\ .
\label{dNdDelta5}
\eea
The normalization condition now reads
\beq
\int_0^{M} d\Delta \left( \frac{dN}{d \Delta} \right) = 1\, .
\eeq

As before, in Fig.~\ref{fig:formLM1}(c) we compare the prediction of our analytical formula 
in eqs.~(\ref{dNdDelta1}-\ref{dNdDelta5}) to the numerical result
obtained earlier in Fig.~\ref{fig:LM1meas}(d) for the LM1 study point, and we
find very good agreement.

\subsection{Distribution of the product $m^2_{j\ell(p)}$}

Finally, for completeness we also list the differential distribution for the 
product variable (\ref{mjlpdef}), for which here we shall use the shorthand notation
$\rho \equiv m^2_{jl(p)}$. To further simplify the notation, we define the function
\beq
X_{\pm}(\rho) \equiv \frac{\sqrt{n}}{2 (f-p)} \Big( \sqrt{n}f\pm\sqrt{f^2 n+4(p-f) \rho^2}\Big),
\eeq
where $n$, $f$ and $p$ are defined as before in (\ref{ndef}-\ref{pdef}).
There are two separate cases:\\

\noindent
If $R_{BC} \le 0.5$, the $\rho$ distribution is made up of two branches
joining at $\rho=\sqrt{n\,p}$ (see, for example the LM1 distribution in 
Fig.~\ref{fig:LM1disc}(c) and the LM6' distribution in Fig.~\ref{fig:LM6disc}(c))
\bea
\frac{dN}{d \rho} 
= \frac{2\,\rho}{n\,f} &\Biggl\{& \theta\left(\sqrt{n\,p}-\rho\right)\,\theta(\rho) \,
\Bigl[ \ln\left(\frac{n}{p}\right)+2 \ln\left(\frac{\rho}{X_{-}(\rho)}\right) \Bigr] \nonumber \\ [2mm]
&+& \theta\left(\frac{f\sqrt{n}}{2\sqrt{f-p}}-\rho\right)\,\theta(\rho-\sqrt{n\,p})\
2 \ln\left(\frac{X_{+}(\rho)}{X_{-}(\rho)}\right) \Biggr\}.
\label{dNdrho1}
\eea
\noindent
If $R_{BC} \ge 0.5$, there is a single branch, as illustrated by the LM1' distribution in 
Fig.~\ref{fig:LM1disc}(c) and the LM6 distribution in Fig.~\ref{fig:LM6disc}(c):
\beq
\frac{dN}{d \rho} 
= \frac{2\,\rho}{n\,f}\,  \theta(\sqrt{n\,p}-\rho)\,\theta(\rho) 
\Biggl\{ \ln\left(\frac{n}{p}\right)+2 \ln\left(\frac{\rho}{X_{-}(\rho)}\right) \Biggr\} .
\label{dNdrho2}
\eeq
In both of those cases, the normalization condition is
\beq
\int_0^{\rho^{max}} d\rho \left( \frac{dN}{d \rho} \right) = 1\, ,
\eeq
where $\rho^{max}$ is the corresponding $m^2_{jl(p)}$ endpoint defined in
(\ref{mjlpmaxdef}).

Fig.~\ref{fig:formLM1}(d) demonstrates that our analytical result (\ref{dNdrho1}) 
agrees well with the numerically derived 
$m^2_{jl(p)}$ distribution in Fig.~\ref{fig:LM1disc}(c) for the LM1 study point.

\end{document}